\documentclass[11pt,a4paper]{article}

\usepackage{t1enc}
\usepackage[utf8]{inputenc}
\usepackage[english]{babel}
\normalfont
\usepackage{amsmath}
\usepackage{amssymb}
\usepackage{amsthm}

\usepackage{mathabx}
\usepackage{bbm}
\usepackage{mathrsfs}
\usepackage{pifont}
\usepackage{hyperref}
\usepackage{pgf}
\usepackage{graphicx}
\usepackage{tikz-cd}

\newcommand{\be}[0]{\begin{equation}}
\newcommand{\ee}[0]{\end{equation}}
\newcommand{\btkz}[0]{\begin{tikzcd}}
\newcommand{\etkz}[0]{\end{tikzcd}}

\numberwithin{equation}{section}

\theoremstyle{plain}
\newtheorem{theorem}{Theorem}[section]

\newtheorem{proposition}[theorem]{Proposition}

\theoremstyle{definition}

\def\be{ \begin{equation} }
\def\ee{ \end{equation}}

\def\ch{{\rm ch}}

\def\cot{{\rm cot}}
\def\det{{\rm det}}
\def\dim{{\rm dim}}

\def\exp{{\rm exp}}

\def\I{{\rm i}}

\def\Tr{{\rm Tr}}

\def\one{{\hbox{ 1\kern-.8mm l}}}

\def\IR{{\mathbb{R}}}

\def\rmk#1{\bigskip\noindent{\bf Remarks} }

\begin{document}


\begin{center}
{\Large
{\bf Remarks on the Green-Schwarz terms of six-dimensional supergravity theories}}
\vspace{2.0cm}

{\large Samuel Monnier$^1$, Gregory W. Moore$^2$}
\vspace*{0.5cm}

$^1$ Section de Math\'ematiques, Universit\'e de Gen\`eve \\
2-4 rue du Li\`evre, 1211 Gen\`eve 4, Switzerland\\
samuel.monnier@gmail.com\\
\vspace*{0.5cm}

$^2$NHETC and Department of Physics and Astronomy\\
Rutgers University\\
Piscataway, NJ 08855, USA\\
gmoore@physics.rutgers.edu

\vspace*{1cm}

{\bf Abstract}
\end{center}

We construct the Green-Schwarz terms of six-dimensional supergravity theories on spacetimes with non-trivial topology and gauge bundle. We prove the cancellation of all global gauge and gravitational anomalies for theories with gauge groups given by products of $U(n)$, $SU(n)$ and $Sp(n)$ factors, as well as for $E_8$. For other gauge groups, anomaly cancellation is equivalent to the triviality of a certain 7-dimensional spin topological field theory. We show in the case of a finite Abelian gauge group that there are residual global anomalies imposing constraints on the 6d supergravity. These constraints are compatible with the known F-theory models. Interestingly, our construction requires that the gravitational anomaly coefficient of the 6d supergravity theory is a characteristic element of the lattice of string charges, a fact true in six-dimensional F-theory compactifications but that until now was lacking a low-energy explanation. We also discover a new anomaly coefficient associated with a torsion characteristic class in theories with a disconnected gauge group.

\newpage

\tableofcontents

\section{Introduction and summary}

Supergravity theories in six dimension contain anomalous chiral fermions and self-dual fields.
Anomaly cancellation imposes strong constraints on the allowed field content of these theories.
 The constraints coming from local anomaly cancellation are well-understood, see for instance \cite{TaylorTASI}. The constraints imposed by global anomaly cancellation are more elusive: global gauge anomaly cancellation has been used in \cite{KMT2, Monnier:2017oqd} to derive constraints on the anomaly coefficients of the theory, but little is known beyond these results. The present work is a step toward a more systematic understanding of global anomalies in 6d supergravity.

The local anomalies of 6d supergravity theories are canceled through a generalization of the Green-Schwarz mechanism. The degree 8 anomaly polynomial $A_8$ of the theory is required to factorize as
\be
\label{EqIntroFactAn}
A_8 = \frac{1}{2} Y \wedge Y
\ee
where $Y$ is a 4-form valued in the Lie algebra $\Lambda_\mathbb{R}$ of   Abelian gauge group of the chiral two-form
gauge potentials. Moreover, $\wedge$ denotes the wedge product of forms tensored with the inner product in $\Lambda_\mathbb{R}$. The Bianchi identity of the field strength $H$ of the self-dual 2-forms is modified to
\be
\label{EqIntroModBianchi}
dH = Y \;,
\ee
and a Green-Schwarz term
\be
\label{EqIntroGSTerm}
\frac{1}{2} \int B \wedge Y
\ee
is added to the action. The standard lore described above is satisfactory for flat spacetimes with trivial gauge bundles, but cannot accommodate non-trivial topologies. For instance the chiral bosonic fields can be described by a 2-form only locally.

In order to obtain a more general definition of the Green-Schwarz terms, it is useful to understand better the anomaly they are supposed to cancel. The anomalies of a $d$-dimensional field theory $\mathcal{F}$ are best pictured as a field theory in dimension $d+1$, the anomaly field theory $\mathcal{A}$ \cite{Freed:2014iua}. For the case of interest to us, $\mathcal{A}$ is \emph{invertible} \cite{Freed:2004yc}, which implies in particular that its partition function is non-vanishing and that its state space has dimension 1. The partition function $\mathcal{F}(M)$ of $\mathcal{F}$ on a $d$-dimensional (Euclidean) spacetime $M$, instead of being a complex number, is an element of $\mathcal{A}(M)$, the one-dimensional Hilbert space/Hermitian line associated by the anomaly field theory to $M$. $\mathcal{F}(M)$ can be seen as a complex number only in a non-canonical way, by picking an isomorphism $\mathcal{A}(M) \simeq \mathbb{C}$. Moreover, if a global symmetry of the background data acts non-trivially on $\mathcal{A}(M)$, $\mathcal{F}(M)$ will transform by a phase: $\mathcal{F}$ has an anomaly with respect to the global symmetry \cite{Monnierd}.

We can now understand conceptually the nature of the Green-Schwarz terms. An exponentiated Green-Schwarz term is a vector in $\mathcal{A}(M)^\dagger$, the Hilbert space complex conjugate to $\mathcal{A}(M)$ \cite{Witten:1999eg}. Adding the Green-Schwarz term to the action amounts to tensoring the partition function with the exponentiated Green-Schwarz term to obtain an element of $\mathcal{A}(M) \otimes \mathcal{A}(M)^\dagger$. The new partition function now takes its value in a Hilbert space canonically isomorphic to $\mathbb{C}$, on which all the symmetries obviously act trivially: the anomalies have been canceled. Implementing the Green-Schwarz mechanism therefore decomposes into two steps:
\begin{enumerate}
\item Identify the anomaly field theory $\mathcal{A}$ of the field theory $\mathcal{F}$ whose anomalies have to be canceled.
\item Construct from the field theory data a vector in $\mathcal{A}(M)^\dagger$ for every spacetime $M$. In order to be able to recast this vector as a Green-Schwarz term, it should take the form of the exponential of an action depending \emph{locally} on the fields of the theory. The Lagrangian of this action is then the Green-Schwarz term to be added to the original action. It is crucial that this vector is constructed from the field theory data, as other constraints such as supersymmetry do not allow us to modify the field content of the theory.
\end{enumerate}

Returning to our original problem in six-dimensional supergravity theories, the anomalies are due to chiral fermions and self-dual fields. Accordingly, the anomaly field theory $\mathcal{A}$ is a product of certain Dai-Freed theories \cite{Dai:1994kq, Freed:2016rqq}, whose partition functions are given by eta invariants of Dirac operators. These theories are difficult to work with, partly because they lack an action principle in terms of fields.

Equations \eqref{EqIntroModBianchi} and \eqref{EqIntroGSTerm} however suggest the following picture. Recall that given a classical action in dimension $d$, there is always a "prequantum" invertible field theory \cite{Freed:1991bn} associated to it, whose partition function is the exponentiated action and whose state space on a $d-1$-dimensional manifold $M$ is the space of boundary values of the action on $d$-dimensional manifolds $U$ such that $\partial U = M$. The latter is generally a Hermitian line non-canonically isomorphic to $\mathbb{C}$. Under (small) gauge transformations, \eqref{EqIntroGSTerm} transforms like the boundary value of a Chern-Simons theory with action
\be
\label{EqNaiveCSAction}
\frac{1}{2} \int A \wedge Y \;,
\ee
where $A$ is a degree 3 Abelian gauge field with field strength $Y$. This suggests that the exponentiated Green-Schwarz term should be a vector in the state space of the prequantum Chern-Simons theory defined by the action \eqref{EqNaiveCSAction}. But problems arise when considering large gauge transformations. \eqref{EqNaiveCSAction} is a Chern-Simons action at half-integer level, so the exponentiated action is not gauge invariant: it can transform by a sign under large gauge transformations.

Fortunately, one can make sense, under certain circumstances, of Chern-Simons theories at half-integer level. In dimension 3, the \emph{spin Chern-Simons theories} \cite{Dijkgraaf:1989pz, 2005math......4524J, 2006math......5239J, Belov:2005ze} have half-integer level and are well-defined on spin manifolds. They play a central role in the effective description of the quantum Hall effect. Their generalizations in dimension $4k+3$, the \emph{Wu Chern-Simons theories}, have been defined and studied recently in \cite{Monnier:2016jlo}. We show indeed that \eqref{EqIntroGSTerm} can be given a precise meaning for topologically non-trivial fields, and that it defines a vector in a Wu Chern-Simons theory. Because of a certain shift in the background field, we call this theory the shifted Wu Chern-Simons (sWCS) theory. Wu Chern-Simons theories require spacetimes endowed with Wu structures, which are higher degree relatives of spin structures. The existence of a Wu structure does not impose constraints on the spacetime in the dimensions of interest to us, and we show that the dependence on the choice of Wu structure drops out thanks to the aforementioned shift. The Green-Schwarz term can therefore really be defined from the 6d supergravity data only.

An interesting feature of the Green-Schwarz mechanism in dimension 6 is that the 2-forms involved are themselves anomalous. This implies additional constraints on the Green-Schwarz term. First, it has to be gauge invariant under the gauge transformation of the anomalous fields (here the self-dual 2-forms). Second, its variation under the non-anomalous fields' gauge transformations has to be independent of the anomalous fields. We discuss the reasons for these constraints in more detail in Section \ref{SecImplSixSugra}. Nontrivially, our construction automatically satisfies these extra constraints.

If the Green-Schwarz term is to cancel the anomaly, we need the shifted Wu Chern-Simons theory to be isomorphic to the complex conjugate of the 6d supergravity anomaly field theory. We show that they are isomorphic up to a bordism invariant of $\Omega^{\rm spin}_7(BG)$  where $G$ is here the vectormultiplet gauge group of the 6-dimensional supergravity theory. It can be any compact Lie group, possibly disconnected.
We show that for $G$ a product of $U(n)$, $SU(n)$ and $Sp(n)$ factors or $G = E_8$, $\Omega^{\rm spin}_7(BG)$ vanishes and the two theories coincide, ensuring the cancellation of all anomalies, local and global. More generally,
the product of the anomaly field theory and the shifted Wu-Chern-Simons theory is a 7-dimensional spin topological field theory $Z_{top}$ whose partition
function is a homomorphism $\Omega^{\rm spin}_7(BG) \to U(1)$.
It seems that the computation of $\Omega^{\rm spin}_7(BG)$ has to be performed gauge group by gauge group, a rather daunting task. There are gauge groups for which $\Omega^{\rm spin}_7(BG) \neq 0$. Indeed, a principal $O(n)$-bundle $P$ over $\mathbb{R}P^7$ with non-trivial first Stiefel-Whitney class provides an example, as $\int w_1(P)^7 = 1$ is a bordism invariant.  Anomaly cancellation using the Green-Schwarz term we construct will only work
if $Z_{top}$ is trivial. Unfortunately, we do not know effective techniques for computing the partition function of $Z_{top}$.
 Finding such effective computational techniques remains an important open problem for the future.

Non-trivial bordism groups also arise when $G$ is finite Abelian. In this case, we can compute the difference between the partition functions of the anomaly field theories of two supergravity theories differing by their matter representations. Computing the partition functions of the associated shifted Wu Chern-Simons theories seems like a difficult task, but we do find  constraints imposed by global anomaly cancellation on the difference of the matter representation. Encouragingly, these constraints are satisfied in known supergravity theories obtained through F-theory.\footnote{It is claimed in the literature that all the six-dimensional supergravity theories realizable in string theory admit a construction in F-theory, see for instance p.77 in \cite{TaylorTASI}.} An important finding of the present paper is therefore that global anomaly cancellation imposes constraints on 6-dimensional supergravity theories beyond the currently known ones.

Along the way, we discover a new anomaly coefficient in theories with a disconnected gauge group. In the standard construction, the anomaly coefficients parametrize the 4-form $Y$ factorizing the degree 8 anomaly polynomial of the theory and acting as a string charge source. When putting the theory on a topologically non-trivial manifold, this 4-form should be promoted to a degree 4 differential cocycle,
the 4-form being the field strength of an Abelian degree 3 gauge field. There can exist non-equivalent differential cocycles that nevertheless share the same field strength. In particular, this is the case for differential cocycles whose topological classes differ by a torsion cohomology class. For a disconnected gauge group, there is precisely a degree 4 torsion characteristic class that can be added to the differential cocycle lifting $Y$. The coefficient of this torsion characteristic class is a new anomaly coefficient, valued in the lattice of string charges $\Lambda$. Because it is associated to a torsion class, it does not appear in $Y$, which is why it has not been noticed until now.

Our construction has other interesting implications for six-dimensional supergravity theories. The field strength of the background field appearing in the Wu Chern-Simons theory is $Y$, as defined in \eqref{EqIntroFactAn}. $Y$ is parametrized by the \emph{anomaly coefficients} of the six-dimensional supergravity theory (see \eqref{EqRefEffAbGaugFFT}), and the consistency of our construction imposes constraints on these anomaly coefficients. As far as the gauge anomaly coefficients are concerned, we recover the strongest constraints obtained in \cite{Monnier:2017oqd}. More interestingly, we find that the gravitational anomaly coefficient has to be a characteristic element of the string charge lattice of the 6d supergravity theory. This constraint is always satisfied in F-theory, but it was unclear until now whether it could arise from low-energy considerations. This constraint excludes 6d supergravity theories that otherwise look perfectly consistent, see the discussion in Section \ref{SecConstrAnCoeff}. Similarly, the consistency of the construction requires the string charge lattice $\Lambda$ of the 6-dimensional supergravity theory to be unimodular, a fact derived previously using reduction to two dimensions \cite{SeibergTaylor}.

Finally, we should end with some words of caution. First, 6d supergravity theories admit self-dual string defects. Tadpole cancellation require the inclusion of such defects in most backgrounds. Our results about anomaly cancellation are conditional on the worldsheet anomalies on the self-dual strings cancelling against anomaly inflow from the supergravity. In the present paper, we simply assume that the cancellation occurs, but conceivably it could lead to further constraints on the supergravity theory or its backgrounds. The anomaly inflow on string defects has been studied in six-dimensional superconformal field theories in \cite{Henningson:2004dh, Berman:2004ew, Kim:2016foj, Shimizu:2016lbw}

Second, when extrapolating an Abelian gauge theory from flat spacetimes to manifolds of arbitrary topology, one has to choose a generalized cohomology theory describing the topologically non-trivial configurations of the gauge fields. The gauge fields are then modeled as cocycle representatives of classes in the corresponding differential cohomology theory \cite{Freed:2000ta, Freed:2006yc}. The relevant generalized cohomology theory is not necessarily the familiar ordinary (integral) cohomology. For example, the topology of the Ramond-Ramond gauge fields of type II string theory are well-known to be described by K-theory \cite{Minasian:1997mm, Witten:1998cd}. A similar choice arises for six-dimensional supergravity theories. In the present work, we make the assumption that the self-dual fields in the gravitational and tensor multiplets are described by (ordinary) differential cohomology. However, the anomaly field theory is expressed in terms of an eta invariant, which can be seen as the integral of a certain KO class. This fact may make the Green-Schwarz anomaly cancellation mechanism more natural when the self-dual fields are differential KO classes. The 2-form Kalb-Ramond field involved in the Green-Schwarz mechanism in type I supergravity is indeed differential KO-valued \cite{Freed:2000ta}. (This is a special case of the differential K-theoretic formulation of RR fields for orientifolds. See, for example, \cite{Distler:2009ri}.) It would be very interesting to understand whether global anomaly cancellation conditions on the field content depend on the choice of generalized cohomology theory used to model the 2-form gauge potentials. 

\cite{MonnierMooreSum2018} summarizes the present paper.\\

The paper is organized as follows. In Section \ref{SecPrelim}, we review six-dimensional supergravity and the cancellation of local anomalies through the Green-Schwarz mechanism. We then review some basic facts about anomalies and determine the anomaly field theory associated to a bare six-dimensional supergravity theory, before the inclusion of the Green-Schwarz terms. In Section \ref{SecMod}, we present a model for the self-dual 2-form gauge fields and their source $Y$ that accommodates fields with non-trivial topology. Section \ref{SecTFT} is devoted to the construction of the Wu Chern-Simons theory as a field theory functor on spacetimes endowed with a Wu structure. In Section \ref{SecRelGSAFT}, we show how the background field of the Wu Chern-Simons theory can be constructed from the 6d supergravity data, yielding the shifted Wu Chern-Simons field theory, whose state space will host the exponentiated Green-Schwarz term. We show in particular that the identification is such that the dependence on the underlying Wu structure drops out. We construct the exponentiated Green-Schwarz term as a vector in the state space of the shifted Wu Chern-Simons theory in Section \ref{SecGSTerms}. We discuss in Section \ref{SecImplSixSugra} the implications of our construction for 6d supergravity, in particular the constraints it imposes on the anomaly coefficients. We also investigate the case where $G$ is finite Abelian, extracting anomaly cancellation constraints and comparing them with F-theory models. Appendix \ref{AppDiffCocCharCl} explains how to refine characteristic classes to differential cocycles by making universal choices on classifying spaces. Appendix \ref{AppEThCalc} reviews (certain generalizations of) the generalized cohomology theory known as E-theory and their cochain models, which play a central role in the definition of the Wu Chern-Simons theories. Appendix \ref{AppGluing} contains a proof of the gluing axioms for the Wu Chern-Simons theory of interest to us. Appendix \ref{AppCompCobGroup} contains the computation of $\Omega^{\rm spin}_7(BG)$ for a few gauge groups $G$. Finally, we compute in Appendix \ref{AppNonTrivGlobAn} certain eta invariants associated to principal bundles of finite Abelian groups on Lens spaces.

\section{Preliminaries}

\label{SecPrelim}

\subsection{Six-dimensional supergravity theories}

We briefly introduce $N = (1,0)$ 6d supergravities. We refer the reader to standard reviews like \cite{Avramis:2006nb, TaylorTASI} for details. We will use the same notations as in Section 2 of \cite{Monnier:2017oqd}. In this section, we follow the standard approach and model all the gauge fields by differential forms. We will present a better model accounting for topologically non-trivial field configurations in Section \ref{SecMod}.

\paragraph{Field content} The spacetime is a six-dimensional spin manifold. For six-dimensional $N = (1,0)$ theories, the R-symmetry is $SU(2) = Sp(1)$. As the spinor representation is quaternionic in dimension 6, a symplectic Majorana-Weyl condition can be imposed on the fermions forming R-symmetry doublets. The $N = (1,0)$ supermultiplets in six-dimensions are the following.
\begin{itemize}
\item The gravitational multiplet $(g_{\mu\nu}, \psi^+_\mu, B^+_{\mu\nu})$. $g_{\mu\nu}$ is the metric tensor, $\psi^+_\mu$ is the gravitino, a spin 3/2 symplectic Majorana-Weyl fermion, and $B^+_{\mu\nu}$ is a self-dual 2-form gauge field.
\item The tensor multiplet $(B^-_{\mu\nu}, \chi^-, \phi)$. $B^-_{\mu\nu}$ is an anti self-dual 2-form gauge field, $\chi^-$ is a negative chirality spin 1/2 symplectic Majorana-Weyl fermion, and $\phi$ is a real scalar.
\item The vector multiplet $(A_\mu, \lambda^+)$. $A_\mu$ is a gauge field associated to a gauge group $G$, and $\lambda^+$ is a positive chirality spin 1/2 adjoint-valued symplectic Majorana-Weyl fermion.
\item The hypermultiplet $(\psi^-, 4\phi)$. $\psi^-$ is a negative chirality spin 1/2 Weyl fermion and a singlet under the R-symmetry, $4\phi$ represents a pair of complex bosons or four real bosons. In general the hypermultiplets take value in a quaternionic representation of the gauge group.
\item The half-hypermultiplet $(\psi^-_\mathbb{R}, 2\phi)$. The half-hypermultiplet can be constructed only if it is valued in a quaternionic representation of $G$. Starting from a hypermultiplet, a symplectic Majorana condition can be applied to $\psi^-$, yielding a negative chirality spin 1/2 symplectic Majorana-Weyl fermion. A corresponding reality condition can be applied to the pair of complex bosons, yielding a pair of real bosons. We emphasize however, that those are degrees of freedom per complex dimension of the original representation $R$. As the latter is quaternionic, those degrees of freedom can only come in pairs: there is no such thing as a single half-hypermultiplet.

We remark that a hypermultiplet valued in a representation $S$ of $G$ can always be seen as a half-hypermultiplet valued in the representation $S \oplus S^\ast$, where $S^\ast$ is the representation complex conjugate to $S$. For practical purpose, we can therefore assume that the matter content is composed of a half-hypermultiplet valued in a certain quaternionic representation $R$ of the gauge group $G$.
\end{itemize}

\paragraph{Gauge group} The vector multiplets contain gauge fields, so 6d supergravity theories generically have a gauge sector based on
a compact Lie group $G$. We always have
\be
1 \rightarrow G_1 \rightarrow G \rightarrow \pi_0(G) \rightarrow 1
\ee
where $G_1$ is the connected component of the identity element. This in turn has the form
\be
\label{EqGlobFormGaugeGroup}
G_1 \cong (\tilde{G}_{\rm ss} \times G_{\rm a})/\Gamma \;,
\ee
where $\tilde{G}_{\rm ss} = \prod_i \tilde{G}_i$ is a semi-simple simply connected compact Lie group with simple factors $\tilde{G}_i$ and $G_{\rm a} \simeq U(1)^r$ is a compact connected Abelian group. Writing $Z$ for the center of $\tilde{G}_{\rm ss}$, $\Gamma$ is a finite subgroup of $Z \times G_{\rm a}$ intersecting $1 \times G_{\rm a}$ trivially. The gauge Lie algebra has a corresponding decomposition
\be\label{gauge}
\mathfrak{g}
= \mathfrak{g}_{\rm ss} \oplus \mathfrak{g}_{\rm a}
= \bigoplus_i \mathfrak{g}_i \oplus \bigoplus_I \mathfrak{u}(1)_I \;.
\ee

\paragraph{Self-dual fields} The gravitational multiplet contains a self-dual 2-form field, while the $T$ tensor multiplets contain each an anti self-dual 2-form field. We can picture the fluxes of the (anti) self-dual fields as taking value in a self-dual lattice $\Lambda$ of signature $(1,T)$ \cite{SeibergTaylor}. (A priori, the lattice $\Lambda$ is just an integral lattice. The argument of Seiberg and Taylor that it is
unimodular is based on global anomalies. Independently of their logic, we will show that our construction of the Green-Schwarz term likewise
requires $\Lambda$ to be unimodular.)
The self-dual fields can then be gathered into a 2-form gauge potential $B$ valued in $\Lambda_\mathbb{R} := \Lambda \otimes \mathbb{R}$. The self-dual field strength $H$ is a $\Lambda_\mathbb{R}$-valued 3-form. (The self-duality constraint depends on the value of tensor-multiplet scalars. These scalars are
constrained to lie on one component of the hyperboloid in $\Lambda_\mathbb{R}$ of vectors of length-squared one. The component of $H$
orthogonal to the hyperboloid must be self-dual and the component tangent to the hyperboloid must be anti-self-dual.)  The vacuum expectation values of the scalars in the tensor multiplet determine an involution $\theta$ of $\Lambda_\mathbb{R}$, and the self-duality condition reads
\be
\label{EqSelfDualCondH}
\ast H = \theta H \;,
\ee
where $\ast$ is the Hodge star operator.

The theory contains instantonic self-dual strings charged under the self-dual and anti self-dual 2-form fields. The lattice $\Lambda$ can alternatively be pictured as the lattice of string charges. The Seiberg-Taylor result about the unimodularity of $\Lambda$ is the completeness hypothesis \cite{Polchinski:2003bq, BanksSeiberg} for string charges.

\paragraph{Green-Schwarz mechanism} A generalization \cite{Green:1984bx, Sagnotti:1992qw, Sadov:1996zm, Riccioni:2001bg} of the Green-Schwarz mechanism \cite{Green:1984sg} is necessary to cancel anomalies. We refer to the (anomalous) supergravity theory obtained before the addition of the Green-Schwarz term as the ``bare supergravity.''  In order for anomaly cancellation to be possible, the degree 8 anomaly polynomial $A_8$ of the bare theory has to factorize as the square of a degree 4 $\Lambda_\mathbb{R}$-valued polynomial $Y$:
\be
A_8 = \frac{1}{2} Y \wedge Y \;,
\ee
where $\wedge$ is the wedge product tensored with the pairing on $\Lambda_\mathbb{R}$ determined by the pairing on $\Lambda$. (We discuss $Y$ in more detail shortly.) Then the Bianchi identity of $H$ is modified to
\be
dH = Y \;,
\ee
and the following factor is included in the path integral
\be
\exp 2\pi i \left( \frac{1}{2} \int B \wedge Y \right) \;,
\ee
where the use of the pairing of $\Lambda_\mathbb{R}$ is again implicit. This factor can be interpreted as coming from a Green-Schwarz term
\be
\label{DefGSDiffFormNaive}
2\pi i \frac{1}{2} \int B \wedge Y \;,
\ee
added to the action. An obvious problem with the expression \eqref{DefGSDiffFormNaive} is that it uses the differential form model for Abelian gauge fields, which captures only the topologically trivial sector. We will remedy this in Section \ref{SecMod}, where we will develop a differential cocycle model for the self-dual fields and the background curvature $Y$.

There is however a much more serious problem, due to the factor $\frac{1}{2}$ in \eqref{DefGSDiffFormNaive}. The standard way of defining Green-Schwarz terms in dimension $d$ is as boundary values of $d+1$-dimensional Chern-Simons term: their gauge variation coincides with the variation of a Chern-Simons term on a $d+1$-dimensional manifold bounded by the spacetime. In the case of \eqref{DefGSDiffFormNaive}, the Chern-Simons term would read up to signs
\be
2\pi i \frac{1}{2} \int_U A \wedge Y \;,
\ee
where $U$ is a 7-dimensional manifold bounded by the 6-dimensional spacetime and $A$ is the degree 3 Abelian gauge field associated to the field strength $Y$. The problem is the following. The factor $\frac{1}{2}$ means that we are dealing with a higher Abelian Chern-Simons theory at a half-integer level, which is not gauge invariant under large gauge transformations. Concretely, this means that even if we could make sense of \eqref{DefGSDiffFormNaive} for topologically non-trivial gauge field configurations, the phase that \eqref{DefGSDiffFormNaive} transform by under a large gauge transformation is defined only up to a sign. Therefore there is no way that \eqref{DefGSDiffFormNaive} can cancel all global gauge anomalies.

It is known how to make sense of half-integer level Chern-Simons theories on spin 3-manifolds: those are the so-called spin Chern-Simons theories \cite{Dijkgraaf:1989pz, 2005math......4524J, 2006math......5239J, Belov:2005ze} that play a central role in the quantum Hall effect. Their higher-dimensional generalization have recently been studied in \cite{Monnier:2016jlo} and the results of that paper will play a central role in the construction of the Green-Schwarz terms in the present paper.

\paragraph{Anomaly coefficients} $Y$ has the general form
\be
\label{EqRefEffAbGaugFFT}
Y = \frac{1}{4} a p_1 - \sum_i b_i c_2^i + \frac{1}{2} \sum_{IJ} b_{IJ} c_1^I c_1^J \,.
\ee
where $a, b_i, b_{IJ} \in \Lambda_\mathbb{R}$ are the \emph{anomaly coefficients} of the theory. Writing $R$ for the curvature of the tangent bundle and ${\rm tr}_{\rm vec}$ for the trace in the vector representation of the orthogonal group, $p_1 := \frac{1}{8\pi^2}{\rm tr}_{\rm vec} R^2$ is the Chern-Weil representative of the first Pontryagin class of the tangent bundle, i.e. the first Pontryagin form. Writing $F = (F^i, F^I)$ for the curvature of the gauge bundle, $c_1^I := \frac{1}{2\pi}F^I$ is the first Chern form associated to the $I$th $U(1)$ component of $G_{\rm a}$, and $c_2^i := \frac{1}{8\pi^2}{\rm tr} (F^i)^2$ is the second Chern form associated to the $i$th simple component of $\tilde{G}_{\rm ss}$. ${\rm tr}$ is normalized so that the dual pairing on the weight space gives length squared 2 to the long roots.

Assuming that the 6d supergravity theory can be defined on any spin spacetime with an arbitrary gauge bundle, it was shown in \cite{Monnier:2017oqd} that
\be
a, b_i, \frac{1}{2}b_{II}, b_{IJ} \in \Lambda \;.
\ee
These constraints supersede the ones previously derived in \cite{KMT2, SeibergTaylor}.

As far as we know, all the known string theory realizations of 6d supergravities can be implemented in F-theory. In F-theory models, $\Lambda$ is the degree 2 homology lattice of the (four-dimensional) base of the elliptic fibration, and $a$ is the homology class of the canonical divisor of the base. This implies (via the adjunction formula) \cite{Monnier:2017oqd} that $a$ is a characteristic element of $\Lambda$, i.e. that it satisfies
\be
\label{EqaCharElLam}
(a,x) = (x,x) \quad \mbox{mod } 2 \quad \forall \; x \in \Lambda \;.
\ee
One naturally wonders whether all consistent six-dimensional supergravities must be such that $a$ is a characteristic element,
but thus far, such a condition has not been derived from the low energy point of view: in fact, there are 6d supergravity theories satisfying all known low energy consistency conditions but violating \eqref{EqaCharElLam}. An example was given in Section 5 of \cite{Monnier:2017oqd}, involving $244$ neutral hypermultiplets, no gauge symmetry, a single tensor multiplet, $\Lambda =  \mathbb{Z}^2$ with bilinear form
\be
\label{EqBilinForm6dSugraNotChar}
\begin{pmatrix}
0&1\\1&0
\end{pmatrix}
\ee
and $a = (4,1)$.

In this paper we will show that our construction of the Green-Schwarz term requires \eqref{EqaCharElLam}. Therefore, unless a more general construction of the Green-Schwarz terms exists, we establish \eqref{EqaCharElLam} as a low energy constraint to be satisfied by any 6d supergravity theory.

\label{SecRev6dSugra}

\subsection{Some facts about anomalies}

\label{SecRevAnomalies}

\paragraph{Generalities} Local and global anomalies of the partition function of a $d$-dimensional quantum field theory can be described by a geometric invariant of $d+1$-dimensional manifolds. This geometric invariant assigns a number mod 1 (or a phase after exponentiation) to any $d+1$-manifold endowed with all the structures necessary to define the $d$-dimensional quantum field theory (metric, principal bundles, connections on principal bundles, spin structure, etc...). The invariant is generally geometric rather than topological because it depends on geometric structures, such as metrics or connections. In the case of chiral fermionic theories in even dimension, the geometric invariant is essentially the eta invariant of a suitable Dirac operator \cite{Witten:1985xe}.

One can extract concrete data from the geometric invariant by evaluating it on certain closed $d+1$-dimensional manifolds $U$. For instance, the phases that the partition function transforms by under an anomalous symmetry transformation is given by the value of the geometric invariant on twisted doubles \cite{Monnier:2014txa}, which are constructed as follows. Take the $d$-dimensional spacetime $M$, and find (if possible) a $d+1$-dimensional manifold $N$ that bounds it. Then construct $U$ by gluing $N$ to $-N$ ($N$ with the opposite orientation) along $M$, using the symmetry transformation to identify the two copies $M$, including their topological/geometrical structures. 

One is often interested in the nature of the partition function obtained by integrating out the anomalous fields as a function of the background values of bosonic non-anomalous fields, such as the scalar fields, the vector-multiplet gauge fields or the metric. Generally, the partition function is a section of a line bundle with connection over this bosonic moduli space. The partition function of the full theory, obtained by integrating the partition function above over the bosonic moduli space, is well-defined only when this line bundle with connection is geometrically trivial, i.e. admits a global trivialization given by the connection. Given the geometric invariant describing the anomaly, one can compute the holonomies of the anomaly connection by evaluating it on mapping tori in a limit where the size of the base circle is large (the "adiabatic limit") \cite{Witten:1985xe, MR861886}. The vanishing of the invariant on all such mapping tori is equivalent to the anomaly bundle being geometrically trivial.

The anomaly geometric invariants are often quite hard to compute, but when the $d+1$-dimensional manifold $U$ is the boundary of a $d+2$-dimensional manifold $W$, they can generally be expressed as the sum $S(W)$ of integrals of top forms on $W$ and of additive topological invariants of $W$. (We will present examples below.) $W$, like $U$, must carry all the structures necessary to define the $d$-dimensional QFT and the structures on $W$ must reduce to the structures on $U$ upon restriction to the boundary. Whether there are some $U$'s that are not the boundaries of some $W$'s is determined by the bordism group of manifolds endowed with the appropriate structures. These groups are known in the simplest cases, but are generally hard to compute.

It is common to guess or construct the anomaly geometric invariant of $U$ as a function $S(W)$ defined on $d+2$-dimensional manifolds bounded by $U$. We then have the following consistency condition. In order for the invariant on $U$ to be well-defined mod 1, $S(W)$ has to be an integer whenever $W$ is a closed $d+2$-dimensional manifold. This follows by a standard argument, which for instance leads to the quantization of the level in the 3d Chern-Simons action. Even if this consistency condition is satisfied, the geometric invariant is then fully determined only if the relevant bordism group vanishes. In general, it is defined only up to a bordism invariant.

\paragraph{Anomaly field theories} The picture of anomalies presented above can be refined by promoting the geometric invariant to be the partition function of a $d+1$-dimensional (usually invertible) quantum field theory, the \emph{anomaly field theory} \cite{Freed:2014iua}. The partition function of the anomalous quantum field theory is then valued in the state space of the anomaly field theory; the anomalous action of the symmetries on the partition function is given by their action on the state space of the anomaly field theory. As an extended field theory, the anomaly field theory also accounts for Hamiltonian anomalies and their analogues associated to higher codimension submanifolds \cite{Monnierd, 2014arXiv1409.5723F}.

In this context, the Green-Schwarz anomaly cancellation can be understood as follows. The partition function of the bare 6d supergravity theory is an element of the state space of its invertible anomaly field theory, which is a Hermitian line $L$. Constructing an exponentiated Green-Schwarz term cancelling the anomaly amounts to constructing a vector in the conjugate Hermitian line $L^\dagger$. The tensor product of the partition function with the exponentiated Green-Schwarz term is then canonically a complex number. Moreover, symmetries act trivially on the tensor product $L \otimes L^\dagger$, showing that the anomalies have been canceled. The idea that the exponentiated Green-Schwarz term should be an element of a suitable Hermitian line appeared already in \cite{Witten:1999eg}, prior to the concept of anomaly field theory.\\

We now review examples of anomalous field theories and their associated geometric invariant/anomaly field theory. As above, the anomalous field theory is $(d+1)$-dimensional, $U$ is a $d+1$-dimensional closed manifold and $W$ is a $d+2$-dimensional manifold whose boundary is $U$.

\paragraph{Complex Weyl fermions} In the case of complex Weyl (i.e. chiral) fermions in even dimension valued in a certain representation $R$ of the gauge symmetry, the geometric invariant computing the anomaly is the modified eta invariant of the Dirac operator in dimension $d+1$ \cite{Witten:1985xe} valued in the same representation $R$, which reads
\be
\frac{1}{2\pi i} \ln {\rm An}_{{\rm Wf}, R}(U) = \xi_R(U) = \frac{\eta_R + h_R}{2} \;,
\ee
where $\eta_R$ is the ordinary eta invariant and $h_R$ is the dimension of the space of zero modes of the Dirac operator on $U$. In the case that $U$ and its gauge bundle extends to a $(d+2)$-dimensional manifold $W$, the Atiyah-Patodi-Singer (APS) theorem \cite{MR0397797} allows one to reexpress the modified eta invariant on $U$ in terms of data on $W$ as follows:
\be
\label{EqAPS}
\xi_R(U) = \int_W I_R - {\rm index}(D_R^{(W)}) \;,
\ee
where $I_R$ is the index density of the Dirac operator $D_R^{(W)}$ on $W$ (the one appearing in the local anomaly formula), and ${\rm index}(D_R^{(W)})$ its index with APS boundary conditions. We see that, modulo 1, we have
\be
\label{EqXiFromW}
\xi_R(U) = \int_W I_R \mbox{ mod } 1 \;.
\ee
This means that if we have a system of complex fermions whose local anomaly vanishes and all relevant $d+1$-dimensional manifolds $U$
together with their gauge bundles are boundaries, then there are no global anomalies.

The corresponding anomaly field theory is the Dai-Freed theory \cite{Dai:1994kq}, which admits the modified eta invariant as its partition function. See Section 9 of \cite{Freed:2016rqq} for a construction of this theory using stable homotopy theory.

\paragraph{Majorana-Weyl fermions} As far as even dimensions are concerned, we can have Lorenz signature Majorana-Weyl fermions in dimensions $8\ell+2$. (This case is not relevant to the present considerations, we include it for completeness.) As the spinor representation is real, we can impose a reality condition on any fermion valued in a real representation $R$. In Euclidean signature, the spinor representation is complex, leading to a factor $\frac{1}{2}$ in the formula for the geometric invariant:
\be
\frac{1}{2\pi i} \ln {\rm An}_{{\rm MWf},R}(U) = \frac{1}{2}\xi_R(U) \;.
\ee
We can still use the APS theorem \eqref{EqAPS}, but as the term involving the index is now a priori a half-integer, we can't immediately express $\xi_R(U)/2$ as the integral of the local index density on $W$ as in \eqref{EqXiFromW}.

However on $W$, i.e. in dimension $8\ell+4$, the spinor representation is quaternionic, so the Dirac operator is quaternionic and the index in \eqref{EqAPS} is necessarily even. So in fact, just as in the case of complex fermions, the index term does not contribute, and the global anomaly reduces to the local anomaly when $U$ bounds.

The anomaly field theory associated to Majorana fermions is a real version of the Dai-Freed theory, which is constructed in Section 9 of \cite{Freed:2016rqq}.

\paragraph{Symplectic Majorana-Weyl fermions} The case of symplectic Majorana-Weyl fermions is very similar. In even dimensions, we can have Lorentz signature symplectic Majorana-Weyl fermions in dimensions $8\ell+6$. As the spinor representation is quaternionic, we can impose a reality condition on any fermion valued in a quaternionic representation $R$. The fermions satisfying such a reality condition are symplectic Majorana fermions. (In Euclidean signature, the spinor representation is complex.) The relevant geometric invariant is
\be
\frac{1}{2\pi i} \ln {\rm An}_{{\rm SMWf},R}(U) = \frac{1}{2}\xi_R(U) \;.
\ee

On $W$, i.e. in dimension $8\ell$, the spinor representation is real, so the Dirac operator on $W$ (being twisted by a quaternionic representation), is quaternionic. The index in \eqref{EqAPS} is necessarily even and again does not contribute. The global anomaly reduces to the local anomaly when $U$ bounds.

Similarly to the case of Majorana fermions, the anomaly field theory associated to symplectic Majorana fermions is a symplectic version of the Dai-Freed theory \cite{Freed:2016rqq}.

\paragraph{Self-dual fields} For a degree $2\ell$ self-dual field in dimension $4\ell+2$ that does not couple to an Abelian degree $2\ell+1$ gauge field, the anomaly geometric invariant is $\frac{1}{4}\xi_\sigma$, where $\xi_\sigma$ is the  modified eta invariant of the $4\ell+3$-dimensional signature Dirac operator \cite{Witten:1985xe}. When the $4\ell+3$-dimensional manifold $U$ bounds, we can rewrite
\be
\label{EqAnSDNeut}
\frac{1}{2\pi i} \ln {\rm An}_{\rm SD0}(U) = \frac{1}{4}\xi_\sigma(U) = \frac{1}{8} \left( \int_W L_{TW} - \sigma_W \right) \;,
\ee
where we used the fact that the index of the signature Dirac operator on $W$ is the signature $\sigma_W$ of the wedge product pairing on the lattice $H^{2\ell+2}_{{\rm DR},\mathbb{Z}}(W,\partial W)$, consisting of relative de Rham cohomology classes with integral periods. \eqref{EqAnSDNeut} is well-defined because on a closed $4\ell+4$-dimensional manifold $Z$, the L-genus $L_{TZ}$ integrates to the signature and the right-hand side vanishes. Here we make use of Novikov's additivity of signature to identify the sum of the signatures of two manifolds with a common boundary with the signature of the closed manifold obtained by gluing them.

Embedding of self-dual field theories in non-chiral theories were studied in \cite{Witten:1996hc, Maldacena:2001ss, Belov:2006jd}, with an action of the form:
\be
\label{EqActSDField}
-\frac{1}{2g^2} \int_M (dB - qA) \wedge \ast (dB - qA) + i\pi p \int_M A \wedge dB
\ee
on a $4\ell+2$-dimensional manifold $M$. In this model, the degree $2\ell$ Abelian gauge field $B$ couples to an Abelian degree $2\ell+1$ gauge fields in two different ways. First, $A$ is a source for $B$, which carries an integer charge $q$. Second, $A$ and $B$ couple through a Green-Schwarz-like term, with an integer coefficient $p$. The action can be rewritten
\be
-\frac{i}{g^2} \int_M \left( (dB)^- \wedge (dB)^+ + q A^- A^+ + (\pi g^2 p + q) A^+ \wedge (dB)^- + (\pi g^2 p - q)  A^- \wedge (dB)^+ \right) \;,
\ee
where the $+$ and $-$ superscripts denote the self-dual and anti self-dual part of $2\ell+1$-forms, respectively. At the special value $g^2 = -q/(\pi p)$ of the gauge coupling, the dependence of the partition function on $A^+$ drops out, allowing to study the dependence of the self-dual field field partition function on $A^-$. It was shown in \cite{Witten:1996hc, Belov:2006jd} that the self-dual field has a gauge anomaly proportional to $k = pq$.

For $k \neq 0$, the anomaly geometric invariant on $U$ is the sum of $\xi_\sigma(U)/4$ and of the Arf invariant of a certain quadratic refinement $q$ of the linking pairing on the degree $2\ell + 2$ torsion cohomology \cite{Monnier2011a}:
\be
\label{EqAnSDka}
\frac{1}{2\pi i} \ln {\rm An}_{\rm SDk}(U) = \frac{1}{4} \xi_\sigma(U) - k {\rm Arf}(q) \;,
\ee
see Section \ref{SecTFT} for a more detailed discussion of the quadratic refinement and its Arf invariant. If $U$ bounds, \eqref{EqAnSDka} can be written more simply as
\be
\label{EqAnSDk}
\frac{1}{2\pi i} \ln {\rm An}_{\rm SDk}(U) = \frac{1}{8} \left( \int_W L_{TW} - \sigma_W \right) - k \left( \frac{1}{2} \int_W Y_W^2 - \frac{1}{8} \sigma_W \right) \;,
\ee
where $L_{TW}$ is the Hirzebruch L-genus of the tangent bundle of $W$. $Y_W$ is the field strength of the extension of the gauge field $A$ from $U$ to $W$, with the following important subtlety. The periods of $Y_W$ on $W$ are integral or half-integral, depending on the value of the degree $2\ell+2$ Wu class $\nu(TW)$. The Wu class is a certain $\mathbb{Z}_2$-valued characteristic class that can be expressed in terms of the Stiefel-Whitney classes, see Section \ref{SecSpinWu}. It has the crucial property that on a closed manifold $Z$ of dimension $4\ell+4$, $x \cup x = x \cup \nu$ for all $x \in H^{2\ell+2}(Z;\mathbb{Z}_2)$. This implies in particular that if $Y_Z$ has periods as above, $F \wedge F = F \wedge 2Y_Z \mbox{ mod } 2$ for $F \in \Omega^{2\ell+2}_{\mathbb{Z}}(Z)$, a $2\ell+2$ differential form on $Z$ with integral periods. The relation above of course passes to de Rham cohomology, which means that $[2Y_Z]_{\rm dR}$ is a \emph{characteristic element} of $H^{2\ell+2}_{\rm dR,\mathbb{Z}}(Z)$, the lattice of de Rham cohomology classes with integral periods.

Consistency requires that the right-hand side of \eqref{EqAnSDk} is an integer when $W = Z$, a closed $4\ell+4$-dimensional manifold. We already explained why this is true for the first term. For the second term, this is due to the fact that the norm square of any characteristic element of a unimodular lattice is equal to the signature modulo 8 (see for instance Remark 2.3 in \cite{Taylor}).

\paragraph{Torus-valued self-dual fields} Consider now self-dual fields valued in a torus. We can describe the torus by means of a lattice $\Lambda$ as $\Lambda \otimes_\mathbb{Z} \mathbb{R}/\Lambda$. The number of self-dual fields and the number of anti self-dual fields in encoded in the signature of the lattice $\Lambda$, which is $(1,T)$ in the case of 6d supergravity theories. We can generalize the action \eqref{EqActSDField} by assuming that the wedge products involve the pairing on $\Lambda \otimes \mathbb{R}$, yielding real-valued forms.

Setting $p = q = 1$, the natural generalization of \eqref{EqAnSDk} to the torus case reads:
\be
\label{EqAnSDgen}
\frac{1}{2\pi i} \ln {\rm An}_{\rm SDgen}(U) = \frac{{\rm sgn}(\Lambda)}{8} \left( \int_W L_{TW} - \sigma_W \right) - \left( \frac{1}{2} \int_W Y_W^2 - t \right) \;,
\ee
where ${\rm sgn}(\Lambda)$ is the signature of $\Lambda$ and $t$ denotes a topological invariant not contributing to the local anomaly, such that second term is well-defined (i.e. independent of $W$).

\eqref{EqAnSDgen} suggests that the anomaly field theory for a torus-valued self-dual field is the product of two distinct field theories. The first one is the product of ${\rm sgn}(\Lambda)$ copies of a "quarter Dai-Freed theory for the signature Dirac operator", described in more detail in Section 4.5 of \cite{Monnier:2017klz}. $t$ must be identified to determine the second quantum field theory. In the present paper, we will construct this theory in the case of interest for 6d supergravities, where $\Lambda$ is a self-dual lattice.

\subsection{Anomalies of six-dimensional supergravities}

\label{SecRevAn6dSugra}

We now turn to six-dimensional supergravity theories. Consistency requires that they are anomaly-free, and a version of the Green-Schwarz mechanism \cite{Green:1984sg, Green:1984bx, Sagnotti:1992qw, Sadov:1996zm} plays a central role in the cancellation of local anomalies. While the cancellation of certain global gauge anomalies has been considered \cite{Bershadsky:1997sb, SuzukiTachikawa, KMT2}, no systematic study of global anomaly cancellation has yet been performed. This paper is a first step in 
that direction. 

We remark that in the bare theory, the self-dual fields have no gauge anomaly, so their anomaly is described by the first term of \eqref{EqAnSDgen}. The anomaly of a bare 6d supergravity is therefore described by the following geometric invariant.
\be
\label{EqAnomBare6dSugra}
\frac{1}{2\pi i} \ln {\rm An}_{\Lambda, G, R}(U) = \frac{1}{2}\xi_{R'}(U) + \frac{{\rm sgn}(\Lambda)}{4} \xi_{\sigma}(U) \;.
\ee
The first term is the modified eta invariant associated to the Dirac operator on $U$ coupled to the virtual $Spin(7) \times G$-representation
\be
R' = (({\rm Vec}Spin(7) \ominus 1) \otimes 1) \ominus (T-1)(1 \otimes 1) \oplus (1 \otimes {\rm Ad}G) \ominus (1 \otimes R) \;,
\ee
where $1$, ${\rm Vec}$ and ${\rm Ad}$ denotes the trivial, vector and adjoint representations. The summands in $R'$ are due respectively to the chiral fermions in the gravitational multiplet (the gravitino), in the tensor multiplets, in the vector multiplets and in the half-hypermultiplets. $(T-1)(1 \otimes 1)$ denotes the direct sum of $T-1$ copies of the trivial representation. The anomaly field theory of the bare six-dimensional supergravity theory is the Dai-Freed theory having \eqref{EqAnomBare6dSugra} as partition function.

We now focus on the case where the 7-manifold $U$ bounds and find a formula for the anomaly of the bare theory. (Let us make clear that here and in the following, when we say that "a 7-dimensional manifold bounds", we really mean that a 7-dimensional spin manifold endowed with a principal $G$-bundle $P$ bounds an 8-dimensional spin manifold over which $P$ extends. Or in short, that the class of the pair $(U,P)$ in the bordism group $\Omega^{\rm spin}_7(BG)$ is trivial.) The anomaly of the bare self-dual fields is given by the first term on the right-hand side of \eqref{EqAnSDgen}. We should add to this term the anomalies of the chiral fermions, which as we saw can be expressed purely in terms of the integrals of local index densities on $W$. In the present paper, we will assume that it is possible to cancel local anomalies, as the associated constraints are well known (see for instance \cite{TaylorTASI}). In particular, this implies that the total local index density factorizes as $\frac{1}{2} Y_W^2$ for some $Y_W$. We immediately deduce that the global anomaly of the bare theory associated to $U$ is given by the following geometric invariant:
\be
\label{EqAnBare6dSugra}
\frac{1}{2\pi i} \ln {\rm An}_{\rm bare}(U) = \frac{1}{2} \int_W Y_W \wedge Y_W - \frac{\sigma_{H^4(W,\partial W;\Lambda)}}{8} \;.
\ee
where we used the multiplicative property of the signature with respect to tensor products to obtain ${\rm sgn}(\Lambda) \sigma_W = \sigma_{H^4(W,\partial W;\Lambda)}$. 

Global anomalies can cancel only if
\be
t = \frac{\sigma_{H^4(W,\partial W;\Lambda)}}{8} \;.
\ee
Therefore, in order to cancel global anomalies associated to 7-dimensional manifolds $U$ that bound, we need to construct a Green-Schwarz term that induces the anomaly
\be
\label{EqAnGSTerms}
\frac{1}{2\pi i} \ln {\rm An}_{\rm GS}(U) = -\frac{1}{2} \int_W Y_W \wedge Y_W + \frac{\sigma_{H^4(W,\partial W;\Lambda)}}{8} \;.
\ee

In order to study the anomaly on 7-dimensional manifolds that do not bound, we would need to study the particular sum of modified eta invariants associated to the field content of the bare theory. We do not know a good way to handle this problem. However, we will be able to rewrite ${\rm An}_{\rm GS}(U)$ in a purely 7-dimensional form, which makes sense on 7-dimensional manifolds $U$ that do not bound. We will also be able to construct explicitly Green-Schwarz terms with anomaly ${\rm An}_{\rm GS}(U)$. As long as the anomaly of the bare 6d supergravity is equal to $-{\rm An}_{\rm GS}(U)$, our construction successfully cancels all global anomalies. This is the case for instance when all 7-dimensional manifolds bound, which as we show in Appendix \ref{AppCompCobGroup}, occurs for a number of gauge groups $G$. In the remaining cases, we will see that the Green-Schwarz terms may fail to cancel all anomalies, leading to new non-trivial constraints on the supergravity theory.

\section{A model for the self-dual fields and their sources}

\label{SecMod}

In this section, we develop a model for the system composed of the self-dual fields and of the effective degree 3 Abelian gauge field appearing in the Green-Schwarz terms. We need to go beyond the usual model in terms of differential forms in order to include situations where the fields are topologically non-trivial. We therefore first review the differential cohomology model for gauge fields \cite{springerlink:10.1007/BFb0075216, hopkins-2005-70}, see also Section 2 of \cite{Freed:2006yc}. We generalize it slightly to accommodate torus-valued gauge fields with shifted quantization laws.

\subsection{Lattice valued differential cochains with shifts}

Let $\Lambda$ be a lattice of dimension $n$, and let $\Lambda_\mathbb{R} := \Lambda \otimes \mathbb{R} \simeq \mathbb{R}^n$ be the corresponding real vector space. $\Lambda_\mathbb{R}/\Lambda \simeq U(1)^n$ is a torus.

\paragraph{Differential cochains} Let us write $C^p(M; \Lambda)$ and $C^p(M;\Lambda_\mathbb{R})$ for the groups of degree $p$ smooth cochains valued in $\Lambda$ or $\Lambda_\mathbb{R}$. Let $\Omega^p(M;\Lambda_\mathbb{R})$ be the group of smooth differential $p$-forms with value in $\Lambda_\mathbb{R}$. Define the group of degree $p$  differential $\Lambda_\mathbb{R}$-valued cochains to be
\be
\check{C}^p(M;\Lambda_\mathbb{R}) = C^p(M;\Lambda_\mathbb{R}) \times C^{p-1}(M;\Lambda_\mathbb{R}) \times \Omega^p(M;\Lambda_\mathbb{R}) \;.
\ee
The group law is just the addition component by component. Note that that we took chains valued in $\Lambda_\mathbb{R}$ rather than $\Lambda$ in the first factor; this is necessary to accommodate the shift in the quantization of the periods of the differential cocycles, as we will see. We write elements of $\check{C}^p(M;\Lambda_\mathbb{R})$ with carons: $\check{c} = (a,h,\omega)$.
We will refer to  $[\check{c}]_{\rm ch} := a$ as the ``characteristic'' of the differential cochain $\check{c}$.
We refer to   $[\check{c}]_{\rm hol} := h$ as the ``connection'' or ``holonomy'' of the differential cochain $\check{c}$.
Finally we refer to
$[\check{c}]_{\rm fs} := \omega$ the ``curvature'' or ``fieldstrength''  of the differential cochain $\check{c}$.

\paragraph{Differential} We define a differential by
\be
d \check{c} = (da, \omega - a - dh, d\omega) \;, \quad d^2 = 0 \;.
\ee
The degree $p$ differential cochains $\check{c}$ such that $d \check{c} = 0$ are called differential cocycles. They form a group written $\check{Z}^p(M;\Lambda_\mathbb{R})$.

\paragraph{Shift and Quantization} Let $\hat\nu$ be any degree $p$ $\Lambda_\mathbb{R}$-valued cochain. We say that a differential cochain $\check{c} \in \check{Z}^p(M;\Lambda_\mathbb{R})$ is \emph{shifted by} $\hat\nu$ if $[\check{c}]_{\rm ch} = \hat\nu$ modulo $\Lambda$. 
\footnote{Later in the text we will use the notation $\nu$ to denote the Wu class of a manifold    and $\hat \nu$ will denote 
an integral cocycle representative. We hope this does not cause confusion.}
The differential cochains shifted by $\hat\nu$ form a subset $\check{C}^p_\nu(M;\Lambda )  \subset \check{C}^p(M;\Lambda_\mathbb{R})$
where $\nu$ is the projection of $\hat \nu$ to the set of $\Lambda_\mathbb{R}/\Lambda$-valued cochains. We call the Abelian group 
 $\check{C}^p_0(M;\Lambda)$ the  group of unshifted integral-quantized differential cochains. Then 
$\check{C}^p_\nu(M;\Lambda)$ is a torsor for $\check{C}^p_0(M;\Lambda)$. The kernel of the differential restricted to $\check{C}^p_0(M;\Lambda)$ is the group of unshifted differential cocycles
$\check{Z}^p_0(M;\Lambda)$ while the kernel restricted to $\check{Z}^p_\nu(M;\Lambda)$ is the group of shifted differential cocycles.

\paragraph{Differential cohomology} $\check{C}^p_\nu(M;\Lambda)$ contains differential cocycles if and only if $\nu$ is itself a cocycle, which we assume now. Let us define the following equivalence relation on differential cocycles. Any two degree $p$ differential cocycles are equivalent if they differ by the differential of an unshifted degree $p-1$ differential cochain with vanishing curvature:
\be
\check{x} \simeq \check{x} + d \check{y} \;, \quad \check{y} = (b, g, 0) \;, \quad b \in C^{p-1}(M; \Lambda) \;, g \in C^{p-2}(M;\Lambda_\mathbb{R}) \;.
\ee
Cocycles with different shifts can never be equivalent. Equivalence classes of cocycles in $\check{C}_\nu^p(M;\Lambda)$ are degree $p$ differential cohomology classes shifted by $\nu$, written $\check{H}_\nu^p(M;\Lambda)$. $\check{H}_\nu^p(M;\Lambda)$ is a torsor on $\check{H}_0^p(M;\Lambda)$ if $\nu$ is exact. The usual short exact sequences satisfied by differential cohomology groups apply; see for instance \cite{Freed:2006yc}.

\paragraph{Cup product} Recall that the lattice $\Lambda$ is endowed with a pairing $\Lambda \times \Lambda \rightarrow \mathbb{Z}$. A cup product on the groups of differential cochains can be defined as follows \cite{springerlink:10.1007/BFb0075216, hopkins-2005-70}:
\begin{align}
\label{EqDefCupProdDiffCohom}
\cup: \: & \check{C}^p(M; \Lambda) \times \check{C}^q(M; \Lambda) \rightarrow \check{C}^{p+q}(M; \mathbb{Z}) \\
& \check{c}_1 \cup \check{c}_2 = (a_1 \cup a_2, \; (-1)^{{\rm deg} a_1} a_1 \cup h_2 + h_1 \cup \omega_2 + H^\wedge_\cup(\omega_1, \omega_2), \; \omega_1 \wedge \omega_2) \notag
\end{align}
$\check{C}^{p+q}(M; \mathbb{Z})$ is the group of $\mathbb{Z}$-valued differential cochains. The cup products on the right-hand side are the ones associated to the pairing on $\Lambda_\mathbb{R}$. $\wedge$ is the wedge product of $\Lambda_\mathbb{R}$-valued forms. The wedge product is homotopically equivalent to the cup product, and $H^\wedge_\cup$ is any choice of equivalence, i.e. a degree -1 homomorphism from $\Omega^\bullet(M; \Lambda_\mathbb{R}) \times \Omega^\bullet(M; \Lambda_\mathbb{R})$ to $C^\bullet(M; \mathbb{R})$ satisfying
\footnote{It is shown in \cite{springerlink:10.1007/BFb0075216} how a homotopy $H^\wedge_\cup$ can be constructed canonically, provided one uses a cubical model for singular cohomology, rather than the more familiar simplicial model. Using a non-canonical homotopy is not a problem, as long as it is used consistently. Note in particular that since $H^\wedge_\cup$ is a homomorphism it vanishes if one of its arguments vanishes. This 
fact will be used quite frequently in computations below.      }
\be
\label{EqPropHomotop}
dH^\wedge_\cup(\omega_1, \omega_2) + H^\wedge_\cup(d\omega_1, \omega_2) + (-1)^{{\rm deg} \omega_1} H^\wedge_\cup(\omega_1, d\omega_2) = \omega_1 \wedge \omega_2 - \omega_1 \cup \omega_2 \;.
\ee
The first term on the right-hand side involves first taking the wedge product and seeing the resulting form as a cochain. The second term is the cup product between $\omega_1$ and $\omega_2$, seen as cochains. One can show that the cup product on differential cocycles satisfies the familiar relation
\be
\label{EqLeibnDiffCoc}
d(\check{c}_1 \cup \check{c}_1) = d\check{c}_1 \cup c_2 + (-1)^{{\rm deg}\check{c}_1} \check{c}_1 \cup d\check{c}_2 \;,
\ee
which ensures that \eqref{EqDefCupProdDiffCohom} passes to a cup product in differential cohomology. The property \eqref{EqPropHomotop} of the homotopy is crucial for \eqref{EqLeibnDiffCoc} to hold on the connection components of the differential cocycles.

\paragraph{Physical interpretation} Physically, we should think of degree $p$ unshifted differential cocycles as representatives of degree $p-1$ Abelian gauge potentials. The curvature/field strength of the differential cocycle corresponds to the physical field strength of the gauge field. The exponential of the connection computes the holonomies of the gauge field along $p-1$-dimensional cycles (Wilson lines/surfaces), and the characteristic contains information about the fluxes of the gauge field, including torsion fluxes undetectable from the field strength. Differential cohomology classes correspond to gauge equivalence classes of gauge fields. Abelian gauge fields with shifted quantization law are modeled as shifted differential cocycles in the formalism above.

The shift is really determined by the \emph{cocycle} $\nu$, and not by its cohomology class. In a physical setup, we can often characterize the shifted quantization of the fluxes by a U(1)-valued cohomology class. (For instance, the fluxes of the M-theory $C$-field are shifted by $\frac{1}{2} w_4$.) As there is no canonical way of picking a canonical cocycle representative of a cohomology class, and that moreover such a choice has to be made on every possible spacetime, this may sound like a serious problem. We are however often (always?) interested in gauge fields whose periods are shifted by the periods of a $U(1)$-valued characteristic class. A characteristic class is the pullback of a cohomology class on a classifying space. We can choose (non-canonically) a "universal" cocycle on the classifying space, and then shift our cocycles by the pull-back of the universal cocycle through the classifying map. There is no canonical way to pick a classifying map either, but as explained in Appendix \ref{AppDiffCocCharCl}, the classifying map can be taken to be part of the gauge data.

\paragraph{Integration} The $\mathbb{R}/\mathbb{Z}$-valued integral of a degree $p$ differential cochain $\check{c} \in \check{C}^p(M)$ over a $p-1$-dimensional manifold is defined as the mod 1 reduction of the integral of the degree $p-1$ real cochain $[\check{c}]_{\rm hol}$ over $M$ \cite{springerlink:10.1007/BFb0075216}. This integration map is a special case of a general construction valid for families \cite{hopkins-2005-70}.

\subsection{Model}

\label{SecModel}

In the presence of a GS term, the gauge invariant field strength $H$ of the self-dual fields is modified to
\be
\label{EqDefGaugInvFT}
H = dB + A \;,
\ee
where $dA = Y$ and $Y$ is 4-form appearing in the factorization of the local anomaly. The Green-Schwarz term
\be
\label{EqDefGSNaive}
2\pi i \frac{1}{2} \int_M B \wedge Y
\ee
is then added to the action on the 6d spacetime $M$. As before, $\wedge$ is the wedge product tensored with the pairing on $\Lambda_\mathbb{R}$, yielding a real-valued differential form. Similarly, the cup products below always include the lattice pairing. There are several puzzles with \eqref{EqDefGSNaive} that we start addressing below.

\paragraph{$Y$ as a differential cocycle} $Y$ is in general not exact on $M$, so we cannot model $Y$ and $A$ as differential forms such that $Y = dA$. This problem is easy to solve. We promote $Y$ to a differential cocycle $\check{Y}$. In fact, there is a universal way of constructing such a differential cocycle from the metric on $M$ and the gauge connection. This is explained in detail in Appendix \ref{AppDiffCocCharCl}, where the notation is also defined. The idea is the following. In addition to the gauge connection for the vectormultiplet gauge group and to the metric, we include in the gauge data a classifying map $\gamma$ from the spacetime into a classifying space $B_{\rm W}\bar{G}$ (whose precise definition is given in Appendix \ref{AppDiffCocCharCl}) classifying the topology of the spacetime and of the vectormultiplet gauge bundle, as well as a choice of Wu structure (see below). A "universal" differential cocycle $\check{Y}_{\rm U}$ is chosen on $B_{\rm W}\bar{G}$, and we take $\check{Y} = \gamma^\ast(\check{Y}_{\rm U})$. This trades the choice of an arbitrary differential cocycle lifting $Y$ on $M$ for the choice of a classifying map $\gamma$, which as we will see will be convenient.  

We then identify $Y$ with the field strength $[\check{Y}]_{\rm fs}$ of $\check{Y}$ and $A$ with its connection $[\check{Y}]_{\rm hol}$. Assuming for a moment that $\check{Y}$ is an unshifted differential cocycle, we can pick a gauge representative for which the characteristic $y = [\check{Y}]_{\rm ch}$ vanishes on a sufficiently small patch, in which case the differential cocycle condition ensures that $d A = Y$ on the local patch. But now $A$ makes sense globally. As we will see momentarily, $\check{Y}$ is actually shifted, but the choice of Wu structure contained in the classifying map $\gamma$ provides a closely-related unshifted differential cocycle $\check{X}$, to which the interpretation above applies straightforwardly.

\paragraph{Shift} Recall that the degree 4 Wu class is a $\mathbb{Z}_2$-valued characteristic class given in terms of the Stiefel-Whitney class by $w_4 + w_2^2$ on oriented manifolds. The differential cocycle $\check{Y}$ constructed from the metric and gauge connection is shifted by $\frac{1}{2} \nu \otimes a$, where $\nu := \gamma^\ast(\nu_{\rm U})$ is a cocycle representative of the degree 4 Wu class pulled back from $B_{\rm W}\bar{G}$, and $a$ is the gravitational anomaly coefficient in \eqref{EqRefEffAbGaugFFT}.


The degree 4 Wu class vanishes on manifolds of dimension 7 and fewer, so $\nu$ is a trivializable cocycle. As a consequence $y$ and $Y$ have integral periods. However, $\nu$ does not necessarily vanish as a cocycle, so $\check{Y}$ is genuinely shifted, in the sense that its characteristic cocycle $y$ is half-integer-valued. (It always takes integer values when evaluated on cycles, but may take half-integer values on chains.)

\paragraph{Wu structure} A Wu structure (discussed in more detail in Appendix \ref{SecSpinWu}) is a certain higher degree generalization of a spin structure. It can essentially be seen as a choice of trivialization $\eta$ of the Wu cocycle $\nu$. Although the GS terms to be constructed are independent of any choice of Wu structure, it is necessary to make such a choices in intermediary steps of the construction. 

A choice of Wu structure is included in the data of the classifying map $\gamma$ to $B_{\rm W}\bar{G}$. Indeed, as explained in Appendix \ref{SecSpinWu}, we pick a trivialization $\eta_{\rm U}$ of the Wu cocycle $\nu_{\rm U}$ on $B_{\rm W}\bar{G}$. Its pullback $\eta$ through the classifying map trivializes the Wu cocycle $\nu$. In Appendix \ref{SecSpinWu}, we also picked an integral lift $\eta_{\mathbb{Z},{\rm U}}$ of $\eta_{\rm U}$. On $M$, we have the pulled back cochain $\eta_\mathbb{Z} := \gamma^\ast(\eta_{\mathbb{Z},{\rm U}})$. $\eta_\mathbb{Z}$ plays a central role in our construction of the Green-Schwarz terms.


\paragraph{Associated unshifted differential cocycle} Given this data, we can construct an unshifted differential cocycle $\check{X}$ from $\check{Y}$ as follows. Let us define $\eta_\Lambda :=  \eta_\mathbb{Z} \otimes a$, where $a$ is the gravitational anomaly coefficient in \eqref{EqRefEffAbGaugFFT}. Define the flat differential cocycle $\check{\nu} := (d\eta_\Lambda, -\eta_\Lambda, 0)$, and then set
\be
\label{EqDefDiffCocX}
\check{X} = \check{Y} - \frac{1}{2} \check{\nu} \;,
\ee
which is unshifted by definition. The differential cohomology class of $\check{X}$ does depend on the choice of Wu structure through the classifying map. Note that the field strength of $X$ coincides with the field strength of $Y$: the distinction between $\check{X}$ and $\check{Y}$ affects only the Wilson observables and the torsion fluxes.

\paragraph{Charge cancellation} Recall that $Y$ carries a string charge, given by the homology class in $H_2(M;\Lambda)$ Poincaré dual to the cohomology class of $Y$ in $H^4(M;\Lambda)$. Like any charge, the total string charge has to vanish on a compact spacetime. So if $Y$ is topologically non-trivial, with charge $c \in H_2(M;\Lambda)$, there must be self-dual strings defects whose worldsheet $\Sigma$ wrap a cycle representing $-c$, so that the total string charge vanishes.

In the following, we will effectively assume that $Y$ is topologically trivial. More precisely, although we will always write $M$ for simplicity, we actually work on $M - \Sigma$ on which $Y$ is trivializable. As this manifold is non-compact, integration by parts should generate boundary terms. However, if anomalies cancel, a necessary condition is that such boundary contributions to anomaly computations cancel through inflow against the worldsheet anomalies of the strings on $\Sigma$. We will not examine the self-dual string worldsheet anomalies in the present paper, and we will simply assume that they cancel the boundary terms in anomaly computations. This gap is certainly worth filling in.

\paragraph{Self-dual fields} On $M - \Sigma$, $Y$ is trivializable. In the differential cocycle language, there is a differential cochain $\check{H} = (h,B,H)$ on $M - \Sigma$ such that
\be
\label{EqTrivChY}
d\check{H} = \check{Y} \;,
\ee
or in components
\be
\label{EqTrivChYComp}
dh = y \;, \quad H - h - dB = A \;, \quad dH = Y \;.
\ee
We will interpret $\check{H}$ as representing the self-dual 2-form gauge fields (or "$B$-fields") of the theory. While $\check{H}$ is ill-defined on $\Sigma$, we do not expect the self-dual 2-forms fields to be well-defined on $\Sigma$ either, due to the fact that $\Sigma$ is a source.

As $\check{Y}$ is a shifted differential cocycle, $\check{H}$ is a shifted differential cochain, with shift given by the $\Lambda_{\mathbb{R}}/\Lambda$-valued cochain $\frac{1}{2} \eta \otimes a$. $\frac{1}{2} \eta \otimes a$ trivializes the shift $\frac{1}{2} \nu \otimes a$ of $\check{Y}$, and is therefore compatible with $d \check{H} = \check{Y}$. We can construct the differential cochain $\check{\eta} = (\eta_\Lambda, 0,0)$, satisfying $d\check{\eta} = \check{\nu}$. We then obtain a natural trivialization of $\check{X}$:
\be
\label{EqDefTrivF}
\check{F} = \check{H} - \frac{1}{2} \check{\eta} \;, \quad d\check{F} = \check{X} \;.
\ee
$\check{F}$ will be useful to define the Green-Schwarz terms in equations \eqref{EqDefGST1} and \eqref{EqDefGST2}.

\paragraph{Gauge transformations} There are gauge transformations associated to the data $(\check{H}, \check{Y}, \check{\eta})$ we defined above. As explained in Appendix \ref{AppDiffCocCharCl}, a subgroup of such transformations is induced by diffeomorphisms, vectormultiplet gauge transformations, $B$-fields gauge transformations, and changes of the classifying map $\gamma$. Those are the transformations under which our constructions should be invariant. Nevertheless, it will be convenient to require that our constructions are invariant under the following larger group of transformations. 

We describe four classes of generators in this group.
\begin{enumerate}
\item Pullbacks by diffeomorphisms: Suppose $f: M \rightarrow M$ is a diffeomorphism. $f$ acts on the data above by
\be
\check{H} \mapsto f^\ast \check{H} \;, \quad \check{Y} \mapsto f^\ast \check{Y} \;, \quad \check{\eta} \mapsto f^\ast \check{\eta} \;, \quad
\ee
This simply means that our constructions may use the data $(\check{H}, \check{Y}, \check{\eta})$ but should otherwise be covariant.
\item B-field gauge transformations: $\check{H}$ can be shifted by the differential of a flat cochain, leaving $\check{Y}$ and $\check{\eta}$ invariant:
\be
\label{EqGaugTransHi}
\begin{aligned}
\check{Y} & \mapsto \check{Y} \;, \\
\check{H} & \mapsto \check{H} + d\check{W} \;, \\
\check{\eta} & \mapsto \check{\eta} \;,
\end{aligned}
\ee
where $\check{W} = (w, W, 0)\in \check{C}_0^2(M;\Lambda)$. In components:
\be
h \mapsto h + dw \;, \quad B \mapsto B - w - dW \;, \quad H \mapsto H \;.
\ee
These transformations include the small B-field gauge transformations discussed in the physical literature ($w = 0$), but also account for large gauge transformations ($w \neq 0$).
\item Gauge transformations of $\check{Y}$: $\check{Y}$ can be shifted   by the differential of a flat cochain. Compatibility with \eqref{EqTrivChY} then requires $\check{H}$ to be shifted by the same flat cochain:
\be
\begin{aligned}
\label{EqGaugTransY}
\check{Y} & \mapsto \check{Y} + d \check{V} \;, \\
\check{H} & \mapsto \check{H} + \check{V} \;, \\
\check{\eta} & \mapsto \check{\eta} \;,
\end{aligned}
\ee
where $\check{V} = (v, V,0)\in \check{C}_0^3(M;\Lambda)$. In components:
\be
\begin{aligned}
\label{EqGaugTransYExpl}
& y \mapsto y + dv \;, \quad A \mapsto A - v - dV \;, \quad Y \mapsto Y \;, \\
& h \mapsto h + v \;, \quad B \mapsto B + V \;, \quad H \mapsto H \;.
\end{aligned}
\ee
The small gauge transformations are those with $v = 0$. We show in Appendix \ref{AppDiffCocCharCl} that a subgroup of these transformations results from diffeomorphisms and vectormultiplet gauge transformations. The transformations above encode the famous fact that, in the Green-Schwarz mechanism, the $B$-field transforms under vector-multiplet gauge transformations and diffeomorphisms.
\item Changes of shift: $\eta_\Lambda$ can be shifted by an arbitrary  $\Lambda$-valued cochain $\rho$, thereby changing the (trivial) representing cocycle representative of the Wu class. The unshifted differential cocycles $\check{X}$ and $\check{F}$ are invariant under such transformations, and we can easily deduce the transformation of $\check{Y}$ and $\check{H}$:
\be
\begin{aligned}
\label{EqGaugChgShft}
\check{Y} & \mapsto \check{Y} + \frac{1}{2}d\check{\rho} \;, \\
\check{H} & \mapsto \check{H} + \frac{1}{2}\check{\rho} \;, \\
\check{\eta} & \mapsto \check{\eta} + \check{\rho}\;,
\end{aligned}
\ee
where $\check{\rho} = (\rho, 0,0)$. In components:
\be
\begin{aligned}
\label{EqGaugChgShftExpl}
& \eta \mapsto \eta + \rho \;, \quad y \mapsto y + \frac{1}{2}d\rho \;, \quad A \mapsto A - \frac{1}{2}\rho \;, \quad Y \mapsto Y \;, \\
& h \mapsto h + \frac{1}{2}\rho \;, \quad B \mapsto B \;, \quad H \mapsto H \;,
\end{aligned}
\ee
$\check{X}$ and $\check{F}$ are obviously invariant under changes of shift. Diffeomorphisms generally do not preserve the Wu cocycle, and therefore induce a change of shift transformation, as explained in Appendix \ref{AppDiffCocCharCl}.
\end{enumerate}

\paragraph{Local degrees of freedom} We should check that the model above has the correct local degrees of freedom. For this purpose, we can assume that all the fields are topologically trivial, so we would like to match the degrees of freedom of the system above with those of a pair of differential forms $(H,Y)$ of degree $(3,4)$ satisfying $dH = Y$.

The shifting differential cocycle $\check{\eta}$ does not contain any degree of freedom. This means that we are free to consider the unshifted differential cocycles $(\check{X}, \check{F})$ instead of $(\check{Y}, \check{H})$, or equivalently to assume that $(\check{Y}, \check{H})$ are unshifted. We work locally on a small open set, on which all the cocycles are topologically trivial. As $\check{Y}$ is topologically trivial, it can be put into the form $\check{Y} = (0,A,Y)$ by a gauge transformation. Here $A$ is a real cochain coming from a smooth differential form satisfying $dA = Y$. The gauge symmetry implies that $A$ is defined only up to shifts by exact forms. Therefore we recover the degrees of freedom of a closed 4-form, $Y$, while $A$ does not contain any local degree of freedom.

Assume that $\check{H}$ is "topologically trivial" as well, which we take to mean that $h = 0$ after a suitable gauge transformation. The second equation of \eqref{EqTrivChYComp} then reads $H - dB = A$, which is exactly \eqref{EqDefGaugInvFT}. As $A$ is fixed by a choice of vectormultiplet connection and metric (see Appendix \ref{AppDiffCocCharCl} for details), the coexact part of $H$ is fixed. The remaining degrees of freedom correspond to the exact part of $H$. Altogether, we see that we reproduce the degrees of freedom of a pair of differential forms $(H,Y)$ satisfying $dH = Y$, as required.

\paragraph{The degree 2 case} The formalism above is valid in any degree. Let us look at the case where $\check{Y}$ has degree 2, where a straightforward geometric interpretation exists. (See Example 2.7 in \cite{hopkins-2005-70}.) Degree 2 differential cohomology classes are in bijection with isomorphism classes of line bundles with connection. $\check{Y}$ therefore describes a line bundle $L$ with connection.

In this case, $\Sigma$ is simply a finite set of points on the surface $M$. The fact that $\check{Y}$ is trivializable over $M - \Sigma$ means that $L|_{M - \Sigma}$ is trivializable.

How can we interpret $\check{H}$? The differential cohomology of degree 1 over a manifold coincides with the $U(1)$-valued functions over this manifold, with the isomorphism given explicitly by the exponential of the connection. So the space of all $\check{H}$ modulo the corresponding gauge transformations is a torsor over the group of $U(1)$-valued functions over $M - \Sigma$. We can think of it as the space of sections of $L$ over $M - \Sigma$. Therefore, the data $(\check{Y}, \check{H})$ correspond to a line bundle $L$ over $M$ and a section of $L$ over $M - \Sigma$. See also Example 2.25 in \cite{Freed:2000ta}.

In the context of interest to us, $\check{Y}$ and $\check{H}$ have respectively degrees 4 and 3.  By analogy with the degree 2 case, degree 3 differential cohomology classes are in bijection with isomorphism classes of gerbes with connections, and degree 4 differential cohomology classes are in bijection with isomorphism classes of 2-gerbes with connection. The previous statement can indeed be taken as a definition of Abelian gerbes and 2-gerbes. $\check{H}$ can therefore be interpreted as defining a section of the Abelian 2-gerbe with connection associated to $\check{Y}$ over $M - \Sigma$.

\paragraph{A model in differential KO-theory?} We should remark here that six-dimensional supergravity theories are usually defined on $\mathbb{R}^{5,1}$, where the differential form model for gauge fields is sufficient. In order to generalize such theories to topologically non-trivial manifolds, one has to choose a generalized cohomology theory whose differential version governs the gauge fields and the self-dual fields. See for instance \cite{Freed:2000ta, Freed:2006yc, Freed:2006ya} for a discussion. In our model, we chose to model the self-dual gauge fields by cochains in the model of ordinary differential cohomology. This is natural if we see them as chiral Abelian gauge fields. However, the cancellation of global anomalies through the Green-Schwarz mechanism in type I string theory \cite{Freed:2000ta} requires the $B$-field to be modeled as a differential KO-theory cochain. We will see that we can prove global anomaly cancellation only up to a certain bordism invariant. Unfortunately, the bordism invariant seems to be difficult to
compute for the case of an arbitrary compact Lie group $G$ with quaternionic representation. It would be very interesting to learn if this
 difficulty could be overcome by modeling the 6-dimensional self-dual fields using a different differential generalized cohomology theory, such as differential KO-theory.

\section{Construction of the Wu Chern-Simons field theory}

\label{SecTFT}

We construct in the present section a 7-dimensional field theory on manifolds with Wu structure endowed with an unshifted degree 4 differential cocycle $\check{X}$. We will see in Section \ref{SecRelGSAFT} that its partition function is a geometric invariant of 7-dimensional manifolds that coincides with the invariant \eqref{EqAnGSTerms} expected to be associated with the Green-Schwarz terms, after a suitable identification of the relation between the differential cocycles $\check{X}$ and $\check{Y}$. The Green-Schwarz terms themselves will then be constructed in Section \ref{SecGSTerms} as vectors in the state space of this field theory.

While we usually consider spin manifolds in the present paper, no spin condition is assumed in this section. The word
"manifold" will refer to a smooth compact oriented manifold, possibly with boundary, endowed with a Wu structure. We write $(M,\check{X})$ for a manifold endowed with a degree 4 unshifted differential cocycle $\check{X}$. Accordingly, instead of equipping the 6- and 7-dimensional manifolds with classifying maps into $B_{\rm W}\bar{G}$, they only have classifying maps into $B_{\rm W}SO$, see Appendix \ref{SecSpinWu}.

We will assume that $\Lambda$ is a unimodular (i.e. self-dual) lattice, because this is actually a constraint on 6d supergravity theories \cite{SeibergTaylor} and it simplifies the construction of the Wu Chern-Simons field theory. At the end of Section \eqref{SecStateSpace}, we will discuss the modifications needed when $\Lambda$ is not unimodular and explain that they result in the Wu Chern-Simons theory not being invertible. As this impairs the construction of the Green-Schwarz terms, our construction rederives the unimodularity of $\Lambda$ as a consistency constraint on the 6d supergravity theory.

\subsection{Linking pairing}

Let $U$ be a 7-manifold, possibly with boundary $\partial U$. The torsion subgroup $H^4_{\rm tors}(U,\partial U; \Lambda) \subset H^4(U,\partial U; \Lambda)$ of the degree 4 relative cohomology of $(U,\partial U)$ valued in $\Lambda$ carries a $\mathbb{R}/\mathbb{Z}$-valued pairing, defined as follows. Let $x_1$ and $x_2$ be cocycle representatives for classes $[x_1], [x_2] \in H^4_{\rm tors}(U,\partial U; \Lambda)$. Suppose $x_2$ has order $k$. Then there is a cochain $y$ such that $dy = kx_2$. Define
\be
\tilde{L}(x_1, x_2) := \frac{1}{k} \int_U x_1 \cup y \quad \mbox{mod } 1 \;.
\ee
$\tilde{L}$ passes to a well-defined pairing on the cohomology, the \emph{linking pairing}.

One can also define an $\mathbb{R}/\mathbb{Z}$-valued pairing on the subgroup $\check{H}^4_{\rm flat}(U,\partial U;\Lambda) \subset \check{H}^4(U,\partial U;\Lambda)$ of flat relative differential cohomology classes on $U$ valued on $\Lambda$ as follows. Let $\check{X}_1 = (x_1, A_1, 0)$ and $\check{X}_2 = (x_2, A_2,0)$ be flat differential cocycles and define
\be
\label{DefPairingFlatDiffCoc}
L(\check{X}_1, \check{X}_2) := \int_U x_1 \cup A_2 = \int_U \check{X}_1 \cup \check{X}_2 \;.
\ee
The pairing $L$ passes to a well-defined pairing on the flat differential cohomology and is closely related to the linking pairing. Indeed, we have a homomorphism $\check{H}^4_{\rm flat}(U,\partial U;\Lambda) \rightarrow H^4_{\rm tors}(U,\partial U; \Lambda)$ sending a flat differential cocycle $\check{X} = (x,A,0)$ to the cohomology class $[x]$. (Note that $[x]$ is torsion because the cocycle condition forces $x = -dA$ in the absence of curvature.) Proposition 4.8 of \cite{Monnier:2016jlo} shows that $L$ is the pull-back of the linking pairing $\tilde{L}$ through the homomorphism above.

\subsection{The action}

\label{SecAction}

\paragraph{Motivation} The first step toward defining a $d$-dimensional Chern-Simons action is to specify a $d+1$-dimensional integral characteristic class. For instance, the standard (level 1) 3-dimensional Chern-Simons theory for a semi-simple gauge group is associated to the second Chern Class $c_2$. Then, on $d$-dimensional manifolds that bound, the Chern-Simons action can be defined as the integral of the associated characteristic form on a bounding $d+1$-dimensional manifold (which is ${\rm Tr} F^2$ in the 3-dimensional case).

We are interested in constructing a Chern-Simons action associated to $\frac{1}{2} (x \cup x)$, where $x$ is the characteristic of the differential cocycle $\check{X}$, seen as a characteristic class associated to the 8-dimensional manifold $(Z,\check{X})$. An immediate problem is that $\frac{1}{2} (x \cup x)$ is not an integral class: in general it has half-integer periods. To remedy this, we modify the characteristic class to
\be
\label{EqModCharClass}
\frac{1}{2} x \cup (x + \nu_\Lambda)
\ee
where $\nu_\Lambda = \nu_\mathbb{Z} \otimes a$. $\nu_\mathbb{Z}$ is a suitable lift of the Wu class to $H^4(Z;\Lambda)$ whose construction will be explained soon. The properties of the Wu class will ensure that \eqref{EqModCharClass} has integral periods, removing the obstruction to constructing a Chern-Simons action. (As an aside, note that we cannot construct $\nu_\mathbb{Z}$ by pulling back the cocycle $\nu_{\mathbb{Z}, {\rm U}}$ described in Appendix \ref{SecSpinWu}. Being an 8-dimensional manifold, $Z$ does not necessarily admit a Wu structure and a classifying map into $B_{\rm W}SO$.)

\paragraph{Lagrangian} Let now $U$ be a 7-manifold. Our next aim is to construct the Chern-Simons Lagrangian on $U$. This can be achieved by refining the characteristic class \eqref{EqModCharClass} to a degree 8 differential cohomology class. The connection part of this differential cohomology class is a degree 7 $\mathbb{R}$-valued cochain. We take the Chern-Simons Lagrangian to be its reduction modulo 1, because integer shifts of the Chern-Simons action are physically irrelevant. Equivalently, we can construct the Chern-Simons action by integrating the differential cocycle over $U$: by definition, this integral is the integral of the connection modulo 1.

$U$ admits a classifying map into $B_{\rm W}\bar{G}$, and therefore comes equipped with a $\mathbb{Z}$-valued cochain $\eta_\mathbb{Z}$. Pick a characteristic element $\tilde{a}$ of $\Lambda$ and define $\eta_\Lambda := \eta_\mathbb{Z} \otimes \tilde{a}$. The reason why $\tilde{a}$ has to be a characteristic element will be explained momentarily. We construct the trivializable differential cocycle
\be
\check{\nu} := (d\eta_\Lambda, -\eta_\Lambda, 0) = d(\eta_{\Lambda},0,0) \in \check{Z}_0^4(M;\Lambda) .   
\ee

Let $\check{X} = (x,A,X)$ be an unshifted differential cocycle of degree 4 on $U$ and define the following degree seven $\mathbb{R}/\mathbb{Z}$-valued cocycle, to be thought of as a Lagrangian for the field theory to be defined:
\begin{align}
\label{EqDefLag}
l(\check{X}) \: & := \frac{1}{2} [\check{X} \cup (\check{X} + \check{\nu})]_{\rm hol} \\
& = \frac{1}{2} x \cup (A - \eta_{\Lambda}) + \frac{1}{2} A \cup X + \frac{1}{2}H^\cup_\wedge(X, X) \quad {\rm mod} \; 1 \notag \;,
\end{align}
where $[...]_{\rm hol}$ denotes as usual the connection part and we used the definition of the cup product of differential cochains on the second line. Note that although $\check{\nu}$ is a trivializable differential cocycle $\frac{1}{2} \check{X} \cup (\check{X} + \check{\nu})$ is not equivalent to $\frac{1}{2} \check{X} \cup \check{X}$, because of the factor $1/2$.

\paragraph{Action} As explained in \cite{Monnier:2016jlo} (see (4.2)-(4.6) there), integrating \eqref{EqDefLag} over a closed $7$-fold $U$ does not yield a gauge invariant action. The reason for this is that $x$ appears in the Lagrangian with a prefactor $\frac{1}{2}$. Under large gauge transformations, which shift $x$ by integer cocycles, the integrated Lagrangian generally changes by a half-integer, which might change the sign of the exponentiated action. The latter is therefore not invariant under large gauge transformations of $\check{X}$.

One should rather proceed as follows. Let us write $x_2$ for the mod 2 reduction of the $\Lambda$-valued cochain $x$. The pair $\bar{l}(\check{X}):=(l(\check{X}), x_2)$ defines a cocycle in a cochain model for a generalized cohomology theory called E-theory, see Appendix \ref{AppEThCalc}. The lattice element $\tilde{a} \in \Lambda$ enters the definition of a certain twist of the E-theory, and consistency requires it to be a characteristic element of $\Lambda$.%
\footnote{$\nu \otimes \tilde{a}$ appears as a twist in the E-theory differential \eqref{EqDefTwDiffEThCoch}. This twist by a characteristic element is required in order for the E-theory integration map on Wu manifolds to be well-defined. The construction appears in Appendix D of \cite{Monnier:2016jlo}, and the characteristic property, hidden in $\tilde{\alpha}$, is required to make (D.28) commute there.}
A more conceptual reason for why $\tilde{a}$ should be a characteristic element will be presented later. $n$-dimensional manifolds with Wu structures, such as $U$ for $n=7$, come with an integration map sending degree $n$ E-theory cochains to $\mathbb{R}/\mathbb{Z}$. We write it $\int^{\rm E}_{U,\omega}$, where we wrote explicitly the dependence on the Wu structure $\omega$. The action is defined by
\be
\label{EqActWCS}
S_\omega(U;\check{X}) := \int_{U,\omega}^{\rm E} (l(\check{X}), x_2) \;,
\ee
and is gauge invariant, as proven in Proposition 4.2 of \cite{Monnier:2016jlo}.

The integration map also makes sense on manifolds with boundary, although as usual, the result then depends on the cocycle $(l(\check{X}),x_2)$ through its boundary values, not just on its E-theory class. See the discussion in Section \ref{SecPreqTh}.

\paragraph{The action as a quadratic refinement} In the following computation, we use freely the calculus of E-cochains described in Appendix \ref{AppEThCalc} (see also Appendix D of \cite{Monnier:2016jlo}). Let $\check{X}$ be as before and let $\check{Z} = (z,Z,0)$ be a flat relative degree 4 differential cocycle on $(U,\partial U)$. $U$ carries a Wu structure $\omega$. The action has the following quadratic property:
\be
\label{EqExprQRy}
\begin{aligned}
S_\omega(U;\check{X} + \check{Z}) \: & = \int^{\rm E}_{U,\omega} \bar{l}(\check{X} + \check{Z}) \\
& = \int^{\rm E}_{U,\omega} \left( l(\check{X} + \check{Z}), (x)_2 + (z)_2 \right) \\
& = \int^{\rm E}_{U,\omega} \left(l(\check{X}) + l(\check{Z}) + \frac{1}{2}(x \cup Z + z \cup A + Z \cup X), (x)_2 + (z)_2 \right) \\
& = \int^{\rm E}_{U,\omega} \left(l(\check{X}) + l(\check{Z}) + \frac{1}{2}(x \cup Z + Z \cup x - d(Z \cup A)), (x)_2 + (z)_2 \right) \\
& = \int^{\rm E}_{U,\omega} \left(l(\check{X}) + l(\check{Z}) + x \cup Z + \frac{1}{2} x \cup_1 z - \frac{1}{2} d(Z \cup A - x \cup_1 Z), (x)_2 + (z)_2 \right) \\
& = S_\omega(U;\check{X}) + \int_U \check{X} \cup \check{Z} + S_\omega(U;\check{Z}) \;.
\end{aligned}
\ee
In this derivation, we use respectively the explicit expression \eqref{EqDefLag} of the Lagrangian, the fact that $\check{X}$ is a closed differential cocycle, the expression \eqref{EqRelHighCupProd} of the commutator for the cup product in terms of higher cup products, the definition \eqref{EqGrpLawCocModETh} of the sum of E-cochains, and the fact that $\int^E_{U,\omega}$ is a group homomorphism. We have been able to drop the exact term on the 5th line because the cocycle $\check{Z}$ is relative to the boundary. A term $\frac{1}{2} x \cup_1 z$ was absorbed when decomposing the E-cochain on the 5th line into a sum of E-cochains according to \eqref{EqGrpLawCocModETh}. In fact, we know from Proposition 4.7 in \cite{Monnier:2016jlo} that $S_\omega(U;\check{Z})$ is a quadratic refinement $q$ of the pairing $L$ when restricted to flat differential cocycles. (I.e. it satisfies the additional relation $S_\omega(U;n\check{Z}) = n^2S_\omega(U;\check{Z})$.) The relation derived above, is however valid for arbitrary differential cocycles $\check{X}$.

\paragraph{The action on boundaries of 8-manifolds} Suppose that $U$ is the boundary of an 8-manifold $W$. In general the Wu class $\nu(W)$ of $W$ is not trivial, and $W$ does not admit a Wu structure. Let us pick an integral lift $\nu_\mathbb{Z}$ of the $\mathbb{Z}_2$-cocycle $\nu(W)$, such that $\nu_\mathbb{Z}|_U = d\eta_\mathbb{Z}$. Extend as well $\eta_\mathbb{Z}$ arbitrarily to $W$ as a real cocycle. We can arrange so that the real cocycle $\lambda'' := -d\eta_\mathbb{Z} + \nu_\mathbb{Z}$ is smooth and represented by a differential form. Is then $\lambda''$ is a differential form lifting the Wu class of $W$ and vanishing on $U$. By construction, $\lambda''$ records the choice of Wu structure $\omega$ on $U$ that was encoded in $\eta_\mathbb{Z}$.

Construct $\nu_\Lambda := \nu_\mathbb{Z} \otimes \tilde{a}$, $\eta_\Lambda := \eta_\mathbb{Z} \otimes \tilde{a}$, $\lambda' = \lambda'' \otimes \tilde{a}$, where $\tilde{a}$ is the given characteristic element of $\Lambda$. We have a relative differential cocycle $\check{\nu} = (\nu_\Lambda, -\eta_\Lambda, \lambda')$. Assume as well that $\check{X}$ extends to a differential cocyle $\check{X}_W = (x_W, A_W, X_W)$ on $W$. Note that the characteristic of $\frac{1}{2} \check{X}_W \cup (\check{X}_W + \check{\nu})$ represents the characteristic class \eqref{EqModCharClass}. Proposition 4.6 of \cite{Monnier:2016jlo} shows that the action is given by the integral of the curvature of $\frac{1}{2} \check{X}_W \cup (\check{X}_W + \check{\nu})$ over $W$:
\be
\label{EqActOnBound}
S_\omega(U;\check{X}) = \frac{1}{2} \int_W X_W \wedge (X_W + \lambda') \quad \mbox{mod } 1 \;.
\ee
\eqref{EqActOnBound} expresses the action in terms of an integral of ordinary differential forms over $W$, rather than an integral in E-theory over $U$.

We see here clearly why $\tilde{a}$ should be a characteristic element of $\Lambda$: this ensures that on a closed manifold $Z$, $\lambda'$ is a characteristic element of the lattice $H^4_{{\rm dR}}(Z;\Lambda) \subset H^4_{{\rm dR}}(Z;\Lambda_{\mathbb{R}})$ composed of de Rham cohomology classes with integral periods. This is necessary for the right-hand side of \eqref{EqActOnBound} to vanish modulo 1 on $Z$, and therefore for \eqref{EqActOnBound} to be independent of the choice of bounding manifold $W$.

\subsection{Prequantum theory}

\label{SecPreqTh}

We can construct out of the action $S$ an invertible quantum field theory, the prequantum theory ${\rm WCS}_\omega^{\rm PQ}$. We only sketch this construction here, see Section 5.2 of \cite{Monnier:2016jlo} for the details.

\paragraph{Partition function} Let $(U,\check{X})$ be a 7-dimensional manifold endowed with a degree 4 differential cocycle and a Wu structure $\omega$. The partition function of the prequantum theory on $(U,\check{X})$ is the exponentiated action:
\be
{\rm WCS}_\omega^{\rm PQ}(U;\check{X}) = \exp 2\pi i S_\omega(U;\check{X}) \;.
\ee

\paragraph{Prequantum state space} Let $(M,\check{X})$ be a 6-dimensional manifold endowed with a degree 4 differential cocycle. Up to a caveat to be described soon, the state space ${\rm WCS}_\omega^{\rm PQ}(M;\check{X})$ is a hermitian line satisfying the following properties. Each cocycle representative $\check{X}$ defines a trivialization of ${\rm WCS}_\omega^{\rm PQ}(M;\check{X})$. The relation between the trivializations associated to $\check{X}_1$ and $\check{X}_2$ is given by the value of the partition function above on a cylinder $M \times I$ with $\check{X}_1$ at one end and $\check{X}_2$ at the other end. Concretely, with $I = [0,1]$, we pick a smooth function from $\rho: I \rightarrow I$, $\rho(0) = 0$, $\rho(1) = 1$ that is constant near $0$ and $1$. We also pick a (necessarily discontinuous) function $\tilde{\rho}: I \rightarrow \{0,1\}$ such that $\tilde{\rho}(0) = 0$, $\tilde{\rho}(1) = 1$. Then, if $\check{X}_2 = \check{X}_1 + d \check{W}$ with $\check{W} = (w,W,0)$, construct the following differential cocycle on $M \times I$:
\be
\check{X}_{12} = \check{X}_1 + dW_I \;, \quad W_I = (\tilde{\rho} w, \rho W,0) \;,
\ee
where the pullbacks of $\check{X}_1, w, W$ from $M$ to $M \times I$ are implicit. $\check{X}_{12}$ interpolates between $\check{X}_1$ and $\check{X}_2$ on a cylinder and the associated trivializations differ by $\exp 2 \pi i {\rm WCS}_\omega^{\rm PQ}(M \times I;\check{X}_{12})$.

If we restrict ourselves to spin manifolds (which is all we will need for applications to supergravity), we can use the following fact to obtain a more intuitive picture of the state space of the prequantum theory. Any spin 6-manifold $M$ endowed with a degree 4 differential cocycle $\check{X}$ can be seen as the boundary of a 7-manifold $U$ endowed with a differential cocycle $\check{X}_U$ extending $\check{X}$. In other words, the bordism groups $\Omega^{\rm Spin}_6({\rm pt})$ and $\Omega^{\rm Spin}_6(K(\mathbb{Z},4))$ vanish (see Stong's appendix in \cite{Witten:1985bt} for a proof of this second fact). Then the state space of the prequantum theory on $(M,\check{X})$ is the vector space of complex-valued functions $f$ on extensions $(U,\check{X}_U)$ such that
\be
f(U_2, \check{X}_{U_2})/f(U_1, \check{X}_{U_1}) = \exp 2\pi i S_\omega(U_{12}; \check{X}_{U_{12}})
\ee
whenever $(U_{12}, \check{X}_{U_{12}})$ is obtained from $(U_1,\check{X}_{U_1})$ and $(U_2,\check{X}_{U_2})$ by flipping the orientation of $U_1$ and gluing it to $U_2$ along $M$.

\paragraph{Torsion anomaly} For the picture above to be consistent, the partition function on the torus $M \times S^1$ (corresponding to a cylinder from a cocycle representative to itself, i.e. to the identity gauge transformation) should be 1. If this condition is not satisfied, there is an anomaly and the state space is simply the zero Hilbert space. This torsion anomaly provides constraints on the set of Wu structures.

We see that if $\check{X}_1 = \check{X}_2$, then $d\check{W} = 0$, so $\check{X}_{12}$ on $M \times S^1$ is the sum of the pullback of $\check{X}_1$ and of a flat differential cocycle $\check{Z}$. $\check{Z}$ is obtained by pushing forward $dW_I$ from $M \times I$ to $M \times S^1$, which is possible because $dW_I|_{M \times \{0\}} = dW_I|_{M \times \{1\}}$. While $dW_I$ is obviously exact, $\check{Z}$ is in general not exact on $M \times S^1$.

We therefore set $U = M \times S^1$, $\check{X}$ a differential cocycle on $U$ pulled back from $M$ and $\check{Z} = (z,Z,0)$ as above. We compute
\be
\label{EqHamAnomCancel}
S_\omega(U;\check{X} + \check{Z}) - S_\omega(U;\check{X}) = S_\omega(U;\check{Z}) + \int_U \check{X} \cup \check{Z} = q_\omega(z) + \int_U x \cup Z \;,
\ee
where we used \eqref{EqExprQRy}, as well as the fact that the action evaluated on flat cocycles is the pullback of a quadratic refinement $q_\omega$ of the linking pairing. $z$ is of the form $\theta \cup z'$, where $\theta$ is a cocycle generating $H^1(S^1;\mathbb{Z})$ and $[z'] \in H^3_{\rm tors}(M;\Lambda)$. The classes represented by such $z$'s form a subgroup 
\be
T := H^3_{\rm tors}(M;\Lambda) \cup \theta \subset H^4_{\rm tors}(M;\Lambda)
\ee
isotropic with respect to the linking pairing. Quadratic refinements are linear on isotropic subgroups, which shows that both terms in the right-hand side of \eqref{EqHamAnomCancel} are linear in $\check{Z}$. Nevertheless, \eqref{EqHamAnomCancel} clearly cannot vanish for arbitrary $\check{X}$: the first term on the right-hand side depends on $z = - dZ$ only, while the second term depends on $Z$. We can have a cancellation for all $\check{Z}$ only if $x = dv$ for $v$ some real cocycle, i.e. $x$ is torsion. In this case the second term on the right-hand side of \eqref{EqHamAnomCancel} can be rewritten
\be
-\int_U v \cup z = - \tilde{L}(x, z) \;.
\ee
We write $x_0 \in H^4_{\rm tors}(M;\Lambda)$ for the class  making the right-hand side of \eqref{EqHamAnomCancel} vanish. As $q_\omega(z)$ is valued in $\{0,1/2\} \subset \mathbb{R}/\mathbb{Z}$, $x_0$ is 2-torsion at most (and therefore the sign is immaterial). It is also not difficult to check that on flat cocycles $z = z' \otimes \alpha$, with $z' \in H^4_{\rm tors}(M;\mathbb{Z})$ and $\alpha \in \Lambda$, we have
\be
q_\omega(z) = q_{\mathbb{Z},\omega}(z') \cdot (\alpha,\alpha)
\ee
where $q_{\mathbb{Z},\omega}$ is the quadratic refinement associated to the action for the lattice $\Lambda = \mathbb{Z}$. This implies that $x_0$ has necessarily the form $x_0 = x_0' \otimes \gamma$, where $x_0 \in H^4_{\rm tors}(M;\mathbb{Z}_2)$ and $\gamma$ is the unique characteristic element of $\Lambda/2\Lambda$. (The unimodularity of $\Lambda$ ensures the uniqueness of $\gamma$.) $x_0'$ is completely determined by the restriction of $q$ on $T$. Analogues of this anomaly have been discussed in various contexts \cite{Witten:1999vg, Diaconescu:2003bm, Belov:2006jd, Monniera, Monnier:2016jlo}.

We now show that the torsion anomaly $x_0$ vanishes for a suitable choice of Wu structure. By definition, $q_\omega$ is the $\mathbb{Z}_2$-character $\tilde{L}(x_0, \bullet)$ on $T$. As the order 2 part of $T$ is mapped injectively into $H^4(M \times S^1; \Lambda/2\Lambda)$ under the reduction mod 2, we can use Poincaré duality on $M \times S^1$ to find a class $\delta \otimes \gamma \in H^3(M \times S^1;\Lambda/2\Lambda)$, $\delta \in H^3(M \times S^1;\mathbb{Z}_2)$ such that for all $z = z' \otimes \alpha \in T$, we have
\be
\tilde{L}(x_0, z) = \langle z_2 \cup (\delta \otimes \gamma), [M \times S^1] \rangle = \langle z'_2 \cup \delta, [M \times S^1] \rangle (\alpha, \gamma) \;.
\ee
Here, $z'_2$ is the reduction mod 2 of $z'$. Given the form of $z$, $\delta$ can be chosen to be the pullback of a class $\delta' \in H^3(M;\mathbb{Z}_2)$. Recall now that Wu structures on $M$ form a torsor for $H^3(M;\mathbb{Z}_2)$. Let us shift the Wu structure on $M$ by $\delta'$. This induces a change of Wu structure on $M \times S^1$ from $\omega$ to $\omega' = \omega + \delta$. Proposition 4.5 of \cite{Monnier:2016jlo} shows that
\be
q_{\omega'}(z) = q_{\omega}(z) - \langle z_2 \cup (\delta \otimes \gamma), [M \times S^1] \rangle \;.
\ee
In order words, $q_{\omega'}$ vanishes on $T$ and therefore the torsion anomaly $x_0$ associated to the Wu structure $\omega'$ vanishes. Note that this procedure does not determine the Wu structure uniquely, due to some freedom in constructing $\delta$. We will call \emph{good} the Wu structures on 6-manifolds that have a vanishing associated torsion anomaly. All the Wu structures used in the present paper from now on will be good.

\subsection{Bordism category}

\label{SecCobCat}

The prequantum theory satisfies the gluing axioms, and therefore is a field theory functor, only with respect to a suitable bordism category $\mathcal{C}_{\rm PQ}$. This is the bordism category whose objects are the 6-dimensional manifolds $(M,\check{X})$ such that $\check{X}$ lies in the torsion class determined by anomaly cancellation. The reason is that given a 7-manifold with a cut along a 6-manifold, the corresponding prequantum amplitude/partition function cannot be factored through the cut if the state space associated to the cut is the zero Hilbert space.

The Wu Chern-Simons theory to be constructed next can be defined on 6-manifolds endowed with arbitrary Wu structures. The subtlety mentioned above then requires certain restrictions on morphisms. To simplify the discussion a bit, we will rather use the fact that in the supergravity context, the Wu structure can be freely chosen, and restrict the definition of the Wu Chern-Simons theory to 6-manifolds carrying good Wu structures.

The Wu Chern-Simons theory is a functor from a bordism category $\mathcal{C}_{WCS}$ defined as follows.
\begin{itemize}
\item The objects of $\mathcal{C}_{WCS}$ are 6-dimensional closed smooth oriented manifolds endowed with a good Wu structure, a trivializable degree 4 differential cocycle $\check{X}$ and a classifying map into $B_{\rm W}SO$.
\item The morphisms of $\mathcal{C}_{WCS}$ are 7-dimensional compact smooth oriented manifolds with boundary, endowed with a Wu structure, a degree 4 differential cocycle $\check{X}$ and a classifying map into $B_{\rm W}SO$. Of course, on the boundary, the induced Wu structure should be good, $\check{X}$ should be trivializable and the classifying maps must match.
\end{itemize}

\subsection{Wu Chern-Simons theory: the partition function}

\label{SecPartFunc}

\paragraph{Definition} The quantum field theories we define below are akin to Dijkgraaf-Witten (DW) theories, in the sense that their configuration space of dynamical fields is finite. The path integral therefore reduces to a finite sum. However, unlike in DW theories, it cannot be interpreted as the gauging of a finite symmetry.

Let $U$ be a morphism in $\mathcal{C}_{\rm WCS}$. The path integral defining the WCS theory on $U$ is essentially a gauss sum for the quadratic refinement $q_\omega$. Recall that $q_\omega$ is the quadratic refinement of the linking pairing $\tilde{L}$ on $H^4_{\rm tors}(U,\partial U;\Lambda)$ defined by the action $S_\omega$. If $\partial U \neq \emptyset$, $\tilde{L}$ has a non-trivial radical $R(U)$ in $H^4_{\rm tors}(U,\partial U;\Lambda)$, i.e. there are $0 \neq x \in H^4_{\rm tors}(U,\partial U;\Lambda)$ such that $\tilde{L}(x,y) = 0$ for all $y \in H^4_{\rm tors}(U,\partial U;\Lambda)$. This is at first sight worrisome, because if $q$ is not \emph{tame}, i.e. if it does not vanish on $R(U)$, then its associated Gauss sum vanishes \cite{Taylor}, which would mean that the WCS theory is not invertible. As our aim is eventually to compare the WCS theory with the anomaly field theory of the six-dimensional supergravity, this would be a problem. Fortunately, the constraint that the Wu structure is good precisely ensures that $q$ vanishes on $R(U)$, as we now show.

The classes in $R(U)$ have representatives supported on a tubular neighborhood $\mathbb{R} \times \partial U$ of the boundary, and take the form $\theta \cup x'$, where $\theta$ generates $H^1_{\rm compact}(\mathbb{R};\mathbb{Z})$ and $x' \in H^3_{\rm tors}(\partial U;\Lambda)$. It is a general fact \cite{Taylor} that the restriction of a quadratic refinement to the radical of the associated pairing is a $\frac{1}{2}\mathbb{Z}/\mathbb{Z}$-valued character. In the case of $q_\omega$, this character is given by $\tilde{L}(x_0, \bullet)$, by the very definition of the torsion anomaly. As the Wu structure on $\partial U$ is assumed to be good, $x_0 = 0$ and $q_\omega$ is tame. Tame quadratic refinements have an Arf invariant, which is the complex argument of the (non-vanishing) associated Gauss sum:
\be
{\rm Arf}(q_\omega) = {\rm arg} \left( \sum_{z \in H^4(U,\partial U;\Lambda)} \exp 2\pi i q_\omega(z) \right) \;.
\ee
If the quadratic refinement were not tame, the Gauss sum would vanish and there would be no associated Arf invariant.

We now define the partition function of the Wu Chern-Simons theory on 7-manifolds as follows:
\be
\label{EqDefPartFuncGS}
\begin{aligned}
{\rm WCS}_\omega(U;\check{X}) \: :& =  N(U) \sum_{[z] \in H^4_{\rm tors}(U,\partial U;\Lambda)} \exp 2\pi i \left( S_\omega(U;\check{X}) - S_\omega(U;\check{Z}) \right) \;, \\
& = \exp 2\pi i (S_\omega(U;\check{X}) - {\rm Arf}(q_\omega)) \;.
\end{aligned}
\ee
The normalization factor $N(U)$ is given by
\be
\label{EqDefNormFact}
N(U) := \frac{1}{\sqrt{|H^4_{\rm tors}(U,\partial U;\Lambda)||R(U)|}} \;.
\ee
It coincides with the modulus of the Gauss sum of $q_\omega$ \cite{Taylor}, which is why it disappears on the second line of \eqref{EqDefPartFuncGS}. The normalization ensures that ${\rm WCS}$ satisfies the relevant gluing relations, proven in Appendix \ref{AppGluing}. The partition function is invariant under equivalences of differential cocycles acting on $\check{X}$ and leaving its boundary value constant. On a closed 7-manifold, it depends only on the differential cohomology class of $\check{X}$.

\paragraph{The partition function when $U$ is a boundary of an 8-dimensional manifold} $U$ is assumed here to be closed. Suppose that $\check{X}$ extends as $\check{X}_W$ to an 8-manifold $W$ such that $\partial W = U$. We know that the action, and therefore the first factor in the partition function, takes the form \eqref{EqActOnBound}. Then standard arguments, detailed for instance in \cite{Brumfiel1973}, allow us to express the Arf invariant in terms of data on $W$:
\be
{\rm Arf}(q_\omega) = \frac{1}{8} \left(\sigma_{H^4(W,U;\Lambda)} - \int_W \lambda'^2 \right) \;,
\ee
where $\lambda'$ is the relative differential form on $W$ defined above \eqref{EqActOnBound}. We therefore obtain:
\be
\label{EqPartFuncGSTermsBound}
\begin{aligned}
{\rm WCS}_\omega(U;\check{X}) \: & =  \exp \frac{2\pi i}{8} \left( 4\int_W X_W \wedge (X_W + \lambda') + \int_W \lambda'^2 - \sigma_{H^4(W,U;\Lambda)} \right) \\
& = \exp 2\pi i \left( \int_{W} \frac{1}{2}(X_W')^2 - \frac{\sigma_{H^4(W,U;\Lambda)}}{8} \right)
\end{aligned}
\ee
where we defined $X_W' = \frac{1}{2} \lambda' + X_W$.

\paragraph{Dependence on the Wu structure} It is also interesting to understand how the partition function ${\rm WCS}_\omega(U;\check{X})$ depends on the Wu structure $\omega$ of $U$. A change in Wu structure $\omega \rightarrow \omega'$ is described by an element $\delta \in H^3(U,\partial U; \mathbb{Z}_2)$. Then Proposition 4.5 of \cite{Monnier:2016jlo} says that
\be
S_{\omega'}(U;\check{X}) = S_{\omega}(U;\check{X}) - \langle x_2 \cup \delta_{\Lambda/2\Lambda}, [U,\partial U] \rangle \;,
\ee
where we recall that $x_2$ is the mod 2 reduction of the characteristic $x$ of $\check{X}$. $\delta_{\Lambda/2\Lambda} = \delta \otimes \gamma$, where $\gamma \in \Lambda/2\Lambda$ is the unique characteristic element of $\Lambda/2\Lambda$, satisfying $(\alpha,\alpha) = (\alpha, \gamma)$ for all $\alpha \in \Lambda/2\Lambda$. The quadratic refinement $q$ changes therefore by
\be
q_{\omega'}(x) = q_{\omega}(x) - \langle x_2 \cup \delta_{\Lambda/2\Lambda}, [U,\partial U] \rangle \;.
\ee
Writing $\beta$ for the Bockstein homomorphism from $H^3(U,\partial U; \Lambda/2\Lambda)$ to $H^4(U,\partial U; \Lambda)$, we can rewrite the second term in terms of the linking pairing
\be
q_{\omega'}(x) = q_{\omega}(x) - \tilde{L}(x,\beta(\delta_{\Lambda/2\Lambda})) \;.
\ee
The corresponding Arf invariant transforms as (see Proposition 1.13 of \cite{Taylor})
\be
{\rm Arf}(q_{\omega'}) = {\rm Arf}(q_{\omega}) - q_{\omega}(\beta(\delta_{\Lambda/2\Lambda})) \;.
\ee
Pick an integral lift $\delta_{\Lambda}$ of $\delta_{\Lambda/2\Lambda}$. Define $\check{\Delta} = \left(d\delta_{\Lambda}, -\delta_{\Lambda}, 0\right)$. Then, by the construction of the Bockstein homomorphism $\beta$ associated to the short exact sequence of groups $\Lambda \stackrel{2 \cdot }{\rightarrow} \Lambda \rightarrow \Lambda/2\Lambda$, $\check{\Delta}/2$
has a characteristic whose cohomology class is $\beta(\delta_{\Lambda/2\Lambda})$. We have $q(\beta(\delta_{\Lambda/2\Lambda})) = S(U; \check{\Delta}/2)$ and $\tilde{L}(x, \beta(\delta_{\Lambda/2\Lambda})) = L(\check{X}, \check{\Delta}/2)$. We can now write
\be
\begin{aligned}
S_{\omega'}(U; \check{X}) - {\rm Arf}(q_{\omega'}) \: & = S_{\omega}(U; \check{X}) - {\rm Arf}(q_{\omega}) + L(\check{X}, \check{\Delta}/2) + S_{\omega}(U;\check{\Delta}/2) \\
& = S_{\omega}(U; \check{X} + \check{\Delta}/2) - {\rm Arf}(q_{\omega}) \quad {\rm mod} \; 1\;,
\end{aligned}
\ee
where we used the fact that $2L(\check{X}, \check{\Delta}/2) = 0 \mbox{ mod } 1$. It follows that
\be
\label{EqDepWuStruct}
{\rm WCS}_{\omega'}(U;\check{X}) = {\rm WCS}_{\omega}(U;\check{X} + \check{\Delta}/2) \;.
\ee
A change of Wu structure can therefore be absorbed by a (torsion) shift of the background field $\check{X}$.

Note that spin Chern-Simons theories have a very similar dependence on the spin structure \cite{Belov:2005ze}.

\subsection{Wu Chern-Simons theory: the state space}

\label{SecStateSpace}

We define the state space ${\rm WCS}_\omega(M;\check{X})$ on a 6-manifold $(M,\check{X})$ as follows:
\be
\label{EqStSpWCS}
{\rm WCS}_\omega(M;\check{X}) := {\rm WCS}_\omega^{\rm PQ}(M;\check{X}) \otimes {\rm WCS}_\omega^{\rm PQ}(M;\check{0}) \;.
\ee
where $\check{0}$ is the zero differential cocycle on $M$. The justification for this definition is that it is designed so that the resulting theory satisfies the gluing laws for the bordism category $\mathcal{C}_{\rm WCS}$. The proof of the gluing laws appears in Appendix \ref{AppGluing}. Note that the state space of ${\rm WCS}$ is always 1-dimensional, as is required for an invertible field theory.

Note that we could have written \eqref{EqStSpWCS} analogously to the partition function \eqref{EqDefPartFuncGS}, summing over torsion classes:
\be
\label{EqStSpWCS2}
{\rm WCS}_\omega(M;\check{X}) :=  {\rm WCS}_\omega^{\rm PQ}(M;\check{X}) \otimes \bigoplus_{[z] \in H^4_{\rm tors}(M;\Lambda)} {\rm WCS}_\omega^{\rm PQ}(M;\check{Z}) \;,
\ee
where as in \eqref{EqDefPartFuncGS}, $\check{Z}$ is a differential cocycle lifting the torsion class $[z]$. Indeed, the torsion anomaly discussed in Section \ref{SecPreqTh} ensures that ${\rm WCS}_\omega^{\rm PQ}(M;\check{Z})$ is the zero Hilbert space unless $\check{Z}$ represents the trivial torsion class.

\subsection{Further remarks}

We can now discuss what would happen had we not imposed the constraint that $\Lambda$ is unimodular. If $\Lambda$ is not unimodular, the constraints imposed by the torsion anomaly are looser and there is a non-trivial subgroup $K$ of $H^4_{\rm tors}(M;\Lambda)$ such that ${\rm WCS}_\omega^{\rm PQ}(M;\check{Z})$ is a Hermitian line if $\check{Z}$ lifts an element in $K$, see Propositions 5.2 and 4.13 in \cite{Monnier:2016jlo}. To satisfy the gluing relations, the state space has to be defined using \eqref{EqStSpWCS2} \cite{Monnier:2016jlo}. Its dimension is therefore higher than 1 and the WCS theory is not invertible. This shows that with $\Lambda$ non-unimodular, there is no way to relate the WCS theory to the anomaly field theory of the 6d supergravity and the construction of the Green-Schwarz term is doomed. In this sense, our construction of the Green-Schwarz term identifies the unimodularity of $\Lambda$ as a consistency condition on the 6d supergravity theory. This fact was previously derived in \cite{SeibergTaylor} using compactification to 2 dimensions.

Note also that the form of the partition function and of the state space shows that ${\rm WCS}$ is the product of two theories. The first factor, depending on the background field $\check{X}$, is nothing but a standard prequantum Wu Chern-Simons theory. We expect the second factor, whose partition function yields the Arf invariant, to coincides with the "quantum" WCS theory, obtained from the WCS theory by promoting the background gauge field to a dynamical field. As the theory is Gaussian, it can be computed exactly as a sum over the critical points of the action, which are the flat gauge fields. Now we saw that on flat gauge fields, the action depends only on the torsion class of the characteristic of the corresponding differential cocycle. This suggests that the full path integral, after a suitable normalization, should reduce to a simple sum over the torsion group $H^4_{\rm tors}(M;\Lambda)$, as in \eqref{EqDefPartFuncGS} and \eqref{EqStSpWCS2}. We did not check these claims formally, however, and the discussion above ignores the potential contribution of the 1-loop determinants associated to the fixed points.

If the conjectural relation above holds, the structure of ${\rm WCS}$ is reminiscent of the spin Chern-Simons theories used to model the fractional quantum Hall effect (see for instance Section 7 of \cite{Belov:2005ze}), with $\check{X}$ playing the role of the Maxwell field and $\check{Z}$ being a statistical Chern-Simons field.

\section{The shifted Wu Chern-Simons theory and its relation to the anomaly field theory}

\label{SecRelGSAFT}

Consider a closed Riemannian spin 7-manifold $U$ with principal $G$-bundle that bounds, i.e. such that $U = \partial W$ with all the structures extending to $W$. From the discussion in Appendix \ref{AppDiffCocCharCl}, we obtain differential cocycles $\check{Y}_U$, $\check{Y}_W$ on $U$ and $W$ respectively, such that $\check{Y}_W$ extends $\check{Y}_U$. We see that the phase of the partition function of the Wu Chern-Simons theory in \eqref{EqPartFuncGSTermsBound} coincides with our expectation for the anomaly of the 6d supergravity \eqref{EqAnBare6dSugra} upon setting $X'_W = Y_W$, or equivalently
\be
\label{RelGSTFTAnomPoly}
X_W = Y_W - \frac{1}{2} \lambda'\;.
\ee
Equivalently, it coincides with minus the anomaly \eqref{EqAnGSTerms} expected for the Green-Schwarz terms.

At the level of differential cocycles, \eqref{RelGSTFTAnomPoly} reads $\check{X}_W = \check{Y}_W - \frac{1}{2} \check{\nu}_W$, which restricts to $U$ as
\be
\label{EqDefXFromY}
\check{X} = \check{Y} - \frac{1}{2} \check{\nu}
\ee
Write $\hat{\lambda}$ for the integral degree 4 characteristic cocycle pulled back from the cocycle $\hat{\lambda}_{\rm BSpin}$ on the classifying space of spin bundles, as defined in Appendix \ref{AppDiffCocFor6dSugra}. $\check{Y}$ is a differential cocycle shifted by $\frac{1}{2}\hat{\lambda} \otimes a \mbox{ mod }1$, where $a$ is the gravitational anomaly coefficient of the 6d supergravity appearing in \eqref{EqRefEffAbGaugFFT}. On spin manifolds, the characteristic class $\lambda$ reduces modulo 2 to $w_4$, which coincides with the Wu class. Moreover, we chose $\hat{\lambda}_{\rm BSpin}$ to coincide with the universal Wu cocycle modulo 2, see Appendix \ref{AppDiffCocFor6dSugra}. This means that the characteristic $x$ of $\check{X}$ is an integer-valued cocycle, provided $a = \tilde{a}$ modulo 2, that is, provided that $a$ is a characteristic element of $\Lambda$. This is an interesting result, because while $a$ is always a characteristic element in F-theory compactifications, it was not clear until now whether or why this condition was required from the low energy supergravity point of view. Note also that we are free to choose $\tilde{a}$ to be any characteristic element of $\Lambda$: a natural choice is obviously $\tilde{a} = a$ if $a$ is a characteristic element of $\Lambda$. 

Thus, when $a$ is a characteristic element of $\Lambda$, we can use \eqref{EqDefXFromY} as a background field for the Wu Chern-Simons theory constructed in Section \ref{SecTFT}. Writing ${\rm WCS}^\dagger$ for the field theory complex conjugate to ${\rm WCS}$, we define
\be
\label{EqIdentBGFieldWCS}
{\rm WCS}^{\rm s}(N;\check{Y}) := {\rm WCS}^\dagger\left(N;\check{Y} - \frac{1}{2} \check{\nu}\right) \;,
\ee
where $N$ is here either a 6- or a 7-dimensional spin manifold endowed with a differential cocycle $\check{Y}$ shifted by $\frac{1}{2} \hat{\lambda} \otimes a$. To make sense of the right-hand side, we choose an arbitrary good Wu structure on $N$. Choosing a different Wu structure shifts $\check{\nu}$ by a 2-torsion differential cocycle, ensuring through \eqref{EqDepWuStruct} that the left-hand side is independent of the Wu structure.

${\rm WCS}^{\rm s}$ is therefore a field theory functor from a bordism category $\mathcal{C}_{{\rm WCS}^{\rm s}}$ into the category of Hilbert spaces, where $\mathcal{C}_{{\rm WCS}^{\rm s}}$ is defined as follows.
\begin{itemize}
\item The objects in $\mathcal{C}_{{\rm WCS}^{\rm s}}$ are 6-dimensional smooth closed oriented spin Riemannian manifolds $M$ endowed with a principal $G$-bundle and a classifying map into $B_{\rm W}\bar{G}$. As explained in Appendix \ref{AppDiffCocCharCl}, this data yields a degree 4 differential cocycle $\check{Y}$ shifted by $\frac{1}{2}\hat{\lambda} \otimes a$. $\check{Y}$ is required to be trivializable. Recall from the discussion in Section \ref{SecModel} that in this context, "trivializable" really means trivializable on the complement of the possible string sources.
\item The morphisms in $\mathcal{C}_{{\rm WCS}^{\rm s}}$ are 7-dimensional smooth spin compact Riemannian manifolds $U$ endowed with a principal $G$-bundle and a classifying map into $B_{\rm W}\bar{G}$. Again the construction of Appendix \ref{AppDiffCocCharCl} yields a degree 4 differential cocycle $\check{Y}$.
\end{itemize}

The comparison of \eqref{EqAnBare6dSugra} and \eqref{EqPartFuncGSTermsBound} show that on 7-manifolds that bound, ${\rm WCS}^{\rm s}$ coincides with the complex conjugate of the anomaly field theory. Moreover, we defined $\mathcal{C}_{{\rm WCS}^{\rm s}}$ precisely so that it coincides with the domain of the anomaly field theory.

This means that for gauge groups such that every spin 7-manifold endowed with a principal $G$-bundle bounds, or more precisely such that $\Omega^{\rm spin}_7(BG) = 0$, the two theories coincide. In Appendix \ref{AppCompCobGroup}, we show that this is the case for $G = U(n), SU(n), Sp(n)$, products of such groups and for $G = E_8$.

As we will see in Section \ref{SecAnCanThDisAbGG}, there are cases where the two theories do \emph{not} coincide on certain 7-manifolds that do not bound.

\section{The Green-Schwarz term}

\label{SecGSTerms}

We will construct the Green-Schwarz term of the 6d supergravity on a manifold $M$ as an element of the state space of the shifted Wu Chern-Simons theory ${\rm WCS}^{\rm s}$:
\be
\label{EqDefGSTermETh}
{\rm GST}(M;\check{Y}, \check{H}) \in {\rm WCS}^{\rm s}\left(M;\check{Y} \right) \;.
\ee
As mentioned in the previous section, the shifted Wu Chern-Simons theory coincides with the complex conjugate of the anomaly field theory for groups such that $\Omega^{\rm spin}_7(BG) = 0$. For such groups, all anomalies are canceled by the Green-Schwarz terms. When $\Omega^{\rm spin}_7(BG) \neq 0$, the anomalies might not completely cancel and there are residual constraints from global anomaly cancellation, see the discussion in Section \ref{SecAnCanThDisAbGG}.

Let us define the Green-Schwarz term on $M$ by
\be
\label{EqDefGST1}
{\rm GST}(M,\check{Y}, \check{H}) := \exp -2 \pi i \int^{\rm E}_{M,\omega} \overline{\rm gst}(M,\check{Y}, \check{H}) \;,
\ee
\be
\label{EqDefGST2}
\overline{\rm gst}(M,\check{Y}, \check{H}) = \left( \frac{1}{2} \left[\left(\check{H} - \frac{1}{2}\check{\eta}\right) \cup \left(\check{Y} + \frac{1}{2}\check{\nu}\right)\right]_{\rm hol} , h_2 - \frac{1}{2} \eta \right) = \left( \frac{1}{2} [\check{F} \cup (\check{X} + \check{\nu})]_{\rm hol} , f_2 \right) \;.
\ee
Here $\left[\left(\check{H} - \frac{1}{2}\check{\eta}\right) \cup \left(\check{Y} + \frac{1}{2}\check{\nu}\right)\right]_{\rm hol}$ is 
projected to a cocycle valued in $\mathbb{R}/\mathbb{Z}$ so that 
$\overline{\rm gst}(M,\check{Y}, \check{H})$ is a degree 6 E-cochain, and as before, $\int^{\rm E}_{M,\omega}$ denotes the integration in E-theory. For this integration map to exist, a Wu structure $\omega$ has to be chosen on $M$. $\check{\eta} = (\eta_\Lambda, 0,0)$ and $\check{\nu} = (d\eta_\Lambda, -\eta_\Lambda, 0)$ are differential cochains constructed from the Wu structure as in Section \ref{SecModel}. $\check{X} := \check{Y} - \frac{1}{2} \check{\nu}$ is an unshifted differential cocycle, and $\check{F} := \check{H} - \frac{1}{2} \check{\eta}$ an unshifted differential cochain trivializing $\check{X}$, see \eqref{EqDefTrivF}. We first prove that ${\rm GST}(M;\check{Y}, \check{H})$ is independent of the choice of Wu structure on $M$, and then prove \eqref{EqDefGSTermETh}.

In the computations below we write $\check{X} = (x,A,X)$, $\check{F} = (f,C,F)$. It will also be important to bear in mind that 
$x$ is integrally quantized.

\paragraph{Independence from the Wu structure} We proceed as in Section \ref{SecPartFunc}. Let $\delta \in H^3(M; \mathbb{Z}_2)$ be the class describing a change of Wu structure $\omega \rightarrow \omega'$ and $\delta_{\Lambda/2\Lambda} = \delta \otimes \gamma \in H^3(M; \Lambda/2\Lambda)$. Under the change of Wu structure, the integral lift $\nu_\Lambda$ changes by a lift $\delta_{\Lambda}$ of $\delta_{\Lambda/2\Lambda}$. If $\check{\Delta} := (d\delta_{\Lambda}, -\delta_{\Lambda}, 0)$, then $\check{\nu}$ changes to $\check{\nu} + \check{\Delta}$ under the change of Wu structure. The construction of the integration map in Appendix D of \cite{Monnier:2016jlo} implies that given a top E-cochain $\overline{s} = (s,y)$,
\be
\label{EqDepIntEthWuStruct}
\int^{\rm E}_{M,\omega'} (s,y) = \int^{\rm E}_{M,\omega} (s,y) + \frac{1}{2} \int_M y \cup \delta_{\Lambda/2\Lambda} \;.
\ee
\eqref{EqDepIntEthWuStruct} is a direct consequence of the construction of the Brown-Comenetz dual of the E-theory spectrum in \cite{Monnier:2016jlo}, see (D.22) there. Note that if $y = 0$, the integration reduces to the ordinary integration of the cochain $s$, and the dependence on the Wu structure disappears, as it should. We can now compute (all equations are understood modulo $\mathbb{Z}$): 
\be
\begin{aligned}
\frac{1}{2\pi i} \ln {\rm GST}_{\omega'}(M,\check{Y}, \check{H}) \: & = -\int^{\rm E}_{M,\omega'} \left( \frac{1}{2} [\check{F} \cup (\check{X} + \check{\nu} + \check{\Delta})]_{\rm hol} , f_2 \right) \\
& = -\int^{\rm E}_{M,\omega} \left(\left( \frac{1}{2} [\check{F} \cup (\check{X} + \check{\nu})]_{\rm hol}, f_2 \right) \boxplus  \left( \frac{1}{2} [\check{F} \cup \check{\Delta}]_{\rm hol}, 0 \right)  \right) \\
& \quad - \frac{1}{2} \int_M f_2 \cup \delta_{\Lambda/2\Lambda} \\
& = \frac{1}{2\pi i} \ln {\rm GST}_{\omega}(M,\check{Y}, \check{H}) + \frac{1}{2} \int_M f \cup \delta_{\Lambda} - \frac{1}{2} \int_M f_2 \cup \delta_{\Lambda/2\Lambda} \\
& = \frac{1}{2\pi i} \ln {\rm GST}_{\omega}(M,\check{Y}, \check{H}) \quad \mbox{mod } 1\;.
\end{aligned}
\ee
This proves that the Green-Schwarz term is independent of the choice of Wu structure on $M$.

\paragraph{Proof of \eqref{EqDefGSTermETh}}
We will compare the gauge transformation of ${\rm GST}(M,\check{Y}, \check{H})$ with the gauge transformation of ${\rm WCS}^{\rm s}(N;\check{Y}_N)$ for $N$ a 7-manifold admitting $M$ as its boundary and $\check{Y}_N$ a degree 4 differential cocycle shifted by $\frac{1}{2} \nu \otimes a$ and restricting to $\check{Y}$ on $M$. The idea is that ${\rm WCS}^{\rm s}(N;\check{Y}_N)$ is a vector in the Hermitian line ${\rm WCS}^{\rm s}(M;\check{Y})$. Any choice of cocycle $\check{Y}$ determines a trivialization of this Hermitian line, and the gauge transformations of $\check{Y}$ relate the trivializations. If the gauge transformations act similarly on ${\rm GST}(M,\check{Y}, \check{H})$ and ${\rm WCS}^{\rm s}(N;\check{Y}_N)$, it means that the Hermitian lines they belong to are canonically isomorphic. (Recall the construction of the state space of the prequantum theory, i.e. of ${\rm WCS}^{\rm s}(M;\check{Y})$, in Section \ref{SecPreqTh}.)

If $M$ does not bound, we use the same reasoning by taking $N$ to be a 7-manifold admitting $M$ as one of its boundary component, and by considering gauge transformation restricting trivially to the other components. ${\rm WCS}^{\rm s}(N;\check{Y}_N)$ is then a tensor product ${\rm WCS}^{\rm s}(M;\check{Y}) \otimes {\rm WCS}^{\rm s}(\partial N - M;\check{Y}_{\partial N - M})$, and the gauge transformations restricting trivially to $\partial N - M$ correspond to changes of trivialization of the first factor.

Note that both the Green-Schwarz term and the shifted Wu Chern-Simons theory are trivially invariant under pullbacks through diffeomorphisms  and under changes of shifts (the transformations 1. and 4. in Section \ref{SecModel}). We will therefore investigate only the $B$-field gauge transformations \eqref{EqGaugTransHi} and the gauge transformations of $\check{Y}$ given in \eqref{EqGaugTransY}.

Let us start by the $B$-field gauge transformations \eqref{EqGaugTransHi} with gauge parameter $\check{W} = (w,W,0)\in 
\check{C}^2_0(M;\Lambda)$. Now,  ${\rm WCS}^{\rm s}(N;\check{Y}_N)$, being independent of $\check{H}$, is obviously invariant. Let us compute the gauge transformation of ${\rm GST}(M,\check{Y}, \check{H})$. It is useful to keep in mind during the computations that all the cochains written in small caps (except for $y$ and $h$, which do not appear) are integer-valued and that the integral is considered modulo $\mathbb{Z}$. 
\footnote{Also, in this equation $w_2$ is the reduction modulo two of some integral cochain $w$ and it is \emph{not} the second 
Stiefel-Whitney class! In the similar computations below $v_2$ will similarly be the reduction of an integral cochain $v$ and 
will \emph{not} refer to the second Wu class!}
\be
\label{EqVarGSGaugeTransW}
\begin{aligned}
\Delta_{\check{W}} \left( \frac{1}{2\pi i} \ln {\rm GST}(M;\check{Y}, \check{H}) \right) \: & = -\int_{M,\omega}^{\rm E} \left( \frac{1}{2}[ (\check{F} + d\check{W}) \cup (\check{X} + \check{\nu})]_{\rm hol}, f_2 + dw_2 \right) \\
& \quad + \int_{M,\omega}^{\rm E} \left( \frac{1}{2}[ \check{F} \cup (\check{X} + \check{\nu})]_{\rm hol}, f_2 \right) \\
& = -\int_{M,\omega}^{\rm E} \left(\left( \frac{1}{2}[ (\check{F} + d\check{W}) \cup (\check{X} + \check{\nu})]_{\rm hol}, f_2 + dw_2 \right) \right. \\
&\quad \quad \quad \quad \left.  \boxminus \left( \frac{1}{2}[ \check{F} \cup (\check{X} + \check{\nu})], f_2 \right) \right) \\
& = -\int_{M,\omega}^{\rm E} \left( \frac{1}{2}[ d\check{W} \cup (\check{X} + \check{\nu})]_{\rm hol} + \frac{1}{2} dw \cup f, dw_2 \right) \\
& = -\int_{M,\omega}^{\rm E} \left( \frac{1}{2}[ d\check{W} \cup (\check{X} + \check{\nu})]_{\rm hol} + \frac{1}{2} (dw \cup f + w \cup \nu_\Lambda + dw \cup dw ), 0 \right)\\
& = -\int_M \frac{1}{2} ( -dw \cup (A - \eta_\Lambda) + (-w - dW) \cup X + dw \cup f \\
& \quad \quad \quad \quad + w \cup \nu_\Lambda )\\
& = -\int_M \frac{1}{2} \left( -w \cup x + dw \cup \eta_\Lambda + dw \cup f + w \cup \nu_\Lambda \right) \\
& = 0 \quad {\rm mod} \; 1 \;.
\end{aligned}
\ee
In the second equality, we used the fact that the E-theory integral is a group homomorphism with respect to the group law $\boxplus$ \eqref{EqGrpLawCocModETh} on E-cochains. In the third equality, we perform the subtraction using \eqref{EqGrpLawCocModEThInv}. We then perform a gauge transformation, by subtracting an E-cochain $d(0,w_2)$,  exact with respect to the differential \eqref{EqDefTwDiffEThCoch}. In the fifth equality, we compute explicitly the connection of the differential cocycle in the bracket. We also use the fact that $\int_{M,\omega}^{\rm E}(s, 0) = \int_M s$ to obtain an ordinary integral. In the sixth equality, we use the closedness of $\check{X}$ and drop exact terms. Finally, by integrating by parts we see that the remaining terms vanish modulo 1. This shows that ${\rm GST}(M;\check{Y}, \check{H})$ and ${\rm WCS}^{\rm s}(N;\check{Y}_N)$ are both invariant under the gauge transformations \eqref{EqGaugTransHi}.

Let us now compute the transformation of ${\rm GST}(M;\check{Y}, \check{H})$ under \eqref{EqGaugTransY}, with gauge parameter $\check{V} = (v,V,0)\in \check{C}^3_0(M;\Lambda)$.
\be
\label{EqVarGSGaugeTransX}
\begin{aligned}
\Delta_{\check{V}} \left( \frac{1}{2\pi i} \ln {\rm GST}(M;\check{Y}, \check{H}) \right) \: & = -\int_{M,\omega}^{\rm E} \left( \frac{1}{2}[ (\check{F} + \check{V}) \cup (\check{X} + d\check{V} + \check{\nu})]_{\rm hol}, f_2 + v_2 \right) \\
& \quad  + \int_{M,\omega}^{\rm E} \left( \frac{1}{2}[ \check{F} \cup (\check{X} + \check{\nu})]_{\rm hol}, f_2 \right) \\
& = -\int_{M,\omega}^{\rm E} \left( \frac{1}{2}[ \check{F} \cup d\check{V} + \check{V} \cup (\check{X} + \check{\nu}) + \check{V} \cup d \check{V}]_{\rm hol} \right.\\
& \quad \quad \quad \left. \phantom{\frac{1}{2}} + \frac{1}{2}(dv \cup_1 f + v \cup f), v_2 \right) \\
& = -\int_{M,\omega}^{\rm E} \left( \frac{1}{2}( -f \cup (-v - dV) - v \cup (A - \eta_\Lambda) + V \cup X \right.\\
& \; \quad \quad \quad \left. \phantom{\frac{1}{2}}  - v \cup (-v -dV) + dv \cup_1 f + v \cup f ), v_2 \right) \\
& = \int_{M,\omega}^{\rm E} \left( \frac{1}{2}( -v \cup_1 x - x \cup V + v \cup A - V \cup X + v \cup \eta_\Lambda \right. \\
& \; \quad \quad \quad \left. \phantom{\frac{1}{2}} - v \cup v - v \cup dV), v_2 \right) \quad {\rm mod} \; 1\;,
\end{aligned}
\ee
where we used the same kind of manipulations as in \eqref{EqVarGSGaugeTransW}. We also used the fact that 
\be
v \cup f + f \cup v = - dv \cup_1 f  + v \cup_1 x + d(v \cup_1 f)
\ee
by the definition of Steenrod's higher cup products $\cup_i$, see \cite{1947} and Appendix \ref{AppEThCalc}.

We now compare \eqref{EqVarGSGaugeTransX} to the variation of ${\rm WCS}^{\rm s}(N;\check{Y}_N)$ under \eqref{EqGaugTransY}. To simplify the notation, we do not distinguish between cochains / differential forms on $N$ and their restrictions to $M = \partial N$. Similarly, we still write $\omega$ for the Wu structure on $N$ restricting to the Wu structure on $M$, using equations \eqref{EqDefLag} and \eqref{EqActWCS} for the case that the 7-manifold is $N$ with $\partial N = M$, we have
\be
\label{EqVarWCSGaugeTransX}
\begin{aligned}
\Delta_{\check{V}} \left( \frac{1}{2\pi i} \ln {\rm WCS}^{\rm s}(N;\check{Y}) \right) \: & = \int_{N,\omega}^{\rm E} \left( \frac{1}{2}[ (\check{X} + d\check{V}) \cup (\check{X} + d\check{V} + \check{\nu})]_{\rm hol}, x_2 + dv_2 \right) \\
& \quad  - \int_{N,\omega}^{\rm E} \left( \frac{1}{2}[ \check{X} \cup (\check{X} + \check{\nu})]_{\rm hol}, x_2 \right) \\
& = \int_{N,\omega}^{\rm E} \left( \frac{1}{2}(x \cup (-v -dV) + dv \cup (A - \eta_\Lambda) + (-v -dV) \cup X \right. \\
& \; \quad \quad \quad \left. \phantom{\frac{1}{2}} + dv \cup (-v - dV) + dv \cup_1 x), dv_2 \right) \\
& = \int_{N,\omega}^{\rm E} \left( \frac{1}{2}d(v \cup A + v \cup_1 x - V \cup X - x \cup V + v \cup \eta_\Lambda \right. \\
& \; \quad \quad \quad \left. \phantom{\frac{1}{2}}  - v \cup dV - v \cup v) + \frac{1}{2}(-v \cup dv + v \cup \hat\nu_{\Lambda}), dv_2 \right) \\
& = \int_{M,\omega}^{\rm E} \left( \frac{1}{2} (v \cup A + v \cup_1 x - V \cup X - x \cup V + v \cup \eta_\Lambda \right. \\
& \; \quad \quad \quad \left. \phantom{\frac{1}{2}}   - v \cup dV - v \cup v ), v_2 \right) \quad {\rm mod} \; 1\;,
\end{aligned}
\ee
where in the third equality we used $d(v\cup_1 x) = dv \cup_1 x + v\cup x - x \cup v$ and, 
in the last line, we used the fact that the integrated E-cochain is exact to reexpress is as the integral of an E-cochain on $M$.

Let us now compare \eqref{EqVarWCSGaugeTransX} and \eqref{EqVarGSGaugeTransX}, recalling that cup products of lower case cochains are integral, and therefore that their sign is irrelevant. We see that ${\rm WCS}^{\rm s}(N;\check{Y})$ and ${\rm GST}(M;\check{Y}, \check{H})$ transform by the same phases under the gauge transformations \eqref{EqGaugTransY}. Together with the fact proven above that they are both invariant under the gauge transformations \eqref{EqGaugTransHi}, we deduce that \eqref{EqDefGSTermETh} holds.

\section{Implications for six-dimensional supergravity theories}

\label{SecImplSixSugra}

\subsection{Anomaly cancellation}

\label{SecAnCan}

Let us start by discussing in broad terms the constraints imposed by global anomaly cancellation.

Assume first that $G$ is such that all 7-dimensional spin manifolds endowed with a principal $G$-bundle bound, i.e. that the spin bordism group $\Omega^{\rm spin}_7(BG)$ vanishes. Then we know that the shifted Wu Chern-Simons theory ${\rm WCS}^{\rm s}$ coincides with the complex conjugate of the anomaly field theory of the six-dimensional supergravity. For a 6-manifold $(M,\check{Y})$, $({\rm WCS}^{\rm s})^\dagger(M;\check{Y})$ is the hermitian line in which the partition function of the bare supergravity is valued. We constructed an exponentiated Green-Schwarz term, written ${\rm GST}(M;\check{Y},\check{H})$ as an element of ${\rm WCS}^{\rm s}(M;\check{Y})$. "Adding the Green-Schwarz term to the action" or more precisely multiplying the exponentiated action by the exponentiated Green-Schwarz term therefore ensures that the partition function of the anomalous fields in the full theory is a complex number, rather than an element of a general hermitian line. Moreover, if symmetries act non-trivially on the bare partition function, i.e. have a non trivial action on $({\rm WCS}^{\rm s})^\dagger(M;\check{Y})$, they necessarily act trivially on $({\rm WCS}^{\rm s})^\dagger(M;\check{Y}) \otimes ({\rm WCS}^{\rm s})(M;\check{Y})$, in which the total partition function takes value. This ensures the cancellation of \emph{all} anomalies, local and global.

Whether there exist non-bounding spin 7-manifolds with a $G$-bundle depends on $G$. We show in Appendix \ref{AppCompCobGroup} that no such manifolds exist for $G = SU(n), U(n), Sp(n)$ and products of such groups, as well as for $G = E_8$. However, we also show that when $G$ is a finite Abelian group, then $\Omega^{\rm spin}_7(BG) \neq 0$. Another related example of a gauge group with non-trivial associated bordism group is $O(n)$. Indeed, $\mathbb{R}P^7$ with a principal $O(n)$ gauge bundle $P$ with non-trivial first Stiefel-Whitney class is a non-bounding 7-manifold. Indeed, the bordism invariant $\int_{\rm \mathbb{R}P^7} w_1(P)^7$ is non-zero.

\subsection{Anomaly cancellation for theories with finite Abelian gauge groups}

\label{SecAnCanThDisAbGG}

We now show that there are cases where the anomaly field theory differs from the shifted Wu Chern-Simons theory on non-bounding 7-manifolds. In such cases, the Green-Schwarz terms do not cancel all anomalies, and there are residual constraints imposed by global anomaly cancellation.

The general idea is the following. Suppose we have two theories with the same tensormultiplet lattice $\Lambda$, and the same 
vectormultiplet gauge group $G$ but different matter representations $R^{(1)}$ and $R^{(2)}$ of $G$.  Assume $R^{(1)}$ and $R^{(2)}$ are such that
\be
\ch(R^{(1)}) = \ch(R^{(2)})
\ee
or, equivalently
\be
\begin{split}
\dim R^{(1)} & = \dim R^{(2)} \\
\Tr_{R^{(1)}} F^2 & = \Tr_{R^{(2)}} F^2 \\
\Tr_{R^{(1)}} F^4 & = \Tr_{R^{(2)}} F^4 \\
\end{split}
\ee
Then there is a "relative anomaly field theory" computing the anomaly difference between the two theories. We see from \eqref{EqAnomBare6dSugra} that on a 7-dimensional manifold $M$, its partition function is given explicitly by
\be\label{eq:RelGlobAnom}
\exp[ \pi i  ( \xi_{R^{(1)}}(U) - \xi_{R^{(2)}}(U) ) ] \;,
\ee
where $\xi_{R}$ is the modified eta invariant of the Dirac operator on $U$ coupled to the vector bundle induced by the matter representation $R$. It is considerably simpler than the "absolute" anomaly field theory. We can then compare the anomaly difference to the anomaly difference associated to the corresponding Green-Schwarz terms. If these differences are not equal, it is impossible that anomalies cancels for both supergravity theories with matter in $R^{(1)}$ and $R^{(2)}$.

It is in general extremely difficult to compute explicitly the exponentiated eta invariant \eqref{eq:RelGlobAnom}. However, when $G = \mathbb{Z}_n$ and $U$ is a certain lens space, we can use results presented in \cite{gilkey1999spectral} to compute \eqref{eq:RelGlobAnom} explicitly, see Appendix \ref{AppNonTrivGlobAn}. More precisely, picture the elements of $\mathbb{Z}_n$ by $n^{\rm th}$ roots of unity and write $\rho_s: z \rightarrow z^s$, $s = 0,...,n-1$ for the distinct complex one-dimensional representations of $\mathbb{Z}_n$. The unitary representation $\pi = \rho_1^{\oplus 4}$ of $\mathbb{Z}_n$ on $\mathbb{C}^4$ has no fixed point on the 7-dimensional unit sphere $S^7$, and the quotient is a spin 7-dimensional lens space $U$. Take the quaternionic matter representation to be of the form $R_s = \rho_s \oplus \rho_{-s}$. Then, as explained in Appendix \ref{AppNonTrivGlobAn},
\be
\label{eq:EtaInv}
\xi_{R_s}(U) = \frac{1}{360\cdot n}  (-11 + 10 n^2 + n^4 - 60 n s + 60 s^2 - 30 n^2 s^2 + 60 n s^3 - 30 s^4) \;.
\ee

Recall that constructing the differential cocycle $\check{Y}$ from the supergravity data is a subtle problem, explained in detail in Appendix \ref{AppDiffCocCharCl}. In the most straightforward (and naive) construction, in which the characteristic of the universal differential cocycle on the classifying space is \eqref{EqCharNaiveY}, $\check{Y}$ and the associated Green-Schwarz terms are \emph{independent} of the matter representation. Clearly, given the obvious dependence of \eqref{eq:EtaInv} on $s$, this means that global anomalies cannot cancel for every choice of matter representation. Determining exactly which matter representations are allowed would require to compute explicitly the shifted Wu Chern-Simons partition function, which looks like a hard problem. We can however describe the difference between the representations of two otherwise consistent 6d supergravity theories.

For this, consider two supergravity theories with matter representations
\be
R^{(i)} = \bigoplus x^{(i)}_s R_s \;, \quad i = 1,2 \;,
\ee
where by $xR$, we mean $R^{\oplus x}$. We assume that they are made anomaly-free by the Green-Schwarz term associated to the " naive" $\check{Y}$ of Appendix \ref{AppDiffCocCharCl}. Let us write $\Delta x_s = x^{(2)}_s - x^{(1)}_s$. Given that the number of hypermultiplet is fixed by local anomaly cancellation, we have the relation $\sum_{s = 0}^{n-1} \Delta x_s = 0$. As the Green-Schwarz terms are independent of $s$, \eqref{eq:EtaInv} also says that
\be
\sum_{s = 0}^{n-1} \Delta x_s p(n,s) = 0 \mbox{ mod } 2 \;,
\ee
where
\be
p(n,s) = \frac{1}{12 n} (- 2 n s + 2 s^2 - n^2 s^2 + 2 n s^3 - s^4)
\ee
For low values of $n$, we can write down the constraints more explicitly:
\be
\label{EqNaiveConstraints}
\begin{aligned}
n = 2 \;: \; & \frac{1}{16} \Delta x_1 = 0 \mbox{ mod } 1 \\
n = 3 \;: \; & \frac{1}{9} (\Delta x_1 + \Delta x_2) = 0 \mbox{ mod } 1 \\
n = 4 \;: \; & \frac{1}{32}(5\Delta x_1 + 8\Delta x_2 + 5\Delta x_3) = 0 \mbox{ mod } 1 \\
n = 5 \;: \; & \frac{1}{5}(\Delta x_1 + 2\Delta x_2 + 2\Delta x_3  + \Delta x_4) = 0 \mbox{ mod } 1 \\
n = 6 \;: \; & \frac{1}{144}(35\Delta x_1 + 80\Delta x_2 + 99\Delta x_3  + 80\Delta x_4 + 35 \Delta x_5) = 0 \mbox{ mod } 1 \\
... &
\end{aligned}
\ee
We can compare these constraints to the observed $\Delta x_s$ among known F-theory models with finite cyclic gauge group. It turns out that F-theory models satisfy similar, but looser constraints
\footnote{The models in question are all obtained through the Higgising of hypermultiplets of charge > 1 in F-theory models with gauge group $U(1)$ \cite{Turner}. We thank Andrew Turner and Wati Taylor for checking that the constraints \eqref{EqFThConstraints} are satisfied in these models.}
:
\be
\label{EqFThConstraints}
\begin{aligned}
n = 2 \;: \; & \frac{1}{4} \Delta x_1 = 0 \mbox{ mod } 1 \\
n = 3 \;: \; & \frac{1}{3} (\Delta x_1 + \Delta x_2) = 0 \mbox{ mod } 1 \\
n = 4 \;: \; & \frac{1}{4} (5\Delta x_1 + 8\Delta x_2 + 5\Delta x_3) = 0 \mbox{ mod } 1 \\
n = 5 \;: \; & \mbox{No apparent constraint} \\
n = 6 \;: \; & \frac{1}{12} (35\Delta x_1 + 80\Delta x_2 + 99\Delta x_3  + 80\Delta x_4 + 35 \Delta x_5) = 0 \mbox{ mod } 1 \\
... &
\end{aligned}
\ee
We can either deduce that some of these F-theory models are inconsistent, or that the construction of the Green-Schwarz term out of the naive $\check{Y}$ is incorrect.

The second option is of course the most plausible. In fact, when constructing $\check{Y}$ out of the gauge data, the only constraint we have is that its field strength $Y$ coincides with \eqref{EqRefEffAbGaugFFT}. This means that one is free to add a torsion differential cocycle to $\check{Y}$. There is indeed a degree 4 torsion class on $U$ (see Appendices \ref{AppDiffCocCharCl} and \ref{SecSpinCobDisAbGrp}). We can add a corresponding differential cocycle representative to $\check{Y}$, multiplied by a new $\Lambda$-valued anomaly coefficient $b_T$ that can be adjusted to cancel anomalies. This introduces a dependence on the representation $R_s$ in $\check{Y}$, and therefore in the shifted WCS theory.  Computing explicitly this dependence looks like a hard problem, but the quadratic property \eqref{EqExprQRy} of the WCS action guarantees that the partition function of the shifted WCS theory will change by multiples of $\frac{1}{2n}$. (A quadratic refinement on $\mathbb{Z}_n$ generically takes value in $\mathbb{Z}_{2n}$.) If we read the constraints \eqref{EqNaiveConstraints} modulo $1/2n$ rather than modulo 1, we get exactly the F-theory constraints \eqref{EqFThConstraints}.

This shows that the most general construction of $\check{Y}$ makes anomaly cancellation possible on $U$ for all known F-theory models, although obviously this is still quite far from a full proof of the anomaly cancellation. 

Let us also recall that the anomaly coefficients can be interpreted as measuring the string charges produced by the background geometry. The new anomaly coefficient is associated to a charge that is always torsion. It leaves no imprint on the 4-form field strength $Y$ of $\check{Y}$, which is why it is completely invisible in the standard framework.

We should note that the formulae from \cite{gilkey1999spectral} apply to more general space forms, including those for nonabelian finite groups, so the above analysis could be considerably extended to many other examples. If F-theory models with nonabelian $\pi_0(G)$ are of interest this exercise would be worth pursuing.

\subsection{Setting the quantum integrand}

We saw that the partition function of the shifted Wu Chern-Simons theory ${\rm WCS}^{\rm s}$ coincides with the partition function of the anomaly field theory on any 7-manifold $U$ that bounds. The tensor product of the shifted Wu Chern-Simons theory ${\rm WCS}^{\rm s}$ and the anomaly field theory is therefore a spin topological field theory $\mathcal{T}$, whose partition function is a spin bordism invariant, i.e. a homomorphism $\mathcal{T}_7: \Omega^{\rm spin}_7(BG) \rightarrow \mathbb{R}/\mathbb{Z}$. As discussed in Section \ref{SecAnCan}, global anomalies cancel whenever $\mathcal{T}_7$ is the trivial homomorphism. 

When $\mathcal{T}_7$ is trivial, the spin topological field theory $\mathcal{T}$ is isomorphic to the trivial field theory. Isomorphisms of field theories are natural isomorphisms of the corresponding field theory functors, whose data can be summarized by an isomorphism $\iota_M: \mathcal{T}(M) \simeq \mathbb{C}$ for each 6-manifold $M$, subject to the standard naturality property $\iota_{M_1} = \iota_{M_2} \circ \mathcal{T}(U)$, for any bordism $U$ from $M_1$ to $M_2$. The fact that $\mathcal{T}(N) = 1$ for any bordism $N$ from $\emptyset$ to itself guarantees that such a collection of isomorphism $\{\iota_{M}\}$ always exists. Nevertheless $\mathcal{T}$ might not be canonically isomorphic to the trivial field theory,i.e. there may not be a preferred collection of isomorphisms $\{\iota_{M}\}$. 

Let us investigate to which extent $\{\iota_{M}\}$ is canonical. Assume that $\mathcal{T}_7$ is trivial, i.e. that $\mathcal{T}_7(U) = 0$ for all closed 7-manifolds $U$, and consider $\mathcal{T}(M)$ for $M$ a 6-manifold. If $M$ bounds, $M = \partial N$, then $\mathcal{T}(N)$ is a vector in $\mathcal{T}(M)$ that is independent of the choice of $N$. This means there is a canonical isomorphism from $\mathcal{T}(M)$ to $\mathbb{C}$. But the same cannot be said if $M$ represents a non-trivial class in $\Omega^{\rm spin}_6(BG)$. There is a priori no canonical way of identifying $\mathcal{T}(M)$ with $\mathbb{C}$. (Although, depending on the details of the theories involved, such a canonical identification might exist.) This means that while $\mathcal{T}$ is isomorphic to the trivial theory, it is not canonically so. A choice of isomorphism involves choices of isomorphisms $\mathcal{T}(M) \simeq \mathbb{C}$ for all bordism classes $[M]$ in $\Omega^{\rm spin}_6(BG)$.

In terms of the 6d supergravity theory, this means that although all anomalies vanish, there is no canonical way of identifying its partition function on $M$ with a complex number. To see the supergravity theory as an ordinary field theory with complex-valued partition functions, as opposed to a relative field theory valued in $\mathcal{T}$, we need to pick an isomorphism of $\mathcal{T}$ with the trivial theory. This means \emph{choosing} the phases of the partition function on representatives $M$ of each bordism class in $\Omega^{\rm spin}_6(BG)$. Different choices yield different anomaly-free supergravity theories. The need to make these choices is an example of a general phenomenon known as "setting the quantum integrand" of the supergravity theory \cite{Witten:1996hc, Freed:2004yc}.

As an example, suppose the vectormultiplet gauge group is $G=U(1)$. In this case \cite{Stong1968}
\begin{equation}
\Omega_6^{\rm spin}(BU(1)) = \Omega_6^{\rm spin}(K(\mathbb{Z},2)) \cong \Omega^{{\rm spin}^c}_4(pt) \cong \mathbb{Z} \oplus \mathbb{Z}
\end{equation}
The independent bordism invariants can be taken to be $\int_{M} c_1(M)^3$ and $\int_{M} c_1(M) p_1(TM)$. Therefore, in this case, the setting simply involves two theta angles associated to these two invariants. It would be interesting to see if these topological terms can be independently supersymmetrized, and whether they have a natural home in F-theory compactifications.

If $\Omega^{\rm spin}_6(BG)$ is pure torsion, the choices involved in setting the quantum integrand are discrete. For instance if $\Omega^{\rm spin}_6(BG) \simeq \mathbb{Z}_2$, generated by $[M]$, then the partition function on $M \sqcup M$ is fixed, and the only choice to be made is a choice of square root, determining the partition function on $M$. A similar situation in dimension 2 is described in \cite{WittenTalkStrings2015}.

\subsection{Consistency constraints on the Green-Schwarz term}

There is a priori a puzzle about the Green-Schwarz mechanism in dimension six. Recall that in the original Green-Schwarz mechanism for 10-dimensional type I supergravity, the $B$-field is non-anomalous. The standard picture is that we perform the path integral over the anomalous fields (which are chiral fermions only) to obtain an anomalous partition function. We then check that its anomaly cancels against the variation of the exponentiated Green-Schwarz term. If this is the case, we can perform the path integral over the remaining bosonic fields.

In dimension 6, the situation is complicated by the fact that the self-dual fields entering the Green-Schwarz terms are themselves anomalous. The Green-Schwarz terms therefore cannot be taken out of the path integral over anomalous fields: they are part of the integrand. Nevertheless, it is sufficient that their anomalous variation can be taken out of the path integral. We deduce from this discussion two consistency constraints on the Green-Schwarz terms:
\begin{enumerate}
\item They should be gauge invariant under the gauge transformations of the anomalous fields, in order for the path integral over the anomalous fields to make sense. In our case, those are the $B$-field gauge transformations \eqref{EqGaugTransHi}. \eqref{EqVarGSGaugeTransW} shows that the Green-Schwarz term we constructed is indeed invariant.
\item Their gauge variation under the gauge transformations of the non-anomalous fields should be independent of the anomalous fields, so that the variation can be taken out of the path integral and cancel the variation of the path integral itself. In our case, those are the transformations \eqref{EqGaugTransY} (i.e. the transformations induced by diffeomorphisms and vectormultiplet gauge transformations). \eqref{EqVarGSGaugeTransX} shows that the variation of our Green-Schwarz term depends only on non-anomalous fields, as required.
\end{enumerate}
It is non-trivial that these two consistency conditions are automatically satisfied by our construction.

\subsection{Constraints on the anomaly coefficients}

\label{SecConstrAnCoeff}

An interesting aspect of the construction of the Green-Schwarz terms is that its consistency provides constraints on the anomaly coefficients $a$, $b_i$ and $b_{IJ}$ in \eqref{EqRefEffAbGaugFFT}.

We already explained that the identification \eqref{EqIdentBGFieldWCS} of $\check{Y}$ with the background field of the Wu Chern-Simons theory requires $a$ to be a characteristic element of the lattice $\Lambda$. In this case, we can assume $\tilde{a} = a$; see the discussion in Section \ref{SecRelGSAFT}. Writing $x$ and $y$ for the characteristics of $\check{X}$ and $\check{Y}$ respectively we have
\be
x + \frac{1}{2} \nu_\Lambda = y = \frac{1}{2}\lambda \otimes a + v \;,
\ee
where
\be
v = - \sum_i b_i c_2^i + \frac{1}{2} \sum_{IJ} b_{IJ} c_1^I \cup c_1^J
\ee
As both $\nu_\Lambda$ and $\lambda$ are integral lifts of the Wu class, we deduce that $v$ is an integral cocycle. The same constraint was inferred in \cite{Monnier:2017oqd} from the fact that background charge represented by $y$ has to be canceled by string instantons. By considering $U = \mathbb{C}P^3 \times S^1$, $W = \mathbb{C}P^3 \times D^2$ and suitable bundles over $\mathbb{C}P^3$, we can recover the constraints of \cite{Monnier:2017oqd} on the anomaly coefficients $b_i$ and $b_{IJ}$. As explained there, $b_i$ and $b_{IJ}$ can be seen as the coefficients of an element $b$ in $H^4(BG_1;\Lambda_\mathbb{R})$, where $G_1$ is the connected component of the identity of $G$. The constraints of \cite{Monnier:2017oqd} read
\be
\frac{1}{2} b \in H^4_{\rm free}(BG_1;\mathbb{Z}) \otimes \Lambda \subset H^4(BG_1;\Lambda_\mathbb{R})\;.
\ee
They imply  
\be
\label{ConstrGaugAnCoeff}
b_i, \frac{1}{2} b_{II}, b_{IJ} \in \Lambda \; ,
\ee
but are generally stronger. See Section 3.3 of \cite{Monnier:2017oqd} for the detailed argument. These constraints include the ones derived in \cite{KMT2} from considerations of global gauge anomaly cancellation. 

An interesting point is that the appearance in the construction of $\check{Y}$ of torsion characteristic classes (see Appendix \ref{AppDiffCocFor6dSugra}) suggests that $b$ should be pictured as an element of $H^4(BG;\mathbb{Z}) \otimes \Lambda \simeq H^4(BG;\Lambda)$ rather than $H^4_{\rm free}(BG_1;\mathbb{Z}) \otimes \Lambda$. (Indeed, \cite{Monnier:2017oqd} focused on the case where $G$ is connected, in which case $H^4(BG;\Lambda) \simeq H^4(BG_1;\Lambda) \simeq H^4_{\rm free}(BG_1;\mathbb{Z}) \otimes \Lambda$.) The generalized condition on the anomaly coefficients therefore reads 
\be
\frac{1}{2} b \in H^4(BG;\Lambda)\;,
\ee
showing at the same time that the fundamental object encoding the gauge anomaly coefficients is $\frac{1}{2} b$ and not $b$. (The factor $\frac{1}{2}$ now matters as 2-torsion may be present in $H^4(BG;\Lambda)$.)

As we explained above, the consistency of the Green-Schwarz terms' construction requires that $a$ is a characteristic element of $\Lambda$. This property is automatically satisfied in F-theory constructions of 6d supergravity theories (see Section 4.1 of \cite{Monnier:2017oqd} for the argument), but up to now it was unclear how it should arise from the point of view of the 6d supergravity. In Section 5 of \cite{Monnier:2017oqd}, we gave an example of a 6d supergravity theory that looks completely consistent, except for the fact that $a$ is not characteristic. It consists of a single tensor multiplet, a string lattice $\Lambda = \mathbb{Z}^2$ with pairing
\be
\begin{pmatrix}
0&1\\1&0
\end{pmatrix}
\ee
with $a =(4,1)$, no gauge symmetry and $244$ neutral hypermultiplets. This theory cannot be realized in F-theory since $a=(4,1)$ is 
not a characteristic element of this lattice. Therefore, the Green-Schwarz term cannot be constructed using the methods of the present paper. This suggests that the supergravity above is inconsistent.

\subsection{Summary}

We summarize the discussion above with two propositions. As already mentioned, a tacit assumption is that string defects are included wherever they are necessary to satisfy the tadpole condition, and that their worldsheet anomalies cancel the boundary contributions to the anomaly of the supergravity theory through the anomaly inflow mechanism.

\begin{proposition}
\label{PropCanAnomZeroBord}
Let $\mathcal{S}$ be a 6-dimensional supergravity theory with gauge group $G$, string charge lattice $\Lambda$ and anomaly coefficients $(a,b, b_T = 0)$, where $b_T$ is the torsion anomaly coefficient in \eqref{EqCharYGen}. Let $A_8$ be the degree 8 anomaly polynomial of the theory and $Y$ the degree 4 form \eqref{EqRefEffAbGaugFFT}. Assume that:
\begin{enumerate}
\item $A_8 = \frac{1}{2} Y \wedge Y$;
\item $\Lambda$ is unimodular;
\item $b \in 2 H^4(BG_1;\Lambda)$;
\item $a \in \Lambda$ is a characteristic element;
\item $\Omega^{\rm Spin}_7(BG) = 0$.
\end{enumerate} 
Then all anomalies of $\mathcal{S}$, local and global, cancel. 
\end{proposition}

\begin{proposition}
Consider the same assumptions as in Proposition \ref{PropCanAnomZeroBord} except for 5. and allowing $b_T \neq 0$. Write $\mathcal{A}$ for the anomaly field theory of $\mathcal{S}$.  Then:
\begin{enumerate}
\item There is a 7-dimensional topological field theory 
\be
\mathcal{T} := \mathcal{A} \otimes {\rm WCS}^{\rm s}
\ee 
which reduces on 7-manifolds to a homomorphism $\mathcal{T}_7: \Omega^{\rm Spin}_7(BG) \rightarrow U(1)$. 
\item The anomalies of $\mathcal{S}$ cancel if and only if $\mathcal{T}_7$ is the trivial homomorphism.
\end{enumerate}
\end{proposition}

\subsection*{Acknowledgements}

We would like to thank Daniel Park for discussions that led to this project. We also thank Dan Freed, Mike Hopkins, Graeme Segal,
Wati Taylor, Andrew Turner, Nathan Seiberg and Edward Witten for useful discussions. G.M. is supported by the DOE under grant DOE-SC0010008 to Rutgers University. S.M. is supported in part by the grant MODFLAT of the European Research Council, SNSF grants No. 152812, 165666, and by NCCR SwissMAP, funded by the Swiss National Science Foundation.

\appendix

\section{Spin structures and Wu structures}

\label{SecSpinWu}

Recall that a spin structure can be defined as follows. The second Stiefel-Whitney class $w_2$ can be seen as a homotopy class of maps from the classifying space of the $n$-dimensional special orthogonal group $BSO(n)$ into the Eilenberg-MacLane space $K(2,\mathbb{Z}_2)$. $BSpin(n)$ can be constructed as the homotopy fiber of this map. (See for instance \cite{Yuan2014}.) A spin structure on a manifold $M$ is then a lift of the classifying map of the tangent bundle from $BSO(n)$ up to $BSpin(n)$.

Wu structures are defined completely analogously, see for instance Appendix C of \cite{Monnier:2016jlo}. We specialize here to the degree 4 case of interest to us. The degree 4 Wu class on a $n$-dimensional oriented manifold is $\nu = w_4 + w_2^2$, and can accordingly be seen as a homotopy class of maps from $BSO(n)$ into $K(4,\mathbb{Z}_2)$. $B_{\rm W}SO(n)$ is defined as the homotopy fiber of this map. $B_{\rm W}SO(n)$ is the classifying space of $n$-dimensional oriented bundles endowed with a degree 4 Wu structure. A Wu structure on $M$ is a lift of the classifying map of the tangent bundle from  $BSO(n)$ up to $B_{\rm W}SO(n)$.

Manifolds of dimension $n$ strictly lower than 8 always admit Wu structures of degree 4. Indeed, $\nu$ is defined by $\nu \cup x = {\rm Sq}^4(x)$, where $x$ has degree $n-4$. As ${\rm Sq}^p$ vanishes on classes of degree strictly smaller than $p$, $\nu$ vanishes on manifolds of dimension strictly smaller than $8$, so those manifolds admit Wu structures. When they exist, Wu structures of degree 4 on $M$ are classified by $H^3(M;\mathbb{Z}_2)$.

It will be useful to us to pick a particular cocycle representative $\nu_{\rm U}$ of $\nu$ on $BSO(n)$, which can be pulled back to $BSpin(n)$ and $B_{\rm W}SO(n)$. On $B_{\rm W}SO(n)$, $\nu = 0$ by definition so $\nu_{\rm U}$ is trivializable. We pick such a trivialization $\eta_{\rm U}$. The pull back of $\eta_{\rm U}$ to the manifold $M$ encodes the Wu structure on $M$. In addition, we lift the $\mathbb{Z}_2$-valued cochain $\eta_{\rm U}$ to a $\mathbb{Z}$-valued cochain $\eta_{\mathbb{Z}, {\rm U}}$, and then define $\nu_{\mathbb{Z}, {\rm U}} := d\eta_{\mathbb{Z}, {\rm U}}$. We make these choices for each $n$, in a way compatible with the maps $BSO(n) \rightarrow BSO(n+1)$, $BSpin(n) \rightarrow BSpin(n+1)$, $B_{\rm W}SO((n) \rightarrow B_{\rm W}SO((n+1)$.

We can define in a completely similar way the classifying space $B_{\rm W}Spin(n)$ of spin manifolds endowed with a Wu structure, which is the homotopy fiber of the map from $BSpin(n)$ to $K(4,\mathbb{Z}_2)$ defined by the Wu class. There is obviously a map $B_{\rm W}Spin(n) \rightarrow B_{\rm W}SO(n)$, corresponding to forgetting the spin structure, and the universal cochains $\eta_{\rm U}$, $\nu_{\rm U}$, $\eta_{\mathbb{Z}, {\rm U}}$ and $\nu_{\mathbb{Z}, {\rm U}}$ defined above pull back to $B_{\rm W}Spin(n)$. To simplify the notation, we will denote these pullbacks by the same letters.

\section{Differential cocycles from characteristic classes}

\label{AppDiffCocCharCl}

In this appendix, we explain how to associate a differential cocycle to a characteristic class of a principal bundle with connection, modulo certain universal choices on classifying spaces. This discussion is adapted from \cite{Freed2005} and generalized to accommodate non-integral characteristic classes. Unlike in the main text, we are careful about distinguishing cocycles from the associated cohomology classes in the present appendix: cocycles carry a hat, while cohomology classes do not.

\subsection{Generalities}

\paragraph{Differential cocycle on the classifying space} Let us fix a compact Lie group $\bar{G}$ with Lie algebra $\bar{\mathfrak{g}}$. As in the main text, $\Lambda$ is a lattice and $\Lambda_\mathbb{R} := \Lambda \otimes \mathbb{R}$ is the associated vector space. Let $y_{\rm U} \in H^{2p}(B\bar{G};\Lambda_\IR)$ be a $\Lambda_\IR$-valued characteristic class of $\bar{G}$ and let $\rho$ be the associated $\Lambda_{\mathbb{R}}$-valued invariant polynomial on the Lie algebra $\bar{\mathfrak{g}}$. (The subscript U is used henceforth for "universal" quantities defined on the classifying space.) We also fix a connection $\theta_{\rm U}$ on a differentiable model of $E\bar{G}$. We write $Y_{\rm U} := \rho(\theta_{\rm U})$ for the Chern-Weil characteristic form obtained by applying $\rho$ to $\theta_{\rm U}$. $Y_{\rm U}$ refines the cohomology class $y_{\rm U}$ to a differential form representative.

By choosing a real cocycle $\hat{y}_{\rm U}$ representing the class $y_{\rm U}$, as well as a real cocycle $\hat{A}_{\rm U}$ satisfying $d\hat{A}_{\rm U} = Y_{\rm U} - \hat{y}_{\rm U}$, we obtain a differential cocycle
\be
\label{EqDiffRefCharClass}
\check{Y}_{\rm U} = (\hat{y}_{\rm U}, \hat{A}_{\rm U}, Y_{\rm U})
\ee
refining further $Y_{\rm U}$. Importantly, the differential cohomology class of $\check{Y}_{\rm U}$ does depend on the choice of cocycle $\hat{y}_{\rm U}$. $\check{Y}_{\rm U}$ is a differential cocycle shifted by $\hat{y}_{\rm U}$ mod 1, and is an unshifted differential cocycle only if $\hat{y}_{\rm U}$ is an integral cocycle.

\paragraph{Gauge data on a manifold} Given a manifold $M$, the gauge data on $M$ consists of a principal $\bar{G}$-bundle $\bar{P}$, a connection $\theta$ on $\bar{P}$ and a classifying map $\gamma: \bar{P} \rightarrow E\bar{G}$. The gauge equivalences are the isomorphisms of principal $\bar{G}$-bundles preserving the connections (but not necessarily preserving the classifying maps). The gauge equivalence classes coincide with the gauge equivalence classes of the more familiar model where the gauge data is given only by the pair $(\bar{P},\theta)$. We write $\rho(\theta)$ for the Chern-Weil characteristic form constructed from $\theta$ and $\rho$.

\paragraph{Relative Chern-Simons term} Let $\bar{P} \rightarrow M$ be a principal $\bar{G}$-bundle over a manifold $M$. Recall that the relative Chern-Simons form $\tau_\rho(\theta_1, \theta_2)$ between two connections $\theta_1$ and $\theta_2$ on $\bar{P}$ is defined as follows. A linear path from $\theta_1$ to $\theta_2$ defines a connection $\Theta$ on the principal bundle $\bar{P} \times [0,1] \rightarrow M \times [0,1]$, and
\be
\label{EqDefTauRho}
\tau_\rho(\theta_2, \theta_1) := \int_{ [0,1]} \rho(\Theta) \;.
\ee
As $\rho(\Theta)$ is closed, the Stokes theorem implies
\be
\tau_\rho(\theta_3, \theta_2) + \tau_\rho(\theta_2, \theta_1) = \tau_\rho(\theta_3, \theta_1) \;.
\ee

\paragraph{Differential cocycle associated to the gauge data} We can now associate to any triple $(\bar{P}, \theta, \gamma)$ a differential cocycle on $M$
\be
\label{EqIndDiffCocFromCharCl}
\check{Y} = \check{Y}(\theta, \gamma) = (\bar{\gamma}^\ast \hat{y}_{\rm U}, \tau_\rho(\theta, \gamma^\ast(\theta_{\rm U})) + \bar{\gamma}^\ast \hat{A}_{\rm U}, \rho(\theta)) \;.
\ee
Here, $\bar{\gamma}$ is the map from $M$ into $B\bar{G}$ induced from $\gamma$.

\subsection{The case of 6-dimensional supergravity}

\label{AppDiffCocFor6dSugra}

In the case of interest to us, $\bar{G} = Spin(n) \times G$, where $G$ is the gauge vectormultiplet group of the 6d supergravity theory, and $n = 6,7,8$ depending on the manifold we are interested in. $\bar{P}$ is the product of the spin lift of the frame bundle of spacetime with the gauge bundle $P$. $\theta$ is the connection on $\bar{P}$ given by the product of the gauge connection with the Levi-Civita connection determined by the Riemannian metric on spacetime. 

For $n = 8$, $\gamma$ is a classifying map into $B\bar{G} = BSpin(d) \times BG$. However, recall that in order to define the shifted Wu Chern-Simons theory, we need a choice of Wu class on manifolds of dimension $6$ and $7$, although eventually nothing depends this choice. Therefore, for $n=6,7$, $\gamma$ is a classifying map to $B_{\rm W}\bar{G} := B_{\rm W}Spin(d) \times BG$. This lift is also necessary in order to be able to pullback to $M$ the cochain $\eta_{\rm U}$ trivializing the Wu cocycle $\nu_{\rm U}$, as the latter is non-trivial on $BSpin(d)$. Of course, any such classifying map $\gamma$ also determines a classifying map to $B\bar{G}$, which we write $\gamma$ as well for simplicity.

Our aim is to associate to the data $(P,\theta, \gamma)$ a differential cocycle $\check{Y}$ whose field strength $Y$ coincides with the expression \eqref{EqRefEffAbGaugFFT} factoring the local anomaly polynomial. The construction of the previous section is exactly what we need.

\paragraph{Naive construction} Recall that $Spin(d)$ has an integral characteristic class $\lambda_{BSpin}$ whose associated characteristic form coincides with half the first Pontryagin form. In addition, $G$ has integral characteristic classes $c_{2,BG}^i$, $c_{1,BG}^I$ corresponding to the second and first Chern classes of the elementary factors in the decomposition \eqref{EqGlobFormGaugeGroup}. These classes pullback to $B_{\rm W}\bar{G}$, and we denote the pullbacks with the same symbols. A natural choice for the characteristic $y_{\rm U}$ would be
\be
\label{NaiveChoiceCharY}
\frac{1}{2}a \lambda_{BSpin} - \sum_i b_i c_{2,BG}^i + \frac{1}{2} \sum_{IJ} b_{IJ} c_{1,BG}^I c_{1,BG}^J \;,
\ee
where $a, b_i, b_{IJ} \in \Lambda_\IR$ are the anomaly coefficients of the supergravity theory and we see the resulting characteristic class as $\Lambda_{\mathbb{R}}$-valued. The associated Chern-Weil form on $M$ is given by \eqref{EqRefEffAbGaugFFT}:
\be
\label{ChernWeilFormY}
Y = \rho(\theta) = \frac{1}{4} a p_1 - \sum_i b_i c_2^i + \frac{1}{2} \sum_{IJ} b_{IJ} c_1^I c_1^J \,,
\ee
as required.

We now need to construct the differential refinement \eqref{EqDiffRefCharClass}. We assume that a $\mathbb{Z}_2$-valued cocycle representative $\hat{\nu}$ of the degree 4 Wu class on $BO(d)$ has been chosen, see Section \ref{SecSpinWu}. $\lambda_{BSpin}$ lifts $w_4 = \nu$, so we pick an integral cocycle representative $\hat{\lambda}_{BSpin}$ lifting $\hat{\nu}$. We also pick integral cocycle representatives $\hat{c}_{2,BG}^i$, $\hat{c}_{1,BG}^I$. All of these cocycles obviously pull back to $B_{\rm W}\bar{G}$. We then set
\be
\label{EqCharNaiveY}
\hat{y}_{\rm U} := \frac{1}{2}a \hat{\lambda}_{BSpin} - \sum_i b_i \hat{c}_{2,BG}^i + \frac{1}{2} \sum_{IJ} b_{IJ} \hat{c}_{1,BG}^I \hat{c}_{1,BG}^J \;.
\ee
We pick a universal connection $\theta_{\rm U}$ on $E_{\rm W}\bar{G}$, the total space of the universal bundle over $B_{\rm W}\bar{G}$, and we set $Y_{\rm U} := \rho(\theta_{\rm U})$. We then fix $A_{\rm U}$ in \eqref{EqDiffRefCharClass} to be an arbitrary solution to the flatness constraint $dA_{\rm U} = Y_{\rm U} - y_{\rm U}$. By the arguments above, we now obtain for each triplet $(\bar{P}, \theta, \gamma)$ a differential cocycle $\check{Y}$ on $M$ whose curvature is $Y$. By construction, $\check{Y}$ is a differential cocycle shifted by
\be
\frac{1}{2} \bar{\gamma}^\ast(\hat{\lambda}_{BSpin}) \otimes a = \frac{1}{2}  \bar{\gamma}^\ast(\hat{\nu}) \otimes a \quad \mbox{ mod } \Lambda \;.
\ee

\paragraph{Problem with the naive construction} We show in Section \ref{SecAnCanThDisAbGG} that for some finite gauge groups $G$, the Green-Schwarz terms cannot cancel all possible global anomalies, and therefore global anomaly cancellation imposes constraints on the matter content. If $\check{Y}$ is constructed as above, these constraints are violated by theories obtained from F-theory compactifications, suggesting that either these F-theory compactifications are somehow inconsistent, or that the naive construction of $\check{Y}$ above is incorrect.

\paragraph{A generalized construction of $\check{Y}$}

We should remember at this point that the lift from $Y$ to $\check{Y}$ involves picking a cocycle $\hat{y}_{\rm U}$. Adding to $\hat{y}_{\rm U}$ in \eqref{EqCharNaiveY} a cocycle representing a degree 4 torsion characteristic class on $B_{\rm W}\bar{G}$ preserves the fact that $\check{Y}$ lifts $Y$. So we should generalize \eqref{EqCharNaiveY} to
\be
\label{EqCharYGen}
\hat{y}_{\rm U} := \frac{1}{2}a \hat{\lambda}_{BSpin} - \sum_i b_i \hat{c}_{2,BG}^i + \frac{1}{2} \sum_{IJ} b_{IJ} \hat{c}_{1,BG}^I \hat{c}_{1,BG}^J + \hat{t}_{4,BG} \;,
\ee
\be
\hat{t}_{4,BG} = \sum_k b^T_k \hat{t}^k_{4,BG} \;,
\ee
where $\hat{t}_{4,BG}$ represents a class in $H^4_{\rm tors}(BG;\Lambda)$. On the second line, $\hat{t}^k_{4,BG}$ are cocycles representing degree 4 $\mathbb{Z}$-valued torsion classes on $BG$, and $b^T_k \in \Lambda$ are new anomaly coefficients. Determining the relevant torsion class $\hat{t}_{4,BG}$ in any given theory would require to compare explicitly the shifted WCS theory to the anomaly field theory and adjusting it to cancel anomalies, something we are not currently able to do.

It is also hard to characterize in full generality the available choices for $\hat{t}_{4,BG}$ without specifying $G$. However, it turns out that for any group $G$, there is a natural torsion characteristic class of degree four. We construct it below and denote it by $u_{2,BG}^2$. When $G$ is connected this class vanishes. Adding (the pullback of) a representing cocycle $\hat{u}_{2,BG}^2$ to $\hat{y}_{\rm U}$ restores the compatibility with F-theory in the examples we inspected, as we discuss in Section \ref{SecAnCanThDisAbGG}.

We now define the characteristic class $u_{2,BG}^2$ and the new form of $\hat{y}_{\rm U}$. Suppose first that $G \simeq \mathbb{Z}_n$. As discussed in Appendix \ref{SecSpinCobDisAbGrp}, $B\mathbb{Z}_n$ can be pictured as an infinite dimensional lens space. Its integral cohomology is $\mathbb{Z}_n$ in even degree (except in degree 0) and 0 in odd degree. As a ring, it is generated by a class $u_{2,B\mathbb{Z}_n}$ in degree $2$.

For any compact Lie group $G$, there is a map $G \rightarrow (G/G_1)^{\rm ab}$ onto the Abelianized group of components of $G$.
(Recall that $G_1$ is the connected component of the identity element of $G$.)  As $G$ is assumed to be compact, $(G/G_1)^{\rm ab}$ is a finite group, and therefore a direct sum of cyclic groups. There is therefore a degree 2 torsion class $u_{2, BG}$ on $BG$ obtained by pulling back from $B(G/G_1)^{\rm ab}$ the sum of the classes $u_{2,B\mathbb{Z}_n}$ for each cyclic component $\mathbb{Z}_n$ of $(G/G_1)^{\rm ab}$.
This is a universal choice available for all compact Lie groups $G$. 

We pick a universal cocycle representative $\hat{u}_{2, BG}$ and take  $\hat{t}_{4,BG} = b_T \hat{u}_{2, BG}^2$, $b_T \in \Lambda$ in \eqref{EqCharYGen}:
\be
\label{EqCharY}
\hat{y}_{\rm U} := \frac{1}{2}a \hat{\lambda}_{BSpin} - \sum_i b_i \hat{c}_{2,BG}^i + \frac{1}{2} \sum_{IJ} b_{IJ} \hat{c}_{1,BG}^I \hat{c}_{1,BG}^J + b_T \hat{u}_{2, BG}^2 \;.
\ee
An interesting point to note is that the gauge groups for which we have been able to prove the cancellation of all anomalies through the vanishing of the corresponding bordism group (see Appendix \ref{AppCompCobGroup}) are all connected, and therefore have a vanishing $u_{2,BG}$.

We then proceed as before. We choose $\hat{A}_{\rm U}$ such that $\check{Y}_{\rm U} := (\hat{y}_{\rm U}, \hat{A}_{\rm U}, Y_{\rm U})$ is a differential cocycle. Then given any manifold $M$ endowed with the gauge data $(P,\theta,\gamma)$, we obtain a differential cocycle $\check{Y}$ on $M$ given by \eqref{EqIndDiffCocFromCharCl} and whose field strength coincides with $Y$.

\paragraph{The gauge transformations of $\check{Y}$} The construction above allows us to characterize the transformation of $\check{Y}$, $\check{H}$ and $\check{\eta}$, as defined in Section \ref{SecModel}, under a change of the gauge data $(\bar{P}, \theta, \gamma)$. In order to study the most general transformation of the gauge data, it is best to decompose it into two transformations:
\begin{enumerate}
\item a transformation given by an automorphism $f$ of $\bar{P}$, under which
\be
\theta \rightarrow f^\ast(\theta) \;,
\ee
together with a covariant change of the classifying map: $\gamma \rightarrow \gamma \circ f$;
\item a change of the classifying map $\gamma$. (Of course, $\gamma$ has to stay a classifying map, so in particular its homotopy class cannot change.)
\end{enumerate}
The natural transformation to make is a combination of the two types of transformations above, pulling back the connection by an automorphism while keeping the classifying map constant. The transformation of $\check{Y}$, $\check{H}$ and $\check{\eta}$ under such transformations are easily deduced from their transformations under the two types of elementary transformations above.

Under the first transformation, we have
\be
\label{EqTransYHetaPullback}
\check{Y} \mapsto \bar{f}^\ast \check{Y} \;, \quad \check{H} \mapsto \bar{f}^\ast \check{H} \;, \quad \check{\eta} \mapsto  \bar{f}^\ast(\check{\eta}) \;,
\ee
which is a transformation of the first type listed in Section \ref{SecModel}. 

The effect of a change of classifying map is slightly more tricky to analyse. It will be useful to define $\eta_{\Lambda, {\rm U}} := \eta_{\rm U} \otimes a$, $\check{\eta}_{\rm U} = (\eta_{\Lambda, {\rm U}}, 0,0)$. 
Then $\check{X}_{\rm U} := \check{Y}_{\rm U} - \frac{1}{2}d\check{\eta}_{\rm U}$ is an unshifted differential cocycle on $B_{\rm W}\bar{G}$, and the pullback of $\check{X}_{\rm U}$ through the classifying map is the differential cocycle \eqref{EqDefDiffCocX} on $M$. 
We write in components $\check{X}_{\rm U} = (\hat{x}_{\rm U}, \hat{C}_{\rm U}, X_{\rm U})$, with 
\be
\hat{x}_{\rm U} = \hat{y}_{\rm U} - \frac{1}{2}d\eta_{\Lambda, {\rm U}} \;, \quad \hat{C}_{\rm U} = \hat{A}_{\rm U} + \frac{1}{2}\eta_{\Lambda, {\rm U}} \;, \quad X_{\rm U} = Y_{\rm U} \;.
\ee 

Under a change of classifying map $\gamma \rightarrow \gamma'$, we have
\be
\label{EqGaugeTransX}
\check{X} \mapsto \check{X} + ((\bar{\gamma}'^{\ast} - \bar{\gamma}^\ast)\hat{x}_{\rm U}, \tau_\rho(\gamma^\ast(\theta_{\rm U}), \gamma'^\ast(\theta_{\rm U})) + (\bar{\gamma}'^{\ast} - \bar{\gamma}^\ast)\hat{C}_{\rm U}, 0) \;.
\ee
As $\gamma'$ and $\gamma$ are homotopic by hypothesis, $\Delta \hat{x} := (\bar{\gamma}'^{\ast} - \bar{\gamma}^\ast)\hat{x}_{\rm U}$ is exact. 
We show below that integrating $\Delta \hat{C} := \tau_\rho(\gamma^\ast(\theta_{\rm U}), \gamma'^\ast(\theta_{\rm U})) + (\bar{\gamma}'^{\ast} - \bar{\gamma}^\ast)\hat{C}_{\rm U}$ on any closed cycle $\Sigma \subset M$ yields an element of $\Lambda$, so we have $\Delta \hat{C} = - v - dV$, where $v$ and $V$ are respectively $\Lambda$-valued and $\Lambda_\mathbb{R}$-valued cochains. We also show that $d\Delta \hat{C} = -\Delta \hat{x} = dv$. This shows that \eqref{EqGaugeTransX} is a gauge transformation 
\be
\label{EqGaugeTransX2}
\check{X} \mapsto \check{X} + d\check{V} \;, \quad \check{V} = (v, V, 0) \;.
\ee
We also have $\check{\eta} \mapsto \check{\eta} + \check{\rho}$, with $\check{\rho} := (\bar{\gamma}'^\ast - \bar{\gamma}^\ast)\check{\eta}$. Therefore
\be
\label{EqGaugeTransY}
\check{Y} \mapsto \check{Y} + d\check{V} + \frac{1}{2} d\check{\rho} \;,
\ee
and we see that the transformation of $\check{Y}$ is a combination of a gauge transformation and a change of shift, according to the terminology of Section \ref{SecModel}. We now easily deduce the transformation of $\check{H}$ under a change of classifying map:
\be
\label{EqGaugeTransH}
\check{H} \mapsto \check{H} + \check{V} + \frac{1}{2} \check{\rho} \;,
\ee

Equation \eqref{EqTransYHetaPullback} and \eqref{EqGaugeTransH} determine the transformation of the $B$-field $\check{H}$ under diffeomorphisms and vectormultiplet gauge transformations.

\paragraph{Technical details} We now prove the two claims we used in deriving \eqref{EqGaugeTransX2}. By the definition \eqref{EqDefTauRho} of $\tau_\rho$, we have
\be
d\tau_\rho(\gamma^\ast(\theta_{\rm U}), \gamma'^\ast(\theta_{\rm U})) = \gamma^\ast(X_{\rm U}) - \gamma'^\ast(X_{\rm U}) \;.
\ee
Combining it with the second term in $d\Delta \hat{C}_{\rm U}$ and using the fact that $\check{X}_{\rm U}$ is a differential cocycle, we get
\be
d\Delta \hat{C}_{\rm U} = (\gamma'^\ast - \gamma^\ast)(d\hat{C}_{\rm U} - X_{\rm U}) = -(\gamma'^\ast - \gamma^\ast) \hat{x}_{\rm U} = -\Delta \hat{x}_{\rm U} \;.
\ee
To prove the second claim, we pick a homotopy $\Gamma$ from $\gamma$ to $\gamma'$, and see it as a map from $M \times I$ into the classifying space. Let $\Sigma \subset M$ be a degree 3 cycle. Using again the definition of $\tau_\rho$ and integration by parts, we can write
\be
\int_\Sigma \Delta \hat{C} = \int_{\Sigma \times I} \left( -\Gamma^\ast X_{\rm U} + d\Gamma^\ast \hat{C}_{\rm U} \right) = \int_{\Sigma \times I} \Gamma^\ast \hat{x}_{\rm U} \;.
\ee
The right-hand side is $\Lambda$-valued because $\hat{x}$ is a $\Lambda$-valued cocycle. 

\section{E-theory calculus}

\label{AppEThCalc}

\paragraph{Higher cup products} Let $M$ be an oriented manifold, possibly with boundary. One can associate to any homomorphism $\Xi_1 \times \Xi_2 \rightarrow \Xi_3$ of Abelian groups higher cousins of the cup product for each non-negative integer $i$:
\be
\cup_i: C^p(M; \Xi_1) \times C^q(M;\Xi_2) \rightarrow C^{p+q-i}(M;\Xi_3) \;.
\ee
The usual cup product is $\cup_0$, and formally $\cup_i = 0$ for $i < 0$. The higher cup products are 
defined in \cite{1947} and    satisfy (\cite{1947}, Theorem 5.1)
\be
\label{EqRelHighCupProd}
d(u \cup_i v) - du \cup_i v - (-1)^p u \cup_i dv = (-1)^{p+q-i} u \cup_{i-1} v + (-1)^{pq + p + q} v \cup_{i-1} u \;,
\ee
where $u$ and $v$ are respectively cochains of degree $p$ and $q$. \eqref{EqRelHighCupProd} equates the failure of $\cup_{i-1}$ to be graded symmetric to the failure of the Leibniz rule for the product $\cup_i$. In the present work, we will mostly be interested in the higher products of integer-valued cochains modulo 2, in which case the signs can be dropped.

\paragraph{The cochain model} A degree $p$ E-cochain on $M$ is a pair
\be
\bar{s} = (s,y) \in \bar{C}^p(M;\Lambda) := C^p(M;\mathbb{R}/\mathbb{Z}) \times C^{p-3}(M; \Lambda/2\Lambda) \;,
\ee
Note that this assumes that $\Lambda$ is unimodular. If not, $\Lambda/2\Lambda$ should be replaced by the group $\Gamma^{(2)}$, defined as the quotient of $\Lambda/2\Lambda$ by the radical of the induced $\mathbb{Z}/2\mathbb{Z}$-valued pairing \cite{Monnier:2016jlo}. The pairing on $\Lambda/2\Lambda$ induces a cup product 
\be
\cup: C^\bullet(M; \Lambda/2\Lambda) \otimes C^\bullet(M; \Lambda/2\Lambda) \rightarrow C^\bullet(M; \mathbb{Z}_2) \;.
\ee
We will often compose this cup product with the embedding $\frac{1}{2}: \mathbb{Z}_2 \rightarrow \mathbb{R}/\mathbb{Z}$. More concretely, given $y_1, y_2 \in C^\bullet(M; \Lambda/2\Lambda)$, we can lift them to $\Lambda$-valued cochain, perform the cup product of $\Lambda$-valued cochain to obtain a $\mathbb{Z}$-valued cochain, see it as a real-valued cochain, divide it by 2 and reduce it modulo $\mathbb{Z}$ to obtain
\be
\frac{1}{2} y_1 \cup y_2 \in C^\bullet(M;\mathbb{R}/\mathbb{Z}) \;.
\ee
The same construction can be applied to the higher cup products.

We now define a non-commutative "addition" on E-cochains by
\be
\label{EqGrpLawCocModETh}
(s_1, y_1) \boxplus (s_2, y_2) = \left(s_1 + s_2 + \frac{1}{2} dy_1 \cup_{p-5} y_2 + \frac{1}{2} y_1 \cup_{p-6} y_2, y_1 + y_2 \right) \;.
\ee
The opposite of $(s, y)$ is
\be
\label{EqGrpLawCocModEThInv}
\boxminus (s, y) = \left(-s + \frac{1}{2} dy \cup_{p-5} y + \frac{1}{2} y \cup_{p-6} y, y  \right) \;.
\ee
We also define a differential
\be
\label{EqDefTwDiffEThCoch}
d(s, y) = \left(ds + y \cup_{p-6} dy + \frac{1}{2} y \cup_{p-7} y + \frac{1}{2} y \cup \hat\nu_{\Lambda/2\Lambda}, dy \right) \;.
\ee
$\hat \nu_{\Lambda/2\Lambda}$ is the cocyle $\hat{\nu} \otimes \gamma$, where $\hat{\nu}$ is the $\mathbb{Z}_2$-valued Wu cocycle on $M$ pulled back from the classifying space, and $\gamma$ is the unique characteristic element of the pairing on $\Lambda/2\Lambda$, i.e. such that $(x,x) = (x,\gamma)$ \cite{Monnier:2016jlo}. It is useful to note that in the main text, we lifted $\hat{\nu}$ to an integral cocycle and tensored it with a characteristic element of $\Lambda$ to obtain a $\Lambda$-valued cocycle $\hat{\nu}_\Lambda$. The reduction to $\Lambda/2\Lambda$ of any such $\hat{\nu}_\Lambda$ necessarily coincides with $\hat \nu_{\Lambda/2\Lambda}$.

In Appendix D of \cite{Monnier:2016jlo}, it was shown that:
\begin{enumerate}
\item $\boxplus$ is associative and forms a group law;
\item $d$ is distributive with respect to $\boxplus$;
\item $d^2 = 0$;
\item Exact cochains form a normal subgroup of the cochain group;
\item The quotient of the degree $p$ closed cochains by the exact cochains is an Abelian group $E[\Lambda/2\Lambda,3]^p(M)$.
\item $E[\Lambda/2\Lambda,3]^\bullet$ is a generalized cohomology theory, fitting into the following long exact sequence:
\begin{align}
\label{EqLongExSeqETh}
...H^p(M;\mathbb{R}/\mathbb{Z}) & \stackrel{i}{\rightarrow} E[\Lambda/2\Lambda,3]^p(M) \stackrel{j}{\rightarrow} H^{p-3}(M;\Lambda/2\Lambda) \\
& \stackrel{{\rm Sq}^{4}}{\rightarrow} H^{p+1} (M;\mathbb{R}/\mathbb{Z}) \stackrel{i}{\rightarrow} E[\Lambda/2\Lambda,3]^{p+1}(M) \stackrel{j}{\rightarrow} H^{p-2}(M;\Lambda/2\Lambda)... \notag
\end{align}
where $i$ and $j$ are the maps induced from the inclusion into the first component and the projection onto the second component at the level of cocycles. $E[\Lambda/2\Lambda,3]^\bullet$ is a natural generalization of a certain generalized cohomology, named E-theory \cite{2005math......4524J, Freed:2006mx}. For this reason, we will also call "E-theory" the generalized cohomology theory $E[\Lambda/2\Lambda,3]^\bullet$ in the present work. See \cite{Monnier:2016jlo} for a discussion of the relations between these generalized cohomology theories.
\end{enumerate}

\paragraph{Integration} If $U$ is a $p$-dimensional manifold endowed with a degree 4 Wu structure $\omega$, with $\partial U = M$, there are integration maps over $U$ and $M$ \cite{Monnier:2016jlo}, given respectively by group homomorphisms 
\be
I^{\rm E}_{U,\omega}: E[\Lambda/2\Lambda,3]^p(U,\partial U) \rightarrow \mathbb{R}/\mathbb{Z} 
\ee 
\be
I^{\rm E}_{M,\omega_M}: E[\Lambda/2\Lambda,3]^{p-1}(M) \rightarrow \mathbb{R}/\mathbb{Z}  \;,
\ee 
where $\omega_M$ is the Wu structure induced on $M$. These integration maps are canonical up to universal choices on classifying spaces. They lift to integration maps on the space of relative cocycles on $U$ and of cocycles on $M$.

For our purpose however, we need to extend them to functions on the cochain groups
\be
\int^{\rm E}_{U,\omega}: \bar{C}^p(U;\Lambda) \rightarrow \mathbb{R}/\mathbb{Z}
\ee
\be
\int^{\rm E}_{M,\omega}: \bar{C}^{p-1}(M;\Lambda) \rightarrow \mathbb{R}/\mathbb{Z}
\ee
This extension is analogous to the choice of a particular cycle representative of the fundamental homology class in ordinary cohomology, and necessarily involves some arbitrariness.

This extension is possible and described in Appendix D of \cite{Monnier:2016jlo}. We record here the following properties of the integration map.
\begin{enumerate}
\item While $\int^{\rm E}_{M,\omega}$ is a group homomorphism, $\int^{\rm E}_{U,\omega}$ is not. This detail is not important to us, as one can show that the homomorphism property holds on E-cocycles, and in the present paper we only integrate cochains on $M$ and cocycles on $U$ (with $p = 7$). Therefore, for all practical purposes, we will consider $\int^{\rm E}_{U,\omega}$ to be a group homomorphism as well. 

\item There is a relation akin to Stokes' theorem between $\int^{\rm E}_{U,\omega}$ and $\int^{\rm E}_{M,\omega}$. Given an E-cochain $\bar{x}$ on $M$, extending to an E-cochain $\bar{x}'$ on $U$, we have
\be
\int^{\rm E}_{M,\omega} \bar{x} = \int^{\rm E}_{U,\omega} d\bar{x}'\;,
\ee
where $d$ is the differential \eqref{EqDefTwDiffEThCoch}.

\item $\int^{\rm E}_{U,\omega}$ is a group homomorphism on E-cochains of the form $(s,0)$. It determines therefore a cycle representative of the fundamental $\mathbb{R}/\mathbb{Z}$-valued homology class of $U$. This cycle can be use to define an integration $\int_U$ on $\mathbb{R}/\mathbb{Z}$-valued cochains on $U$ and we have
\be
\int^{\rm E}_{U,\omega} (s,0) = \int_U s \;.
\ee

\item An interesting fact is that we always have  
\be
2 \int^{\rm E}_{U,\omega} (s,y) = \int^{\rm E}_{U,\omega} (s,y) \boxplus (s,y) = \int_U 2s + \int_U \frac{1}{2} y \cup \hat{\nu}_{\Lambda/2\Lambda} \quad \mbox{mod } 1 \;,
\ee
so up to an ordinary integral $\int_U \frac{1}{2} y \cup \hat{\nu}_{\Lambda/2\Lambda}$, all the subtleties in the E-theory integration translate into signs (after exponentiation).

\item Define $f(y) := \int^E_{U,\omega} (0,y)$ for $y \in C^{p-3}(M;\Lambda/2\Lambda)$. We have
\be
\int^E_{U,\omega} (s,y) =  \int_U s  + f(y)
\ee
If $y_1, y_2 \in C^{p-3}(M;\Lambda/2\Lambda)$ is closed, use the homomorphism property of $\int^E_{U,\omega}$ to check that
\be
f(y_1 + y_2) - f(y_1) - f(y_2) =  \int_U y_1 \cup_{p-6} y_2
\ee
so $f$ is a quadratic refinement of the pairing defined by the right-hand side. This property still holds when $y_1$ and $y_2$ are arbitrary cochains, but the pairing of the right-hand side then has additional terms.
\end{enumerate}
Properties 3-5 also hold for $\int^{\rm E}_{M,\omega_M}$.

\section{Proof of the gluing axioms of the shifted Wu Chern-Simons theory}

\label{AppGluing}

In this appendix, we prove that the Wu Chern-Simons theory ${\rm WCS}$ is a field theory functor. Its domain is the bordism category $\mathcal{C}_{\rm WCS}$ defined in Section \ref{SecCobCat}.

${\rm WCS}$ is multiplicative on disjoint unions, and transforms by complex conjugation under changes of orientation. To show that ${\rm WCS}$ is a field theory functor, we only need to show that the gluing axioms hold (see for instance Proposition 5.1 of \cite{Monnier:2016jlo} for a proof of this claim). The latter are formulated as follows. Let $U$ be a 7-manifold, possibly with boundary, $M$ a codimension 1 closed submanifold disjoint from the boundary, and $U_M$ the manifold obtained by cutting $U$ along $M$. There is a surjective gluing map $g: U_M \rightarrow U$ identifying the two boundary components of $U_M$ created by the cut. Let $\check{X}_U$ be a $\Lambda$-valued differential cocycle on $U$, $\check{X}_{U_M} := g^\ast(\check{X}_U)$ and $\check{X}_M := \check{X}_U|_M$. $\check{X}_M$, by the assumption that $(M,\check{X}_M)$ is an object of $\mathcal{C}_{\rm WCS}$, is a trivializable differential cocycle. By the same assumption, the Wu structure induced on $M$ is good. The gluing axioms state that there is a canonical isomorphism
\be
\label{EqGlueingAxioms}
{\rm Tr}_{{\rm WCS}(M;\check{X}_M)} {\rm WCS}(U_M; \check{X}_{U_M}) \simeq {\rm WCS}(U;\check{X}_U)\;.
\ee
The trace should be understood as follows. Write $\check{X}_{\partial U} := \check{X}_U|_{\partial U}$ and $\check{X}_{\partial U_M} := \check{X}_{U_M}|_{\partial U_M}$. ${\rm WCS}(U;\check{X}_U)$ is a vector in ${\rm WCS}(\partial U;\check{X}_{\partial U})$, while ${\rm WCS}(U_M; \check{X}_{U_M})$ is a vector in
\be
{\rm WCS}(\partial U_M; \check{X}_{\partial U_M}) \simeq {\rm WCS}(\partial U; \check{X}_{\partial U}) \otimes {\rm WCS}(M; \check{X}_M) \otimes ({\rm WCS}(M; \check{X}_M))^\dagger \;,
\ee
where we use the decomposition $\partial U_M \simeq \partial U \sqcup M \sqcup -M$ and $()^\dagger$ denotes complex conjugation. ${\rm Tr}_{{\rm WCS}(M;\check{X}_M)}$ is simply the canonical pairing between ${\rm WCS}(M; \check{X}_M)$ and $({\rm WCS}(M; \check{X}_M))^\dagger$.

Let us prove \eqref{EqGlueingAxioms}. We first remark that the partition functions of ${\rm WCS}$ always have norm 1. As the trace is taken in a 1-dimensional state space, the norm of each side are both equal to 1. It remains to study the phase.

Using the property of the trace with respect to tensor products, we can write the left hand side of \eqref{EqGlueingAxioms}
\be
{\rm Tr}_{{\rm WCS}^{\rm PQ}(M; \check{X}_M)} {\rm WCS}^{\rm PQ}(U_M; \check{X}_{U_M}) \otimes \sum_{z \in H^4_{\rm tors}(U_M,\partial U_M;\Lambda)} {\rm Tr}_{{\rm WCS}^{\rm PQ}(M; \check{0})} {\rm WCS}^{\rm PQ}(U_M; \check{Z}_{U_M}) \;,
\ee
where $\check{Z}_{U_M}$ is any differential cocycle representative of $z$ vanishing on $M$, and $\check{0}$ is the zero differential cocycle on $M$. We can use the fact that the prequantum theory satisfies the gluing relation to obtain
\be
\label{EqProofGlueingAxioms1}
{\rm WCS}^{\rm PQ}(U; \check{X}_{U}) \otimes \sum_{z \in H^4_{\rm tors}(U_M,\partial U_M;\Lambda)} {\rm WCS}^{\rm PQ}(U; \check{Z}_{U}) \;,
\ee
where $\check{Z}_U$ is the differential cocycle obtained by pushing forward $\check{Z}_{U_M}$ through the gluing map. (This is possible because $\check{Z}_{U_M}$ vanishes on $M$.) Now we need to replace the sum over $H^4_{\rm tors}(U_M,\partial U_M;\Lambda)$ by a sum over $H^4_{\rm tors}(U,\partial U;\Lambda)$ to obtain the right hand side of \eqref{EqGlueingAxioms}.

Let us write $g_\ast: H^4_{\rm tors}(U_M,\partial U_M;\Lambda) \rightarrow H^4_{\rm tors}(U,\partial U;\Lambda)$ for the pushforward through the gluing map, at the level of torsion cohomology. The value of the action on $\check{Z}_U$ and $\check{Z}_{U_M}$ coincide. We can therefore replace the sum in \eqref{EqProofGlueingAxioms1} by a sum over ${\rm im}(g_\ast)$, up to a prefactor given by the order of ${\rm ker}(g_\ast)$. We do not care about this prefactor because we already showed that \eqref{EqGlueingAxioms} holds in norm.

To see that the sum over ${\rm im}(g_\ast)$ can be replaced by a sum over $H^4_{\rm tors}(U,\partial U;\Lambda)$, we need to understand a bit better the structure of ${\rm im}(g_\ast)$. We remark that we have $H^4_{\rm tors}(M \times I, \partial(M \times I); \Lambda) \simeq H^3_{\rm tors}(M; \Lambda)$. We have therefore a homomorphism
\be
H^3_{\rm tors}(M; \Lambda) \stackrel{h}{\rightarrow} H^4_{\rm tors}(U,\partial U;\Lambda) \;,
\ee
obtained by identifying a tubular neighborhood of $M$ with a cylinder $M \times I$. The classes in the image of $h$ are represented by cocycles supported in the tubular neighborhood. This makes it clear that ${\rm im}(h)$ is an isotropic subgroup of $H^4_{\rm tors}(U,\partial U;\Lambda)$ with respect to the linking pairing. Recall that $q$ is the quadratic refinement of the linking pairing such that $q(x) = S(\check{X})$ for any flat differential coycle $\check{X}$ lifting the torsion class $x$. The isotropy of ${\rm im}(h)$ implies that $q$ restricts to a character of ${\rm im}(h)$. On the other hand, any class in ${\rm im}(g_\ast)$ can be represented by a cocycle vanishing in a tubular neighborhood of $M \subset U$. We deduce that ${\rm im}(h)$ and ${\rm im}(g_\ast)$ are orthogonal with respect to the linking pairing $\tilde{L}$. Moreover, any class in $H^4_{\rm tors}(U,\partial U;\Lambda)$ that does not belong to ${\rm im}(g_\ast)$ is represented in the tubular neighborhood of $M$ by a non-trivial cocycle pulled back from $M$, and cannot be orthogonal to ${\rm im}(h)$.

By the very definition of the torsion anomaly and the fact that it vanishes, we also have that $q(y) = 0$ for any $y \in {\rm im}(h)$. Now if $x \in H^4_{\rm tors}(U,\partial U;\Lambda)$, then
\be
q(x + y) - q(x) = \tilde{L}(x,y)\;,
\ee
and $q(x + y) - q(x)$ is a character that is non-trivial whenever $x \notin {\rm im}(g_\ast)$. This shows that the sum in \eqref{EqProofGlueingAxioms1} can be replaced by a sum over $H^4_{\rm tors}(U,\partial U;\Lambda)$, and therefore proves the gluing axioms for ${\rm WCS}$.

\section{Computations of certain bordism groups}

\label{AppCompCobGroup}

We identified the anomaly field theory of the bare 6d supergravity theory as the shifted Wu Chern-Simons theory only up to a bordism invariant. The anomaly field theory coincides with the shifted Wu Chern-Simons theory if the relevant bordism group vanishes. In this Appendix, we compute the relevant cobordism group for certain compact groups $G$ that might show up as vectormultiplet gauge groups in 6d supergravity theories. Note that very similar computations appeared in the recent paper \cite{Garcia-Etxebarria:2018ajm}, which contains also more details about the Atiyah-Hirzebruch spectral sequence.

The bordism group of interest is $\Omega^{\rm spin}_7(BG)$, where $G$ is the gauge group of the 6d supergravity theory. $G$ is a priori any compact Lie group. We compute this bordism group using the Atiyah-Hirzebruch spectral sequence (AHSS):
\be
E^2_{p,q} = H_p(BG, \Omega^{\rm spin}_q({\rm pt.})) \Rightarrow \Omega^{\rm Spin}_{p+q}(BG) \;.
\ee
Recall also that the spin bordism group of the point reads
\be
\Omega^{\rm spin}_\bullet({\rm pt.}) = \left( \begin{array}{ccccccccc}
0 & 1 & 2 & 3 & 4 & 5 & 6 & 7 & 8\\
\mathbb{Z} & \mathbb{Z}_2 & \mathbb{Z}_2 & 0 & \mathbb{Z} & 0 & 0 & 0 & \mathbb{Z}^2
\end{array} ... \right) \;.
\ee
Surprisingly, the integral homology of $BG$ is far from being known for all compact Lie groups. The computations below cover all the cases for which it is known, as far as we are aware of.

\subsection{$U(1)$}

The integral homology of $BU(1) = \mathbb{C}P^\infty$ is $\mathbb{Z}$ in even degrees and zero in odd degrees. The second page of the AHSS is
\renewcommand\arraystretch{1.2}
\be
\label{EqSecPagAHSSU1}
\begin{array}{c|ccccccccc}
8 & \mathbb{Z}^2 & 0 & \mathbb{Z}^2 & 0 & \mathbb{Z}^2 & 0 & \mathbb{Z}^2 & 0 & \mathbb{Z}^2\\
7 & 0 & 0 & 0 & 0 & 0 & 0 & 0 & 0 & 0\\
6 & 0 & 0 & 0 & 0 & 0 & 0 & 0 & 0 & 0\\
5 & 0 & 0 & 0 & 0 & 0 & 0 & 0 & 0 & 0\\
4 & \mathbb{Z} & 0 & \mathbb{Z} & 0 & \mathbb{Z} & 0 & \mathbb{Z} & 0 & \mathbb{Z}\\
3 & 0 & 0 & 0 & 0 & 0 & 0 & 0 & 0 & 0\\
2 & \mathbb{Z}_2 & 0 & \mathbb{Z}_2 & 0 & \mathbb{Z}_2 & 0 & \mathbb{Z}_2 & 0 & \mathbb{Z}_2\\
1 & \mathbb{Z}_2 & 0 & \mathbb{Z}_2 & 0 & \mathbb{Z}_2 & 0 & \mathbb{Z}_2 & 0 & \mathbb{Z}_2\\
0 & \mathbb{Z} & 0 & \mathbb{Z} & 0 & \mathbb{Z} & 0 & \mathbb{Z} & 0 & \mathbb{Z}\\
\hline q/p  & 0 & 1 & 2 & 3 & 4 & 5 & 6 & 7 & 8
\end{array}
\ee

The only non-vanishing term relevant to the computation of $\Omega^{\rm Spin}_{7}(BU(1))$ is $E^2_{6,1} = \mathbb{Z}_2$, generated by the dual of $\rho_2(c_1^3) = w_2^3$. $\rho_2$ is here the reduction mod 2. (Note that the characteristic classes are characteristic classes on $BU(1)$, or equivalently characteristic classes of the $U(1)$ bundle. They have nothing to do with the characteristic classes of the spacetime, so in particular, $w_2 \neq 0$.)

There is a sequence
\be
\btkz
E^2_{8,0} \arrow[r, "d^2_{8,0}"] & E^2_{6,1} \arrow[r, "d^2_{6,1}"] & E^2_{4,2}
\etkz
\ee
where $E^2_{8,0} = \mathbb{Z}$, generated by the dual of $c_1^4$, and $E^2_{4,2} = \mathbb{Z}_2$, generated by the dual of $\rho_2(c_1^2)$. The second differential $d^2_{p,q}$ of the AHSS coincides at $q = 0$ and $q = 1$ with the dual of the second Steenrod square composed with reduction mod 2 and with the dual of the second Steenrod square, respectively \cite{Zhubr2001}:
\footnote{We use the perfect pairing of homology and cohomology with $\mathbb{Z}_2$ coefficients to 
define the dual.}
\be
d^2_{p,0} = ({\rm Sq}^2)^\ast \circ \rho_2 \;, \quad d^2_{p,1} = ({\rm Sq}^2)^\ast \;.
\ee
Equivalently, $(d^2_{p,0})^\ast = \epsilon \circ {\rm Sq}^2$, where $\epsilon : H^p(BU(1);\mathbb{Z}_2) \rightarrow {\rm Hom}(H_p(BU(1),\mathbb{Z}), \mathbb{Z}_2)$ is given by the evaluation of representing cocycles on representing cycles.

We can now compute, using $\rho_2(c_1) = w_2$ and the known action of the Steenrod squares on the Stiefel-Whitney classes:
\be
{\rm Sq}^2(w_2^2) = 2w_2^3 + w_3^2 = 0
\ee
which means that $d^2_{6,1} = 0$. We also have
\be
{\rm Sq}^2(w_2^3) = w_2^4 = \rho_2(c_1^4) \;.
\ee
$c_1^4$ generates a non-trivial $\mathbb{Z}_2$ character of $H_8(BU(1),\mathbb{Z})$, so $(d^2_{8,0})^\ast \neq 0$. This means that $d^2_{8,0} \neq 0$, hence is surjective, meaning that $E^2_{6,1}$ is killed on the second page.

We deduce that
\be
\Omega^{\rm spin}_7(BU(1)) = 0 \;.
\ee

\subsection{$U(2)$ and $SU(2)$}

The integral cohomology of $BU(2)$ is the freely generated ring on the generators $c_1$ in degree 2 and $c_2$ in degree 4 (the first two Chern classes) \cite{Brown1982}. We therefore have the following homology
\be
\label{EqHomFIntCoeffU2}
H_\bullet(BU(2);\mathbb{Z}) = \left( \begin{array}{ccccccccccc}
0 & 1 & 2 & 3 & 4 & 5 & 6 & 7 & 8\\
\mathbb{Z} & 0 & \mathbb{Z} & 0 & \mathbb{Z}^2 & 0 & \mathbb{Z}^2  &  0 & \mathbb{Z}^3\\
1 & - & c_1^\ast & - & (c_1^2)^\ast, c_2^\ast & 0 & (c_1^3)^\ast, (c_1c_2)^\ast & - & (c_1^4)^\ast, (c_1^2c_2)^\ast, (c_2^2)^\ast
\end{array} ... \right)
\ee
where the first line is the degree, the second line the homology groups and the third line the (additive) generators, expressed as the duals of products of Chern classes.

The second page of the AHSS looks therefore very similar to \eqref{EqSecPagAHSSU1}, except that the groups are squared for $p = 4,6$, and cubed for $p = 8$. We see that the only possible contribution to the 7-dimensional bordism group is again from $E^2_{6,1} = \mathbb{Z}_2^2$, generated by the duals of $w_2^3$ and of $\rho_2(c_2 c_1) = w_4 w_2$. We already know that $w_2^3$ gets killed, and we compute
\be
{\rm Sq}^2(w_4) = w_2 w_4 + w_6 = w_2 w_4 \;,
\ee
because $w_6 = \rho_2(c_3)$ vanishes on $BU(2)$. The dual of $w_4 w_2$ is therefore not in the kernel of $d_{6,1}$, and gets killed on the second page. Therefore the bordism group vanishes:
\be
\Omega^{\rm spin}_7(BU(2)) = 0 \;.
\ee

Note that in the $SU(2)$ case, $c_1 = 0$, so $E^2_{6,1} = 0$ showing that $\Omega^{\rm spin}_7(BSU(2)) = 0$.

\subsection{$U(n)$ and $SU(n)$, $n \geq 3$}

The integral cohomology ring of $BU(n)$ is obtained from the one of $BU(2)$ by adding a generator $c_k$ in degree $2k$, $k \leq n$, the higher Chern classes \cite{Brown1982}. Using $\rho_2(c_k) = w_{2k}$, we have
\be
E^2_{6,1} = {\rm span}_{\mathbb{Z}_2} \left((w_2^3)^\ast, (w_2 w_4)^\ast, (w_6)^\ast \right)
\ee
We now have
\be
{\rm Sq}^2(w_2^3) = w_2^4 = \rho_2(c_1^4)
\ee
\be
{\rm Sq}^2(w_2 w_4) = w_2 w_6 = \rho_2(c_1 c_3)
\ee
\be
{\rm Sq}^2(w_6) = w_2 w_6 = \rho_2(c_1 c_3)
\ee
\be
\label{EqSq2w4}
{\rm Sq}^2(w_4) = w_2 w_4 + w_6 \;,
\ee
Again we see that $d^2_{8,0}$ has image generated by $(w_2^3)^\ast$ and $(w_2 w_4)^\ast + (w_6)^\ast$. However, we also have $d^2_{6,1}((w_6)^\ast) = d^2_{6,1}((w_2 w_4)^\ast) = (w_4)^\ast + ... \neq 0$, so the bordism group vanishes:
\be
\Omega^{\rm spin}_7(BU(n)) = 0 \;.
\ee

In the $SU(n)$ case, $c_1$, and therefore $w_2$ vanish. $E^2_{6,1}$ is generated by $(w_6)^\ast$, but as before $d^2_{6,1}((w_6)^\ast) \neq 0$, so we have again $\Omega^{\rm spin}_7(BSU(n)) = 0$.

\subsection{$Sp(n)$}

The cohomology of $BSp(n)$ is described in \cite{Mimura1991}, p.137. It is generated as a ring by the symplectic Pontryagin classes $q_i \in H^{4i}(BSp(n);\mathbb{Z})$, $i = 1,...,n$.

From this structure, we can directly see that there is no obstruction on the second page of the AHSS, and we readily have
\be
\Omega^{\rm spin}_7(BSp(n)) = 0 \;.
\ee

\subsection{Arbitrary products of $U(n)$, $SU(n)$ and $Sp(n)$ factors}

As the cohomologies/homologies of $BU(n)$ and $BSp(n)$ have no torsion, the K\"unneth formula shows that the arguments above can be applied factor by factor, hence that the bordism group of 7-manifold endowed with an arbitrary product of $U(n)$, $SU(n)$ and $Sp(n)$ factors vanishes.

\subsection{$E_8$}

$BE_8$ and $K(\mathbb{Z},4)$ are homotopically equivalent in degrees less than 16, so $\Omega^{\rm spin}_7(BE_8) \simeq \Omega^{\rm spin}_7(K(\mathbb{Z},4))$. The homology of $K(\mathbb{Z},4)$ is given by
\cite{Breen2016}
\be
H_\bullet(K(\mathbb{Z},4);\mathbb{Z}) = \left( \begin{array}{ccccccccc}
0 & 1 & 2 & 3 & 4 & 5 & 6 & 7 & 8\\
\mathbb{Z} & 0 & 0 & 0 & \mathbb{Z} & 0 & \mathbb{Z}_2 & 0 & \mathbb{Z}
\end{array} ... \right) \;.
\ee
\be
H_\bullet(K(\mathbb{Z},4);\mathbb{Z}_2) = \left( \begin{array}{ccccccccc}
0 & 1 & 2 & 3 & 4 & 5 & 6 & 7 & 8\\
\mathbb{Z}_2 & 0 & 0 & 0 & \mathbb{Z}_2 & 0 & \mathbb{Z}_2 & \mathbb{Z}_2 & \mathbb{Z}_2
\end{array} ... \right) \;.
\ee
We write $\iota$ for the generator of $H^4(K(\mathbb{Z},4);\mathbb{Z}) \simeq \mathbb{Z}$. The second page of the AHSS is
\be
\label{EqSecPagAHSSE8}
\begin{array}{c|ccccccccc}
8 & \mathbb{Z}^2 & 0 & 0 & 0 & \mathbb{Z}^2 & 0 & 0 & \mathbb{Z}_2^2 & \mathbb{Z}^2\\
7 & 0 & 0 & 0 & 0 & 0 & 0 & 0 & 0 & 0\\
6 & 0 & 0 & 0 & 0 & 0 & 0 & 0 & 0 & 0\\
5 & 0 & 0 & 0 & 0 & 0 & 0 & 0 & 0 & 0\\
4 & \mathbb{Z} & 0 & 0 & 0 & \mathbb{Z} & 0 & \mathbb{Z}_2 & 0 & \mathbb{Z}\\
3 & 0 & 0 & 0 & 0 & 0 & 0 & 0 & 0 & 0\\
2 & \mathbb{Z}_2 & 0 & 0 & 0 & \mathbb{Z}_2 & 0 & \mathbb{Z}_2 & \mathbb{Z}_2 & \mathbb{Z}_2\\
1 & \mathbb{Z}_2 & 0 & 0 & 0 & \mathbb{Z}_2 & 0 & \mathbb{Z}_2 & \mathbb{Z}_2 & \mathbb{Z}_2\\
0 & \mathbb{Z} & 0 & 0 & 0 & \mathbb{Z} & 0 & \mathbb{Z}_2 & 0 & \mathbb{Z}\\
\hline q/p  & 0 & 1 & 2 & 3 & 4 & 5 & 6 & 7 & 8
\end{array}
\ee
There is a single potential obstruction in $E^2_{6,1} \simeq H_6(K(\mathbb{Z},4);\mathbb{Z}_2)$, with generator the dual of ${\rm Sq}^2 \iota$. This generator is therefore not in the kernel of $d^2_{6,1}$ and is killed by the spectral sequence. We deduce that
\be
\Omega^{\rm spin}_7(BE_8) = 0 \;.
\ee
Note that $\Omega^{\rm spin}_p(BE_8)$, $p = 1,...,11$ was computed in Stong's appendix to \cite{Witten:1985bt}, but the present derivation of $\Omega^{\rm spin}_7(BE_8) $ is more straightforward.

\subsection{Finite Abelian groups}

\label{SecSpinCobDisAbGrp}

Consider now the case where $G$ is a finite Abelian group. Any such group is a product of $\mathbb{Z}_n$ factors, so we simply focus here on the case $G = \mathbb{Z}_n$. 

$B\mathbb{Z}_n$ is the Eilenberg-MacLane space $K(\mathbb{Z}_n,1)$, which can be seen as an infinite dimensional lens space.
Its integral cohomology is generated as a ring by a single element of order $n$ in degree 2. Its integral homology has a generator of order $n$ in each odd degree. The homology with coefficients in $\mathbb{Z}_2$ is zero if $n$ is odd or $\mathbb{Z}_2$ in odd degrees if $n$ is even.

If $n$ is odd, it is easy to see that no cancellation can occur in the AHSS and we conclude that $\Omega^{\rm spin}_7(B\mathbb{Z}_n) \;, n \mbox{ odd.}$ is an extension of $\mathbb{Z}_n$ by $\mathbb{Z}_n$ (so in particular is non-zero).

If $n$ is even, the second page of the AHSS is:
\be
\label{EqSecPagAHSDiscr}
\begin{array}{c|ccccccccc}
8 & \mathbb{Z}^2 & \mathbb{Z}_n^2 & 0 & \mathbb{Z}_n^2 & 0 & \mathbb{Z}_n^2 & 0 & \mathbb{Z}_n^2 & 0\\
7 & 0 & 0 & 0 & 0 & 0 & 0 & 0 & 0 & 0\\
6 & 0 & 0 & 0 & 0 & 0 & 0 & 0 & 0 & 0\\
5 & 0 & 0 & 0 & 0 & 0 & 0 & 0 & 0 & 0\\
4 & \mathbb{Z} & \mathbb{Z}_n & 0 & \mathbb{Z}_n & 0 & \mathbb{Z}_n & 0 & \mathbb{Z}_n & 0 \\
3 & 0 & 0 & 0 & 0 & 0 & 0 & 0 & 0 & 0\\
2 & \mathbb{Z}_2 & \mathbb{Z}_2 & \mathbb{Z}_2 & \mathbb{Z}_2 & \mathbb{Z}_2 & \mathbb{Z}_2 & \mathbb{Z}_2 & \mathbb{Z}_2 & \mathbb{Z}_2\\
1 & \mathbb{Z}_2 & \mathbb{Z}_2 & \mathbb{Z}_2 & \mathbb{Z}_2 & \mathbb{Z}_2 & \mathbb{Z}_2 & \mathbb{Z}_2 & \mathbb{Z}_2 & \mathbb{Z}_2\\
0 & \mathbb{Z} & \mathbb{Z}_n & 0 & \mathbb{Z}_n & 0 & \mathbb{Z}_n & 0 & \mathbb{Z}_n & 0 \\
\hline q/p  & 0 & 1 & 2 & 3 & 4 & 5 & 6 & 7 & 8
\end{array}
\ee
Writing $u_1$ for the degree 1 generator of the $\mathbb{Z}_2$-valued cohomology and $u_n = (u_1)^n$ for the generator in degree $n$, we have ${\rm Sq}^k(u_n) = \binom{n}{k} u_{n+k}$, where $\binom{n}{k}$ are the mod 2 binomial coefficients \cite{Smirnov2006}.

We have potential contributions from $E^2_{7,0}$, $E^2_{6,1}$, $E^2_{5,2}$ and $E^2_{3,4}$. They all survive to the third page. $E^2_{7,0}$ can potentially get killed on the third page, while $E^2_{3,4}$ could get killed on the fourth page. $E^2_{6,1}$, $E^2_{5,2}$ survive through the whole spectral sequence.

We immediately see that ${\rm Sq}^2(u_5) = 0$, so $d_{7,0}: E^2_{7,0} \rightarrow E^2_{5,1}$ vanishes. $E^2_{7,0}$ therefore survives the second page. We deduce that $\Omega^{\rm Spin}_7(B\mathbb{Z}_n)$ is non-zero as well when $n$ is even, although we cannot compute it exactly.

We note that $\Omega^{\rm spin}_7(B\mathbb{Z}_2) = \mathbb{Z}_{16}$ has been computed using different methods in \cite{Kapustin:2014dxa} (see Table 1 there).

\subsection{Summary}

When the gauge group $G$ is an arbitrary product of $SU(n)$, $Sp(n)$ and $U(1)$ factors, the bordism group of 7-dimensional spin manifolds endowed with a principal $G$-bundle vanishes. The same is true for $E_8$.

The AHSS computation is inconclusive for $SO(3)$ and it only gets worse for higher rank orthogonal groups. We have not been able to find a computation of the integral homology of $BG$ for the other compact simple Lie groups in the literature.

For Abelian finite groups, we found non-vanishing spin bordism groups. This suggests way of constructing non-trivial bordism classes for disconnected Lie groups: find an embedding of $\mathbb{Z}_n \subset G$ that passes to a non-trivial homomorphism into the group of components $G/G_0$. We can push forward $\Omega^{\rm Spin}_7(B\mathbb{Z}_n)$ into $\Omega^{\rm Spin}_7(BG)$ and potentially get non-trivial bordism classes. For instance,  when $n$ is odd,  using the standard embedding of $\mathbb{Z}_2 \subset O(n)$ with image $ {\rm diag}(\pm 1, 1,...,1)$, $u_1$ is sent to $w_1$ and we can construct non-trivial bordism classes in $\Omega^{\rm Spin}_7(BO(n))$ involving non-orientable bundles.

\section{Eta invariants for finite Abelian gauge groups}

\label{AppNonTrivGlobAn}

Eta invariants associated to the covering $S^{2k-1} \rightarrow S^{2k+1}/G$ for $G = \mathbb{Z}_n$ can be computed explicitly (\cite{gilkey1999spectral}, Theorem 1.8.5). In the present section, we describe this computation, which is used in Section \ref{SecAnCanThDisAbGG} to compute the partition function of the relative anomaly field theory associated to two different matter representations. Analogous computations appeared recently in  \cite{Garcia-Etxebarria:2018ajm}.

Let $G = \mathbb{Z}_n$ and let $\tau: G \rightarrow U(k)$ be a complex representation of $G$ inducing a free action of $G$ on the unit sphere $S^{2k-1} \subset \mathbb{C}^k$. Let $U := S^{2k-1}/\tau(G)$ be the corresponding Lens space. $U$ comes endowed with a metric inherited from the round metric on the sphere. A square root $\sqrt{{\rm det}(\tau)}$ of the determinant representation of $\tau$ determines a spin structure on $U$. Pick a (virtual) representation $R$ of $G$. Consistent with our previous notation, $R$ should be thought of as the matter representation in the case of interest to us. We will write $R$ both for the representation vector space and for the representation map. $R$ determines a vector bundle $V$ over $U$ by quotienting $S^{2k-1} \times R$ by the action of $G$. Then the modified eta invariant of the Dirac operator on $U$ coupled to $V$ is given by (\cite{gilkey1999spectral}, Theorem 1.8.5):
\be
\xi_R(U) = \frac{1}{|G|} \sum_{g \in G - \{1\}} {\rm Tr}(R(g)) \frac{\sqrt{{\rm det}(\tau(g))}}{{\rm det}(\tau(g) - I)}
\ee

We see $\mathbb{Z}_n$ as realized as the multiplicative group of $n^{th}$ roots of $1$. Let us write $\rho_s$ for the representation of $\mathbb{Z}_n$ in $\mathbb{C}$ sending $z \in \mathbb{Z}_n$ to $z^s$. We take $k = 4$ and
\be
\tau = \rho_1^{\oplus 4} \;,
\ee
so that
\be
\det(\tau(z)) = z^{4}
\ee
has a natural square root
\be
\sqrt{\det(\tau(z))} = z^{2} \;.
\ee
$U$ is therefore spin, with a given spin structure. Similarly,
\be
\det(\tau(z) - I) = (z-1)^4 \;,
\ee
so if we write $z = e^{ \frac{2\pi\I}{n} j } $  then
\be
\frac{\sqrt{\det(\tau(z))}}{\det(\tau(z) - I)} = \frac{1}{16 (\sin (\frac{\pi}{n} j))^4 } \;.
\ee
In the case of interest to us, the representation $R$ must be quaternionic, so that we must take $R$ to be a (virtual) direct sum of representations of the form
\be
R_s := \rho_s \oplus \rho_{-s} \;.
\ee
Taking the difference of two such representations, we obtain
\be
\xi_{R_{s_1} \ominus R_{s_2}}(U) = f(s_1) - f(s_2)
\ee
where
\be
f(s) = \frac{1}{8n} \sum_{j=1}^{n-1} \frac{\cos( \frac{2\pi}{n} js ) }{(\sin \frac{\pi}{n} j )^4 }
\ee
A contour integral argument using $g(z) = \cot(\pi n z) \frac{\cos(2\pi s z)}{(\sin \pi z)^4} $ shows that we can rewrite this expression as a polynomial
\be\label{eq:FunnyResult}
f(s) = \frac{1}{8\cdot 45\cdot n}  (-11 + 10 n^2 + n^4 - 60 n s + 60 s^2 - 30 n^2 s^2 + 60 n s^3 -
   30 s^4)
\ee
for $-1\leq s \leq n+1$. Note that 
\be
f(s+n) = f(s) - \frac{1}{3} s (s^2-1)
\ee
so $f(s) \mbox{ mod } \mathbb{Z}$ descends to a function on $ \mathbb{Z}_n$.  

The fact that we get a polynomial in $s$ can be derived from applying the APS index theorem to a suitable 
line bundle over a suitable disk bundle over $\mathbb{C}\mathbb{P}^3$.

{
\small

\providecommand{\href}[2]{#2}\begingroup\raggedright\endgroup

}

\end{document}